\documentclass[twocolumn, tighten, table]{aastex63}
\pdfoutput=1

\usepackage{amsmath,amstext}
\usepackage[T1]{fontenc}
\usepackage[figure, figure*]{hypcap}
\usepackage[tight,footnotesize]{subfigure}
\usepackage[toc,page]{appendix}

\graphicspath{{./}{./figures/}}

\accepted{February 26, 2021}

\reportnum{FERMILAB-PUB-20-337-E}
\begin{document}

\title{Hubble  Frontier Field Clusters and their Parallel Fields: Photometric and Photometric Redshift Catalogs}
\author{A. Pagul}
\shortauthors{A. Pagul et al.}
\affiliation{Department of Physics and Astronomy, University of California Riverside, Pierce Hall, Riverside, CA., USA}
\author{F. J. S\'{a}nchez}
\affiliation{Fermi National Accelerator Laboratory, Wilson Hall, Batavia, IL., USA}
\author{I. Davidzon}
\affiliation{The Cosmic Dawn Center, VIbenshuset, Lyngbyvej 2, Copenhagen, Denmark}
\author{Bahram Mobasher}
\affiliation{Department of Physics and Astronomy, University of California Riverside, Pierce Hall, Riverside, CA., USA}

\shorttitle{Photometric Data Catalogs for the Hubble Frontier Fields}

\begin{abstract}

We present a multi-band analysis of the six \textit{Hubble} Frontier Field  clusters and their parallel fields, producing catalogs with measurements of source photometry and photometric redshifts\footnote{Catalogs can be found \href{https://drive.google.com/drive/u/1/folders/17fbz4BAk1gZ2TuEppdQWsH2pyU7pBL_X}{here.}}. We release these catalogs to the public along with maps of intracluster light and models for the brightest galaxies in each field. This rich data set covers a wavelength range from 0.2 to 8\,$\mu m$, utilizing data from the \textit{Hubble} Space Telescope, Keck Observatories, Very Large Telescope array, and \textit{Spitzer} Space Telescope. We validate our products by injecting into our fields and recovering a population of synthetic objects with similar characteristics as in real extragalactic surveys. The photometric catalogs contain a total of over 32,000 entries with 50\% completeness at a threshold of $\mathrm{mag_{AB}}\sim 29.1$ for unblended sources, and  $\mathrm{mag_{AB}}\sim 29$ for blended ones, in the IR-Weighted detection band. Photometric redshifts were obtained by means of template fitting and have an average outlier fraction of 10.3\% and scatter $\sigma = 0.067$ when compared to spectroscopic estimates. The software we devised, after being tested in the present work, will be applied to new data sets from ongoing and future surveys. 

\end{abstract}

\section{Introduction}
\label{sec:introduction}
To develop a comprehensive picture of the primordial building blocks of the universe, identification and study of properties of the youngest galaxies ($< 1$ Gyr) soon after the Big Bang is essential. Given their enormous distance, these systems are best detected at near-infrared wavelengths. However, because of high sky background at these wavelengths, such observations need to be done from space- i.e. using the \textit{Hubble} Space Telescope (HST) or \textit{James Webb} Space Telescope (JWST). This also requires multi-waveband observations spanning a range of wavelengths from optical and near-infrared. Furthermore, study of galaxies at these redshifts is often biased as we mainly sample the intrinsically brighter populations. 

To accomplish the above, very deep multi-waveband observations over a large area are needed to detect statistically representative samples of very high redshift galaxies. This is currently beyond the reach of the largest available telescopes. To achieve such depths, one could leverage the natural phenomenon of gravitational lensing by targeting rich clusters of galaxies \citep[][and references thereon]{schneider84,blandford_narayan86} \citep[see][for a review]{kneib_natarajan11}. This magnifies fluxes from high redshift sources located behind massive galaxy clusters, helping to probe deeper into the universe. This will be highly beneficial for even after the largest space telescopes (i.e. JWST), with their unprecedented sensitivity, are commissioned. It has been demonstrated that JWST will be able to shed light on the mechanisms that re-ionized the intergalactic medium at $6 < z < 10$ only by exploiting the cluster lensing phenomenon, to observe galaxies with $M < 10^9 \ M_\sun$ \citep{atek18}.

The \textit{Hubble} Frontier Field program \citep[HFF,][]{lotz17} is a survey, designed with these objectives in mind. With an allocation of 630 HST orbits, it performed deep observations of six very massive clusters and their parallel fields in optical and near-IR bands, with Advanced Camera for Surveys (ACS) and Wide Field Camera 3 (WFC3), respectively. In addition, each field has been extensively observed in the mid-IR regime (between 3 and 5\,$\mu$m) with the Infrared Array Camera (IRAC) on board the \textit{Spitzer} Space Telescope. 

The six clusters span the redshift range $0.3 < z < 0.55$ and their magnification power  allows detection of galaxies up to $z\sim 9$, i.e.\ at the re-ionization epoch \citep[see][]{barkana01,miralda-escude03}. 
HFF followed the successful tradition of deep, pencil-beam HST observations as in \textit{Hubble} Deep Field \citep{williams95}, \textit{Hubble} Ultra-Deep Field \citep[HUDF,][]{2006AJ....132.1729B} and CANDELS \citep{Koekemoer:2011p12718}, as well as programs covering wider areas like the Cosmic Evolution Survey \citep[COSMOS,][]{Scoville:2007p12720} and Galaxy Evolution from Morphology and SED \citep[GEMS,][]{hanswalter04}. With respect to galaxy cluster observations, the Cluster Lensing And Supernovae survey with \textit{Hubble} \citep[CLASH,][]{postman12} paved the way for the HFF. 

The HFF clusters will be the reference fields for the exploration of the distant universe. At present, additional HST coverage is provided by the ``Beyond Ultra-deep Frontier Fields and Legacy Observations''  \citep[BUFFALO,][]{steinhardt20} which aims at covering the outskirts of the HFF clusters over the same wavebands and to the same depth, as well as increasing the depth in the centers of the clusters.  Most notably, sources in the HFF clusters will also be JWST targets, with some of them already selected by both Guarateed Time Observation teams\footnote{\href{https://jwst-docs.stsci.edu/jwst-opportunities-and-policies/jwst-cycle-1-guaranteed-time-observations-call-for-proposals/jwst-gto-observation-specifications}{as listed at this http URL.}} and an Early Release Science Program\footnote{Treu et al., program 1324 ``Through the Looking GLASS: A JWST Exploration of Galaxy Formation and Evolution from Cosmic Dawn to Present Day''.}. Eventually, the knowledge acquired from the HFF shall be applied to analyze data from an unprecedented number of galaxy clusters that \textit{Euclid} and \textit{Roman} space telescopes  will discover by surveying thousands of square degrees of the sky \citep{laureijs11,spergel15}.  


Apart from providing access to the high redshift Universe, the HFF will also allow the study of properties of dark matter  \citep[e.g.][and references therein]{2015MNRAS.452.1437J} and the role of `environment' in the evolution of galaxies at $z<1$  \citep[e.g.][]{2005MNRAS.363L..66G, 2018MNRAS.476.3631P}. It will also provide a standard sample of cluster galaxies, to be compared with nearby systems. The HFF parallel fields provide similar data to the same depth, that would minimize selection effects and biases in any environmental study of galaxy populations.    
The cornerstone of all these studies is a multi-waveband photometric catalog, containing self-consistent photometry for every detected source. Producing such a catalog is particularly challenging, given the depth of the imaging data, analysis of images in the crowded fields, the wide wavelength baseline (which imply substantial changes in the point spread function) required, and contamination from the intra-cluster light (ICL). 
The task has already been fulfilled by two distinct teams (AstroDeep\footnote{\url{http://www.astrodeep.eu/frontier-fields/}} and DeepSpace\footnote{\url{http://cosmos.phy.tufts.edu/~danilo/HFF/Home.html}}) which released their respective HFF catalogs to the astronomical community~\citep{2016A&A...590A..31C, 2017A&A...607A..30D, 2018ApJS..235...14S, 2019MNRAS.489...99B}. 
Given the challenges mentioned above in identifying individual sources, in performing accurate photometric measurements for the HFF galaxies, and in simultaneously dealing with a multiple parameter space, it is important to apply completely independent techniques to generate catalogs for the same clusters. This allows a better understanding of the final data products, selection effects and the caveats in the data and the data processing pipelines. In particular, the different data reduction steps could introduce systematic effects in the final results, which need to be studied. Such problems and inaccuracies in source identification and photometry will also be reflected into selection of different populations of galaxies. For example, comparing two independent estimates of photometric redshifts can be instrumental to selecting the most robust  galaxy candidates at $z>8$ \citep{2016MNRAS.459.3812M}. 
The existing catalogs produced from the above mentioned studies contain serious discrepancies in the photometry of the same galaxies. An independent analysis will help us study the origins of these differences.



In this study, we generate  new galaxy catalogs for the six HFF clusters and their parallel fields. The results include photometric catalogs, photometric redshifts, Intra-Cluster Light (ICL) maps, and surface brightness models to remove the brightest galaxies from our images in order to detect fainter galaxies to deeper levels. The parallel fields provide ancillary deep-field data and will also serve as control samples. The main differences between the strategy here and the previous methods is that, we provide models of the intra-cluster light maps, as well as perform forward modelling of photometry to characterize biases and uncertainties in flux measurements. Moreover, the pipelines developed for this study will be applied to the new BUFFALO data (Pagul et al., in prep.) to extend the HFF catalogs to wider areas, providing self-consistent data for the HFF and BUFFALO galaxies. 
%

 The paper is organized as follows. In Section 2 we introduce the data sets. Section 3 presents a detailed study of the data reduction, followed by source extraction in Section 4. In Section 5, we study the completeness and photometric uncertainties in HFF clusters, as well as include comparisons between our results and previous works. In Section 6 we present the photometric redshifts for the HFF measurements. In Section 7, we discuss the lensing measurements. Section 8 presents information about the released data products. Section 9 summarizes our results. 
 
 Throughout this paper we assume standard cosmology with $\Omega_M = 0.23$, $\Omega_\Lambda = 0.76$ and $H_0 = 73$ Km/sec/Mpc. Magnitudes are in the AB system.
 

\begin{figure*}
    \centering
    \includegraphics[width=\textwidth]{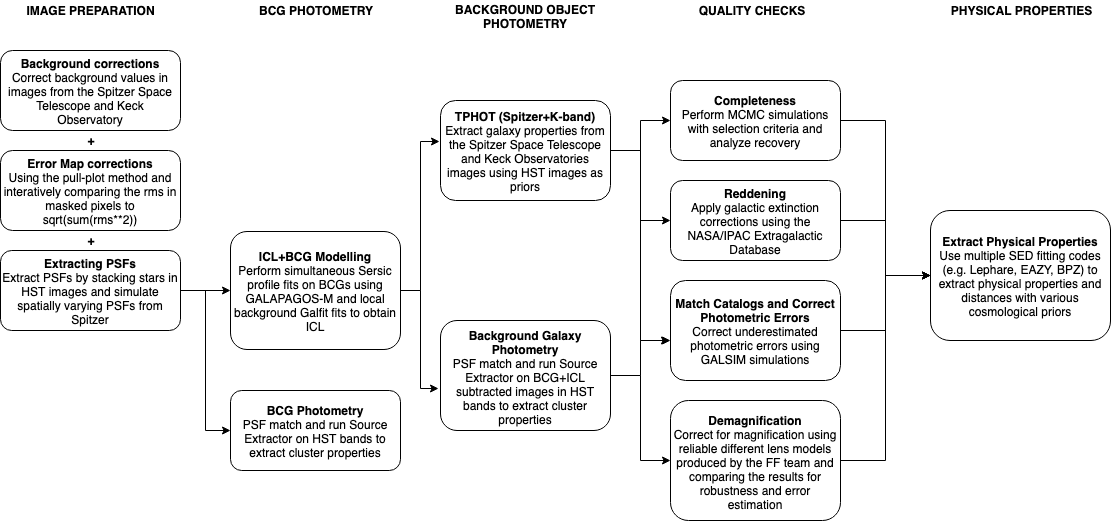}
    \caption{Workflow of the data processing as performed in the present work.}
    \label{fig:workflow}
\end{figure*}

\begin{deluxetable*}{lcccccc}
\tablecaption{\label{tab:surveyprop}
	Frontier Field cluster and parallel field positions, along with clusters' mean redshift ($z_\mathrm{clu}$), virial mass ($M_\mathrm{vir}$), and X-ray luminosity ($L_X$) \citep{lotz17}}

\tablehead{%
Field	        	& Cluster Center (J2000) &	Parallel Center (J2000)	& $z_\mathrm{clu}$  & $M_{vir}$ & $L_{X}$ }
\startdata
Abell 370           &  02:39:52.9, -01:34:36.5 & 02:40:13.4, -01:37:32.8 & 0.375 & $\sim 1\times10^{15}$ & $1.1\times10^{45}$\\
					&		    &   	&   	        	\\
MACS J0717.5+3745   &  07:17:34.0 +37:44:49.0 & 07:17:17.0 +37:49:47.3 & 0.545 & $\sim 2-3\times10^{15}$ & $3.3\times10^{45}$\\
					&		    &   	&   	        	\\
MACS J0416.1-2403	&  04:16:08.9, -24:04:28.7 & 04:16:33.1, -24:06:48.7   & 0.396 & $1.2 \times 10^{15}$ &  $1.0\times10^{45}$   &\\       	      	
					&	    	&   &	&     \\
Abell S1063		    & 22:48:44.4, -44:31:48.5 & 22:49:17.7, -44:32:43.8  & 0.348 & $1.4\times10^{15}$ & $1.8\times10^{45}$\\
   &		    &   	&   	        	\\
Abell 2744		    &  00:14:21.2, -30:23:50.1 & 00:13:53.6, -30:22:54.3  & 0.308  & $1.8 \times 10^{15}$  &  $3.1\times10^{45}$ \\
        			&	    	&   	&   			        	\\
MACS J1149.5+2223	&  11:49:36.3, +22:23:58.1 & 11:49:40.5, +22:18:02.3   & 0.543 & $2.5\times10^{15}$  &  $1.8\times10^{45}$	\\
	& & & & 
\enddata
\end{deluxetable*}
\section{The Data}
\label{sec:dataset}

\subsection{HST observations}
The HFF observations are the deepest of galaxy clusters to date, and second only to the HUDF when also considering the blank fields. This results from a total of 840 orbits, performed through coordinated HST parallel observations in the following filters:  F435W, F606W, F814W in ACS, and F105W, F125W, F140W, F160W in WFC3. These seven bands reached a depth of $m_{AB}\sim 29$ mag for point sources within a 0.4" diameter aperture \citep{lotz17}. Through a different program, UV observations in the F275W and F336W filters of two HFF clusters, Abell 2744 and MACS J0717, were carried out with details presented in ~\citep{2016ApJ...832...56A}. 

\begin{deluxetable}{lccc}[t]
\tablecaption{The Point Spread Function and effective wavelengths for different photometric bands used for the HFF.}
\label{tab:hstdata}
\tablehead{\colhead{Band} & \colhead{\hspace{.75cm}$FWHM$}\hspace{.5cm} & \colhead{$\lambda_{pivot}$ (\AA)}}
\startdata
F275W & 0.075" & 2707\\
F336W & 0.109" & 3355\\
F435W & 0.109" & 4329\\
F606W & 0.112" & 5922\\
F814W & 0.111" & 8045\\
F105W & 0.175" & 10551\\
F125W & 0.176" & 12486\\
F140W & 0.172" & 13923\\
F160W & 0.173" & 15369\\
Ks & 0.364" & 21524\\
I1 & 1.29" & 35634\\
I2 & 1.42" & 45110\\
I3 & 1.50" & 57593\\
I4 & 1.84" & 79595\\
\enddata
\tablecomments{Representative PSF FWHM values for the photometric bands in this study. These values were calculated for the Abell 2744 cluster. Bands F275W and F336W are only available for Abell 2744 and MACS J0717. IRAC Channels 3 and 4 are only available for Abell 2744, Abell 370, Abell S1063, and MACS J0717.}
\end{deluxetable}

The gravitational potential of the clusters' halo, besides binding together the galaxies in the system, produces a lensing magnification that could detect background objects to an \textit{intrinsic} brightness of 30--33 mag, i.e.\ 10--100 times fainter than previous surveys. Details of the HFF survey design are provided in \citet{lotz17}.  In Table \ref{tab:surveyprop} we report the main characteristics of the six clusters, with a summary of the observations in Table~\ref{tab:hstdata}. The filter throughput for the ACS and WFC3 instruments are shown in Figure~\ref{fig:filters} (upper panel). We use mosaics that have been reduced by the Frontier Fields team \footnote{https://archive.stsci.edu/prepds/frontier/}, with a pixel scale of 0.06"/pix. These images have been reduced using the HST science data products pipeline \citep{Koekemoer:2011p12718}. Data from other HST programs using the same
filters have also been included by the HFF team, with all the exposures aligned relative to each other using \texttt{Tweakreg}\footnote{part of the DrizzlePac software suite: \url{https://www.stsci.edu/scientific-community/software/drizzlepac.html}.}. Other steps in the reduction process include: correction of standard imaging artifacts; bad pixel and cosmic ray rejection; geometric distortion correction and image stacking using \texttt{AstroDrizzle} \citep{drizzlepac}. Throughout this work we will focus on the analysis and detection of these previously reduced and calibrated data.

\subsection{Ancillary data}
\label{subsec{ancillary}}

A number of independent observations of the HFF have generated complementary data to that available from the HST.  The \textit{Spitzer} Space Telescope dedicated more than 1,000 hours of Director's Discretionary time to obtain IRAC 3.6\,$\mu$m (channel 1) and 4.6\,$\mu$m (channel 2) imaging down to the depths of 26.5 and  26.0\,mag., in cluster and parallel fields respectively. These observations are crucial for photometric redshift measurement, for identifying low-redshift interlopers, and are beneficial in constraining galaxy properties since they provide a good proxy for galaxy stellar mass. In addition, there are legacy IRAC observations  at 5.8\,$\mu$m (channel 3) and 8.0\,$\mu$m  (channel 4) which are also included in our analysis\footnote{retrieved from the Spitzer Hertiage Archive \url{https://sha.ipac.caltech.edu/applications/Spitzer/SHA/}}. To produce the Frontier Fields mosaics, the following \textit{Spitzer} Program IDs were used:
\begin{itemize}
    \item {Abell 2744: 83, 90275}
    \item {MACS J0416.1-2403: 80168, 90258}
    \item {MACS J0717.4+3745: 40652, 60034, 90009, 90259}
    \item {MACS J1149.4+2223: 60034, 90009, 90260}
    \item {Abell S1063 (RXC J2248.7-4431): 83, 10170, 60034}
    \item{Abell 370: 137, 10171, 60034}
\end{itemize}


Another follow-up program is the $K_s$-band Imaging of the Frontier Fields \citep[KIFF][]{KIFF}, carried out with the High Acuity Wide Field K-band Imager (HAWK-I) at the Very Large Telescope (VLT). This reached a depth of 26.0\,mag (5$\sigma$, point-like sources) for Abell 2744, MACS-0416, Abell S1063, and Abell 370 clusters in the southern hemisphere. The K-band imaging campaign of the HFF also used the Multi-Object Spectrometer for Infrared Exploration (MOSFIRE) at Keck to observe MACS-0717 and MACS-1149 in the northern hemisphere to a 5$\sigma$ depth of 25.5 and 25.1\,mag respectively.

\begin{figure}
\includegraphics[width=0.5\textwidth]{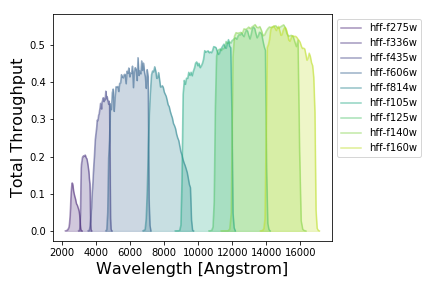}
\includegraphics[width=0.5\textwidth]{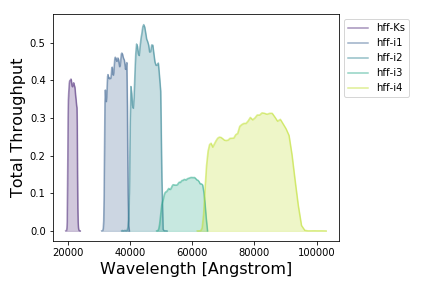}
\caption{Throughput of the nine HST filters (\textit{upper panel}) and ancillary Ks-band and IRAC bands (\textit{lower panel}) used in this analysis. }
\label{fig:filters}
\end{figure}

Table \ref{tab:otherdata} provides a summary of the available ancillary data,  with their filter transmission curves shown in Figure~\ref{fig:filters} (lower panel). 



\begin{deluxetable*}{lccccc}
\tablecaption{\label{tab:otherdata}
	Existing multi-wavelength HFF coverage from follow-up programs, as used in the present work. The 5-$\sigma$ point-source depth was estimated by integrating the noise in a Gaussian PSF aperture with the values of FWHM given in Table 2.}

\tablehead{%
Field	        	& Observatory	& Wavelengths	& Depth & Reference }
\startdata
Abell 370           & VLT/HAWK-I   & 2.2$\mu m$  & $\sim$ 26.18  &  \citet{KIFF}	\\
          & \textit{Spitzer} IRAC 1,2  & 3.6$\mu m$, 4.5$\mu m$ & $\sim$ 25.19, 25.09  & (PI: T. Soifer and P. Capak)	\\
          & \textit{Spitzer} IRAC 3,4 (cluster-only)  & 5.8$\mu m$, 8.0$\mu m$ & $\sim$ 23.94, 23.39    &	See Section \ref{subsec{ancillary}}\\
					&		    &   	&   	        \\
					MACS J0717.5+3745	& Keck/MOSFIRE   & 2.2$\mu m$  & $\sim$ 25.31   &	\citet{KIFF}\\
	& \textit{Spitzer} IRAC 1,2,3,4   & 3.5$\mu m$, 4.5 $\mu m$  &  $\sim$ 25.04, 25.17  &(PI: T. Soifer
and P. Capak)	\\
& \textit{Spitzer} IRAC 3,4   & 5.8$\mu m$, 8.0$\mu m$  &  $\sim$ 23.94, 23.39  & See Section \ref{subsec{ancillary}}	\\
					&		    &   	&   	        	\\
MACS J0416.1-2403	& VLT/HAWK-I   & 2.2$\mu m$  & $\sim$ 26.25   & \citet{KIFF}	\\
	& \textit{Spitzer}   & 3.5$\mu m$, 4.5 $\mu m$  & $\sim$ 25.31, 25.44    &	(PI: T. Soifer
and P. Capak)\\        	      	
					&	    	&   &	&     \\
Abell S1063		    & VLT/HAWK-I   & 2.2$\mu m$  & $\sim$ 26.31   & \citet{KIFF} \\
   & \textit{Spitzer} IRAC 1,2  & 3.6$\mu m$, 4.5$\mu m$ & $\sim$ 25.04, 25.04     &(PI: T. Soifer
and P. Capak)	\\
   & \textit{Spitzer} IRAC 3,4 (cluster-only)  & 5.8$\mu m$, 8.0$\mu m$   & $\sim$ 22.96, 22.64  &  See Section \ref{subsec{ancillary}} &	\\ 
   &		    &   	&   	        	\\
Abell 2744		    & VLT/HAWK-I   & 2.2$\mu m$  & $\sim$ 26.28  &  \citet{KIFF} \\
	    & \textit{Spitzer} IRAC 1,2  & 3.6$\mu m$, 4.5$\mu m$ & $\sim$ 25.32, 25.08  &   (PI: T. Soifer
and P. Capak)	\\
        & \textit{Spitzer} IRAC 3,4 (cluster-only)  & 5.8$\mu m$, 8.0$\mu m$   & $\sim$ 22.78, 22.45    & See Section \ref{subsec{ancillary}}	\\
        			&	    	&   	&   			        	\\
MACS J1149.5+2223	& Keck/MOSFIRE   & 2.2$\mu m$  & $\sim$ 25.41   & \citet{KIFF} \\
	& \textit{Spitzer}   &  3.5$\mu m$, 4.5 $\mu m$ &  $\sim$ 25.24, 25.01 &  (PI: T. Soifer
and P. Capak)	\\
	& & & & 
\enddata
\end{deluxetable*}

\section{Data reduction}
The workflow followed for the data processing in this work is presented in Figure~\ref{fig:workflow}.

\subsection{Modeling the point spread function in different bands}
\label{sec:psf_matching}

\begin{figure*}
\includegraphics[width=\textwidth]{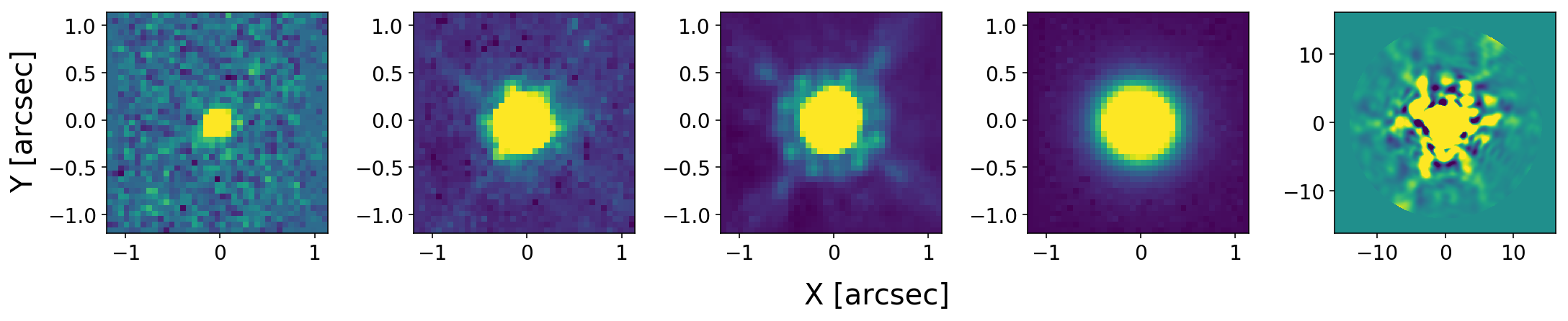}
\caption{Representative examples of PSF for the instruments used in this study, corresponding to a 0.06"/pixel scale normalized with the \texttt{ZScale} algorithm. From left to right, panels show F336W (WFC3), F606W (ACS), F125W (WFC3), $K_s$ (VLT), IRAC Channel 1 (\textit{Spitzer}). See Section \ref{sec:psf_matching} for more details} 
\label{fig:psf_fig}
\end{figure*}

Accurate knowledge of the point spread function (PSF) as a function of wavelength and optical system is crucial to perform consistent photometry within a `panchromatic' baseline. Knowledge of the PSF for HFF galaxies is needed to reconstruct their intrinsic morphology as well as mapping the ICL (Section \ref{sec:ICL_fitting}). To perform consistent multi-waveband photometry, taking into account band-to-band variations, we need to convolve images to the same PSF. This task, known as ``PSF matching'', is performed via kernel convolution. Images in multi-wavelength surveys are often affected by different diffraction levels, which make it difficult to obtain homogeneous measurements e.g.\ of aperture photometry. Figure~\ref{fig:psf_fig} illustrates the broad range of PSF sizes by showing postage stamps extracted from a few HFF images from $\sim0.3$ to $3.6\mu$m. Besides the optical performance of the instrument itself, the final PSF model also depends on the specific observing strategy of the survey (depth, dithering, etc.) and in the case of ground-based facilities also on the seeing conditions (in the case of Figure~\ref{fig:psf_fig}, VLT/HAWK-I). 

For HST data (divided between cluster vs.\  parallel fields), PSFs were estimated by stacking unsaturated stars. We use \texttt{IRAF} to visually inspect stellar light curves and determine whether an object is saturated. Then we stack unsaturated stars using the \texttt{IDL} routine \texttt{Star Finder}~\citep{2000SPIE.4007..879D}. The resulting PSF models used in PSF-matching are $2.34\arcsec\times 2.34\arcsec$ postage stamps in the HST and Ks bands, and $28.56\arcsec$ for \textit{Spitzer}. These sizes are chosen in order to enclose diffraction spikes (as one can see in Figure~\ref{fig:psf_fig}).


Our procedure works well for the HST images and the $K_\mathrm{s}$ band as the variation of the PSF across the image is small. However, stacking unsaturated stars does not produce a robust result in the mid-IR \emph{Spitzer} channels. This is due to large variations of the PSF as a function of the position on the image as well as the asymmetry of its shape\footnote{
See \href{https://irsa.ipac.caltech.edu/data/Spitzer/docs/irac/calibrationfiles/psfprf/}{the \emph{Spitzer}/IRAC handbook}}. This makes IRAC PSFs depend on the orientation of the camera.  Moreover, IRAC Channels 1 and 2 pixels undersample the response of a point source\footnote{More information in the  \href{https://irsa.ipac.caltech.edu/data/Spitzer/docs/files/Spitzer/simfitreport52_final.pdf}{the \emph{Spitzer}/IRAC handbook}.}. Thus, instead of \texttt{Star Finder}, we use a synthetic pixel response functions  (PRF) that combines the information on the PSF, the detector sampling, and the intrapixel sensitivity variation in response to a point-like source. A PRF model for a given position on the IRAC mosaic is generated by the code \texttt{PRFMap} (A.\ Faisst, private communication)  by taking into account the single-epoch frames contributing to that mosaic. To do so, \texttt{PRFMap} stacks individual PRF models  with the same orientation of the frames, resulting in a realistic, spatially-dependent PSF model (an example for IRAC channel 1 is shown in Figure~\ref{fig:psf_fig}). 

We use \texttt{GalSim} in order to calculate the FWHM for each PSF/PRF listed in Table~\ref{tab:hstdata}. More specifically, we use the \texttt{calculateFWHM} function, which computes the maximum intensity of the PSF, $I_{0}$, the centroid of the PSF intensity, the first pixel from the centroid at which the intensity is $I < I_{0}/2$ and the last pixel at which the intensity is $I > I_{0}/2$, and then linearly interpolates between these two to estimate the value of the FWHM. For the Spitzer bands, we input into this function an averaged stack of all the normalized PRFs modelled in the field for a more conservative estimate.

\subsection{Modeling the intra-cluster light}
\label{sec:ICL_fitting}

Given the richness of HFF clusters, those fields are particularly crowded with a significant probability of having cluster members aligned along the same line of sight of more distant background galaxies. The analysis of background systems is also impaired by the ICL, i.e.\ the residual emission from stars that are generally not bound to any cluster galaxy \citep{montes19, 2017ApJ...846..139M, sampaio-santos20}.  Blending between ICL and bright cluster galaxies is also a source of uncertainty.  
In an effort to alleviate these effects we attempt to model ICL and cluster members using GALFIT~\citep{2010AJ....139.2097P} and GALAPAGOS-2~\citep{2012MNRAS.422..449B,2013MNRAS.430..330H} respectively, following a similar procedure as in~\citet{2017ApJ...846..139M}. Because the severity of the ICL-galaxy blending decreases towards the bluer wavelengths, we perform these fits for all but the bluest HST bands used here (i.e., $F275W$ and $F336W$). 

In this Section we focus on ICL modeling, while in Section \ref{sec:gal_morpho} we will describe the procedure to fit the brightest cluster members. Subtracting their flux from the images will allow us to detect some of the faintest objects that otherwise would remain hidden by their over-shining neighbors (see discussion in Section \ref{sec:source_extraction}). 


\begin{figure*}
    \centering
    \includegraphics[width=\textwidth]{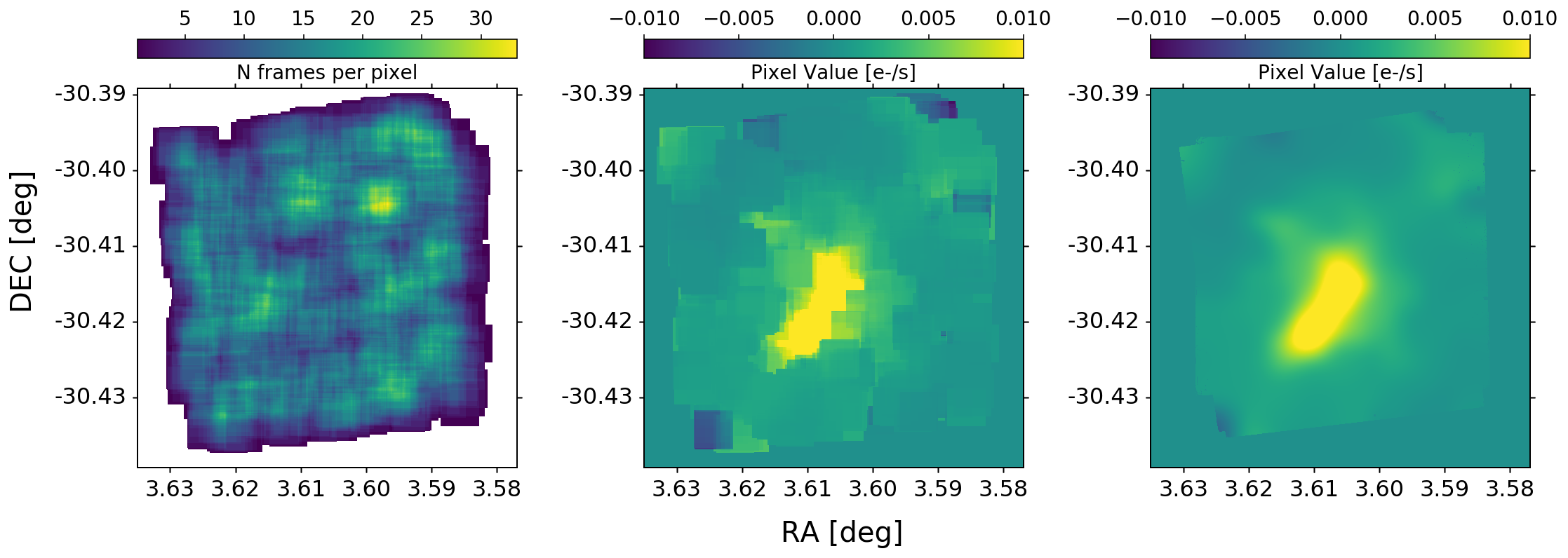}
    \includegraphics[width=\textwidth]{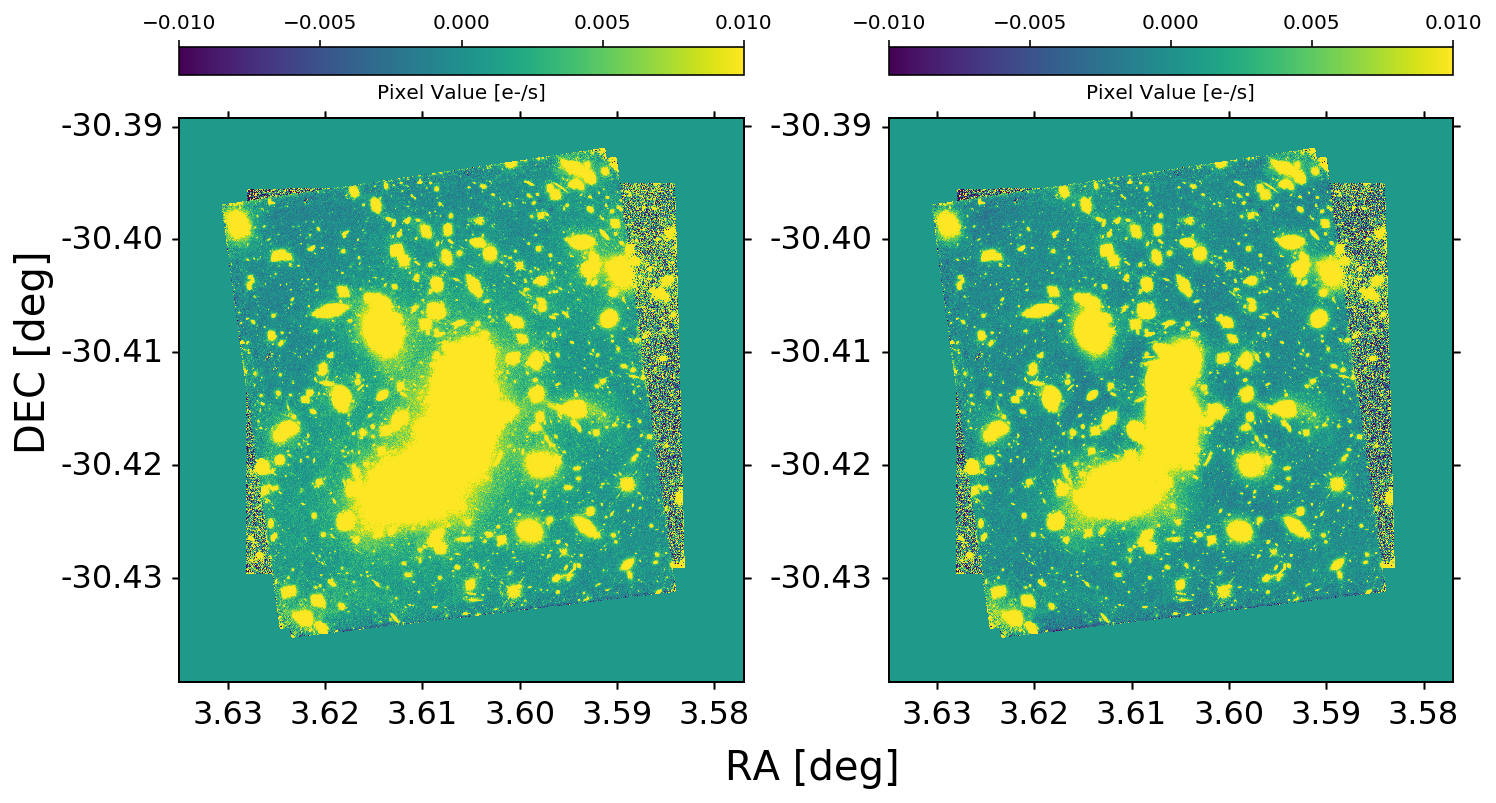}
    \caption{We illustrate the steps to generate the ICL map (as described in Section \ref{sec:ICL_fitting}). Here the example is presented for the F160 band in Abell 2744. \textbf{Top Row:} \textit{Left panel:} the ``coverage map'' showing the number of fit cutouts that overlap with each pixel. \textit{Middle panel:}  the resulting ICL map for the same band and cluster created by combining the fit background value of each cutout. Overlapping stamps are stacked using Eq. \ref{eq:FICL}.   \textit{Right panel:} Final ICL map after smoothing the map in the Middle panel with a representative Gaussian kernel. More details about kernel creation as provided in section \ref{sec:ICL_fitting}. \textbf{Bottom Row:} \textit{Left panel:} Original Abell 2744 F160W science image. \textit{Right panel:} F160W image after smoothed ICL (top row, right panel) subtraction.}
    \label{fig:ICL_gen}
\end{figure*}

\begin{figure}
    \centering
    \includegraphics[width=\columnwidth]{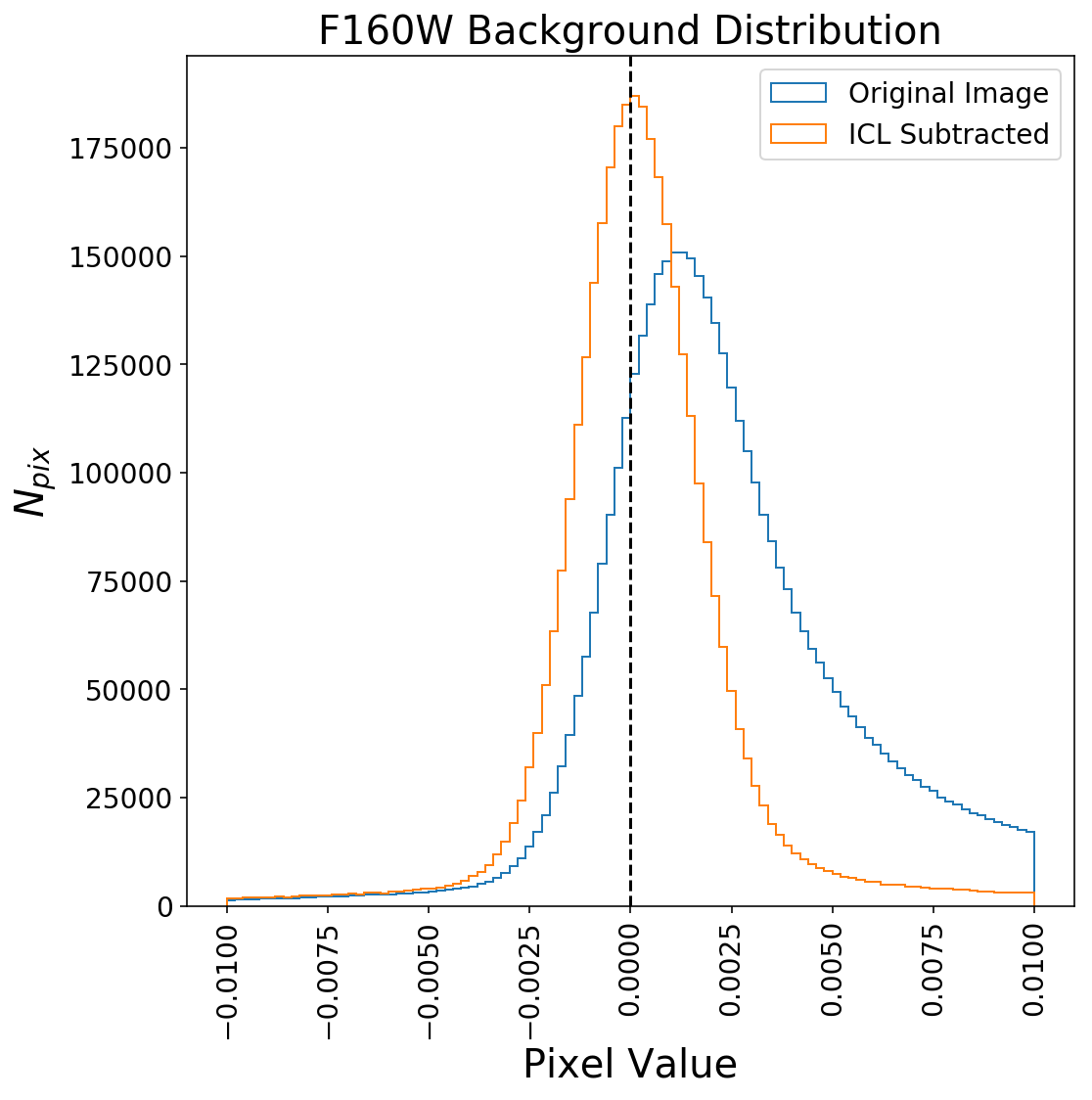}
    \caption{Pixel values of the original background (blue histogram) and after ICL subtraction (orange histogram); the vertical dashed line marks the zero value.}
    \label{fig:iclhist}
\end{figure}

 To model the ICL we follow the methodology presented in~\citet{2017ApJ...846..139M}. We first run \texttt{Source Extractor} \citep[][hereafter \texttt{SE}]{1996A&AS..117..393B} on each image/band to get a first-pass estimate of the morphological parameters of detected galaxies, and their segmentation map. In this way we can create a mask removing the detected pixels of sources fainter than 26\,mag, as the fitting procedure (described below) is less reliable below that threshold. Then we produce a $18 \times 18$ arcsecond ($300 \times 300$ pixel) patch for every remaining source, with the patch centered at the coordinate identified by \texttt{SE} as the centroid of the galaxy.  By means of \texttt{GALFIT}, we simultaneously fit single S\'{e}rsic profiles to all the objects included in that patch.  For patches whose \texttt{GALFIT} fits failed in the first instance, we randomly move the cutout center, creating five additional $18 \times 18$ arcsecond patches on which we run the same algorithm.  
 
 It should be noted that the number of patches across the field depends on the projected density of sources. This is illustrated in Figure~\ref{fig:ICL_gen} (left panel) by showing a  representative ``coverage map'' for Abell 2744. In case a given pixel with coordinates $(x, y)$ is only included in one cutout, the ICL emission ($F_\mathrm{ICL}$) is defined as the local background measurement found by \texttt{GALFIT} (namely, the \texttt{sky value} parameter). If there are overlapping cutouts in $(x, y)$, we use the inverse $\chi^2$-weighted mean of their background measurements:
\begin{eqnarray}
\label{eq:FICL}
F_\mathrm{ICL}(x,y)=\frac{\Sigma_{i} s_i(x,y)/\chi^2_i(x,y)}{\Sigma_{i} 1/\chi^2_i(x,y)} \;,
\end{eqnarray}
where $s_{i}$ and $\chi_i^2$ are the \texttt{sky fit}  and $\chi^2$ values from \texttt{GALFIT} for the $i$-th cutout. The overall ICL map, in the case of Abell 2744, is shown in Figure~\ref{fig:ICL_gen} (middle panel).


The merged ICL map, represented by $F_{ICL}(x,y)$, may still contain some sharp features due to the finite nature of the cutouts used to build it. In order to avoid these features we introduce an additional step to the method presented in~\citet{2017ApJ...846..139M} by smoothing the merged ICL map with a Gaussian kernel. The size of this kernel is chosen by analyzing the (radially averaged) power spectrum of the coverage map (see the left panel of Figure~\ref{fig:ICL_gen}) defined as $P(k)= \sum_{x, y} I_{r(x,y)} e^{-ikr}$, where $r$ is the radial distance from the center of the image, $r = \sqrt{(x-x_{0})^{2} + (y-y_{0})^{2}}$; $x, y$ are the coordinates of each pixel, $I_{r(x,y)}$ is the value of the coverage map at the pixel $x,y$ at a distance $r$ from the center, and $x_{0}, y_{0}$ are the coordinates of the center of the image. We find this scale to be at $k\simeq0.08$, i.e.\ about 72 pixels, or 4.32 arcseconds. In the right panel of Figure~\ref{fig:ICL_gen} we can see that this strategy successfully mitigates the impact of sharp features in the merged ICL map.  
As a sanity check, we generate a histogram of the sky background by masking sources using the corresponding segmentation map. We then compare this between the original image and the ICL subtracted image. We find that the resulting histogram of background pixels is both more symmetric, narrower, and centered at zero in the ICL subtracted image (Figure \ref{fig:iclhist}). 

The resulting map, shown in Figure~\ref{fig:ICL_gen} for the Abell 2744 cluster, not only serves as a representative model of the ICL, but also as a background correction for the entire HFF mosaic. Figure~\ref{fig:ICL_gen} also shows the difference between the original map (middle image) and the one after kernel filtering (right-most image): The sharp flux gradients caused by finite box numbers and sizes is mitigated.

To model the ICL in the $K_s$ and \textit{Spitzer} images we use a local background routine built into the \texttt{T-PHOT} software \citep{2016A&A...595A..97M}, which calculates a background template for each object and merges them into a single image. As done in the optical and other near-infrared bands, we smooth the image with a representative kernel and subtract the result from the HFF mosaic. The smoothed ICL mosaics will be released as two-dimensional arrays in the same units (ADU/s) as the HFF mosaics.

\subsection{Removal of the Bright Galaxies in Clusters}
\label{sec:gal_morpho}

After correcting the cluster images for ICL and sky background, we  model the brightest galaxy members of each cluster (i.e. galaxies with $\mathrm{MAGAUTO_{F160W}}$ $<$19). These objects are selected via a first-pass \texttt{SExtractor} run in the F160W band.  The aim is to remove them and their diffuse light so that one could push the observations deeper. We use the publicly available code  \texttt{GALAPAGOS-M}\footnote{\url{https://github.com/MegaMorph/galapagos}}~\citep{2012MNRAS.422..449B,2013MNRAS.430..330H}, which is a software that automates source detection and bulge-disk S\'{e}rsic modeling, and takes advantage of multi-wavelength information in its parameter fits.


The advantage of using \texttt{GALAPAGOS-M} is in its ability to perform single and multi-component (bulge/disk decomposition) fits and to input galaxies using information from multiple bands \textit{simultaneously}. In order to robustly measure color gradients in large galaxies, we restrict the degrees of freedom in the \texttt{GALAPAGOS-M} fits by imposing a wavelength dependency (with a quadratic function) for the half-light radius and S\'{e}rsic index. Compared to fitting morphology independently in each band, this approach is more stable and tightens the constraints of the morphological parameters measured for each galaxy \citep[see][]{2013MNRAS.430..330H}.    However, a multi-component model is susceptible to overfitting the images; therefore, we use the residual flux fraction  \citep[RFF, as in][]{2012MNRAS.419.2703H} to assess the number of components that most effectively models the light profile of bright cluster members; such residual flux is obtained by subtracting the model from the input image. The residual flux fraction is a measure of signal excess in the residual image, not due to background fluctuations, and is defined as
\begin{eqnarray}
\mathrm{RFF} = \frac{\sum \vert I_{x,y} - I_{x,y}^\mathrm{model} \vert - 0.8 \times \sum {\sigma_{\mathrm{BKG},x,y}}}{\sum I_{x,y}}\;,
\label{eq:rff}
\end{eqnarray}
where $I_{x,y}$ and $I_{x,y}^{model}$ are respectively the observed and model fluxes for a given pixel with coordinates $(x,y)$, while $\sigma_{\rm{BKG}, x,y}$ is the background RMS in the same location. We sum over the pixels associated to a given galaxy. The 0.8 factor is included to ensure that the mean RFF of an image is null when it is  exclusively affected by Gaussian noise with constant variance. For a given object to fit, we favor either a single- or two-component model depending on which of them results in the smallest RFF.  Generally, studies of galaxy morphology require the RFF relative difference between two models, i.e.\ $(\mathrm{RFF}_1-\mathrm{RFF}_2)/\mathrm{RFF}_1$, to be larger than  1.0--1.6~\citep[e.g.,][]{2012MNRAS.419.2703H}. This is a conservative threshold for the selection of the multi-component fit ($\mathrm{RFF}_2$) to prevent over-fitting. Here, we are more concerned  about effectively removing the flux of a given galaxy (to get a smaller residual) rather than  providing a realistic description of its morphology. For this reason we select the multi-component solution when $\mathrm{RFF}_2<\mathrm{RFF}_1$. 

\begin{figure*}[ht]
\centering
\includegraphics[width=\textwidth]{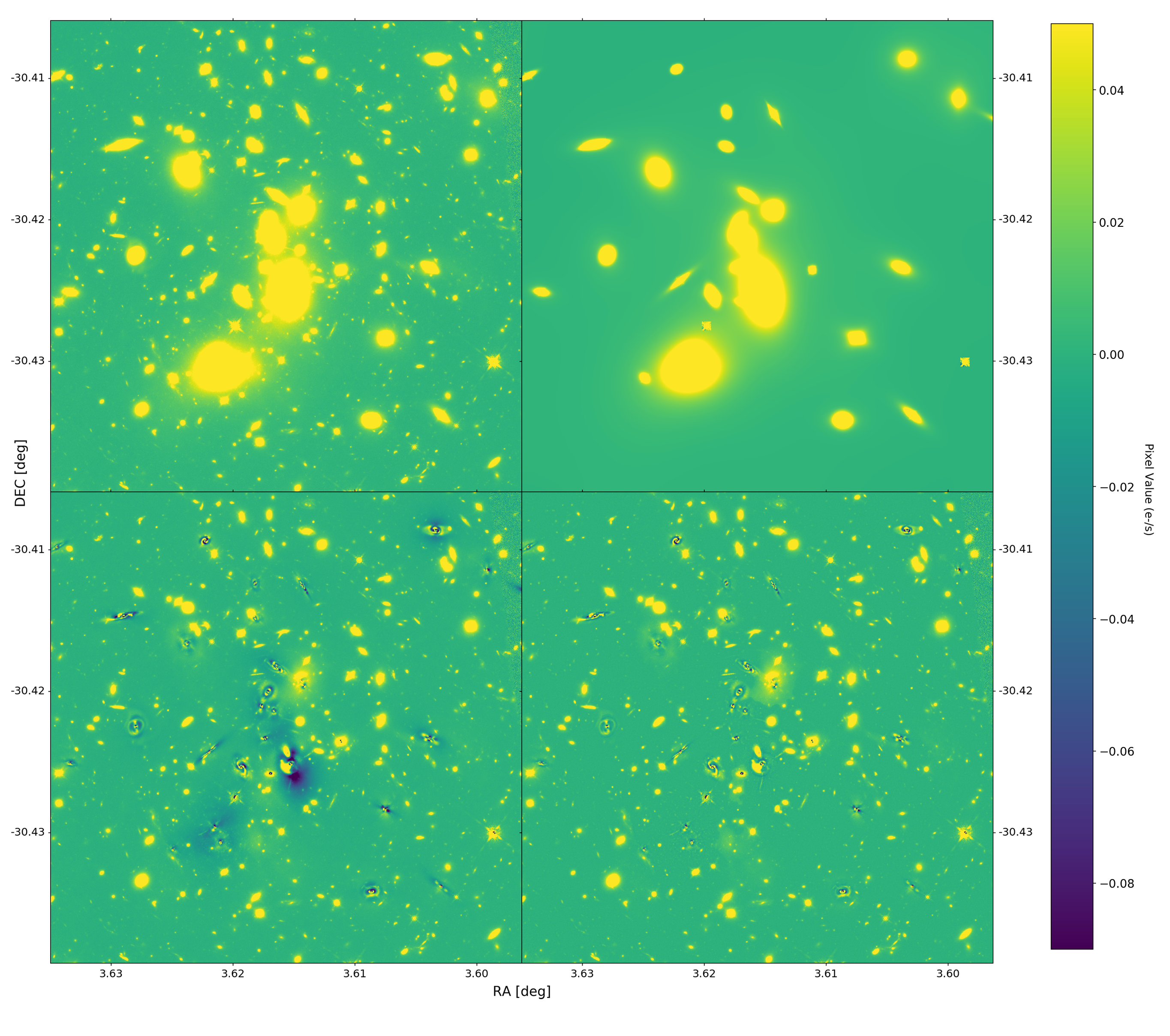}
\caption{ A summary of the various steps in bright cluster + ICL modelling (in this case for cluster Abell 2744). Upper panels show the original image (left) and the galaxy/ICL models (right). Lower panels show the residual image before and after median filtering (left- and right-hand panel respectively). The colorbar denotes the pixel intensity in counts/s. See Sections from \ref{sec:ICL_fitting} to \ref{sec:gal_morpho} for more details.} 
\label{fig:ICL_model_residual}
\end{figure*}

After the modelling process and subtracting both ICL and bright cluster members, we apply a median filter to the cleaned image. This is a well-known smoothing technique that replaces the value of a given pixel by the median of its neighboring pixels  \citep[see][]{2016A&A...595A..97M}. We use a filter with a box size of $1$\,deg per side. We apply this filtering only to pixels with a flux within 1$\sigma$ of the background level in order to reduce the effects of over-subtraction in the residual. The resulting improvement can be seen in the bottom-right panel of Figure \ref{fig:ICL_model_residual}. Note that this process does not affect the outskirts of the cluster.


\begin{figure}
\includegraphics[width=\columnwidth]{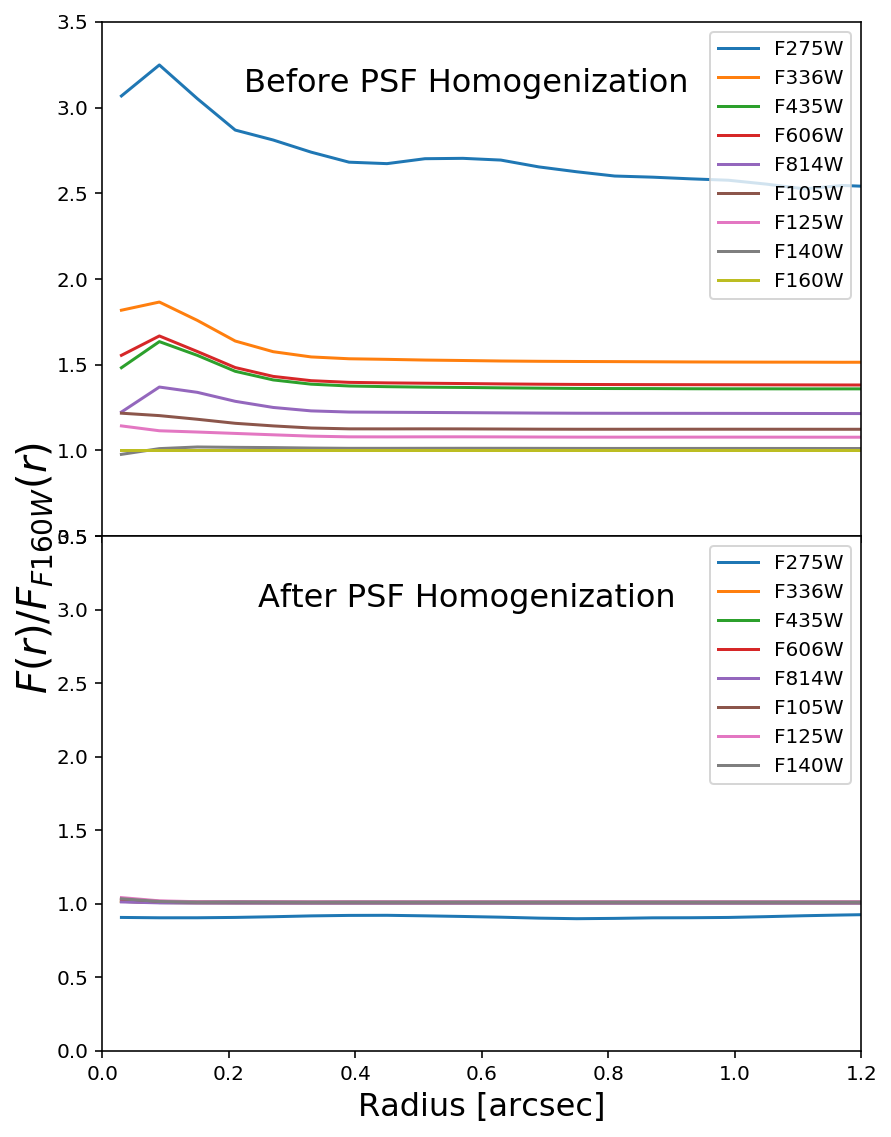}
\caption{Growth curves for each HST-derived PSF normalized by the F160W growth curve before (top) and after (bottom) homogenization.} 
\label{fig:psf_homogenization}
\end{figure}

\begin{figure}
    \centering
        \includegraphics[width=0.5\textwidth]{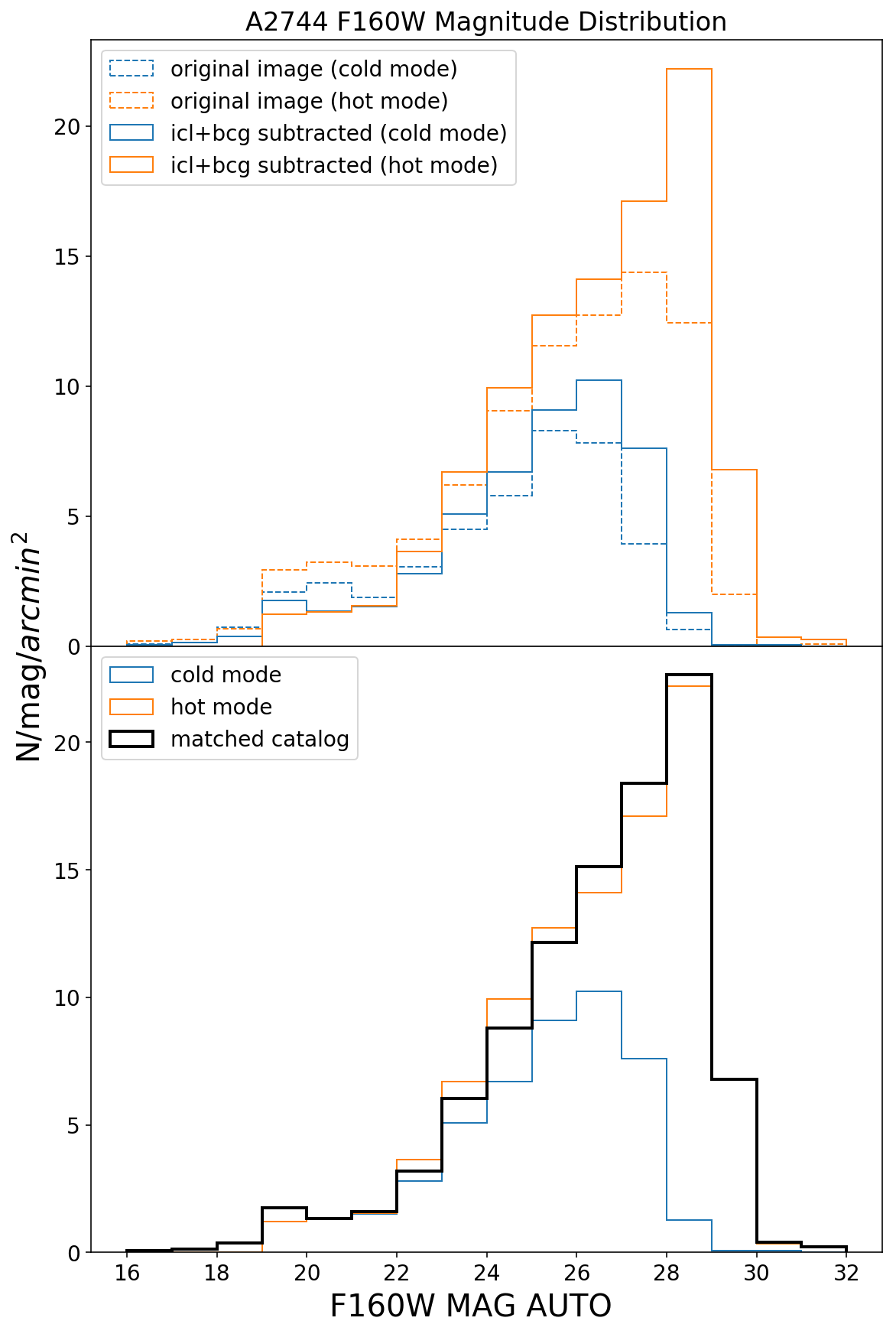}
        \label{fig:mag_hists}
\caption{  Magnitude distributions for the objects detected using the cold (blue) and hot (orange) modes for the original image v. ICL+BCG subtracted image (top) and ICL+bcg subtracted image (bottom); matched (black) refers to total number of galaxies detected and kept in the catalog per magnitude bin. We see a significant increase in the number of detected galaxies in the cluster subtracted image, as well as a slight shift in mean magnitude toward fainter objects.}
\end{figure}

\begin{figure*}
    \centering
        \includegraphics[width=0.90\textwidth]{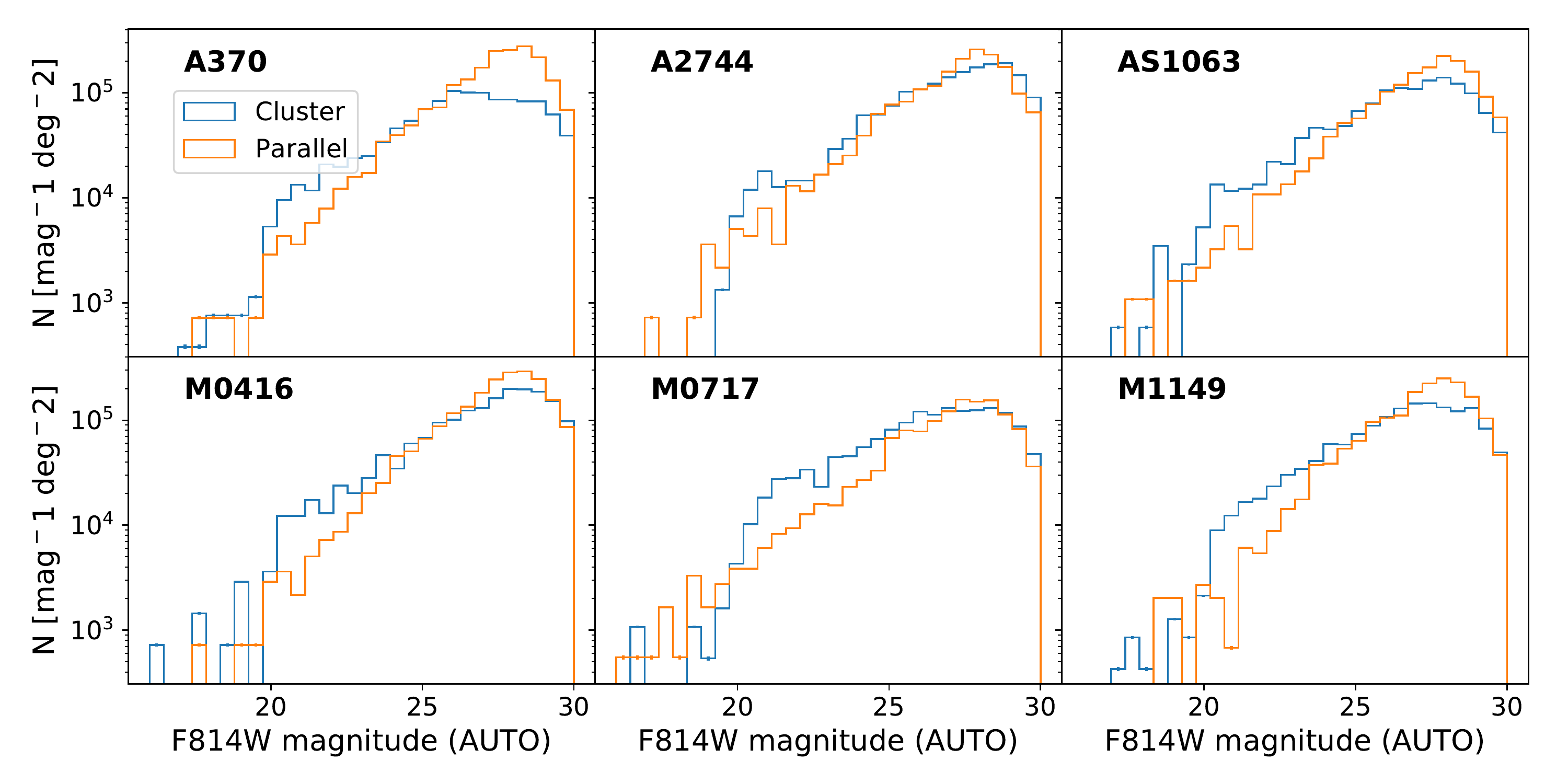}
        \label{fig:allfield_maghists}
\caption{Comparison of magnitude distributions for the HFF cluster and their parallel fields. This clearly shows the excess in number of galaxies in clusters compared to the field. It also shows the field and cluster samples have similar depths.}
\end{figure*}

\begin{figure}
    \centering
     \includegraphics[width=0.45\textwidth]{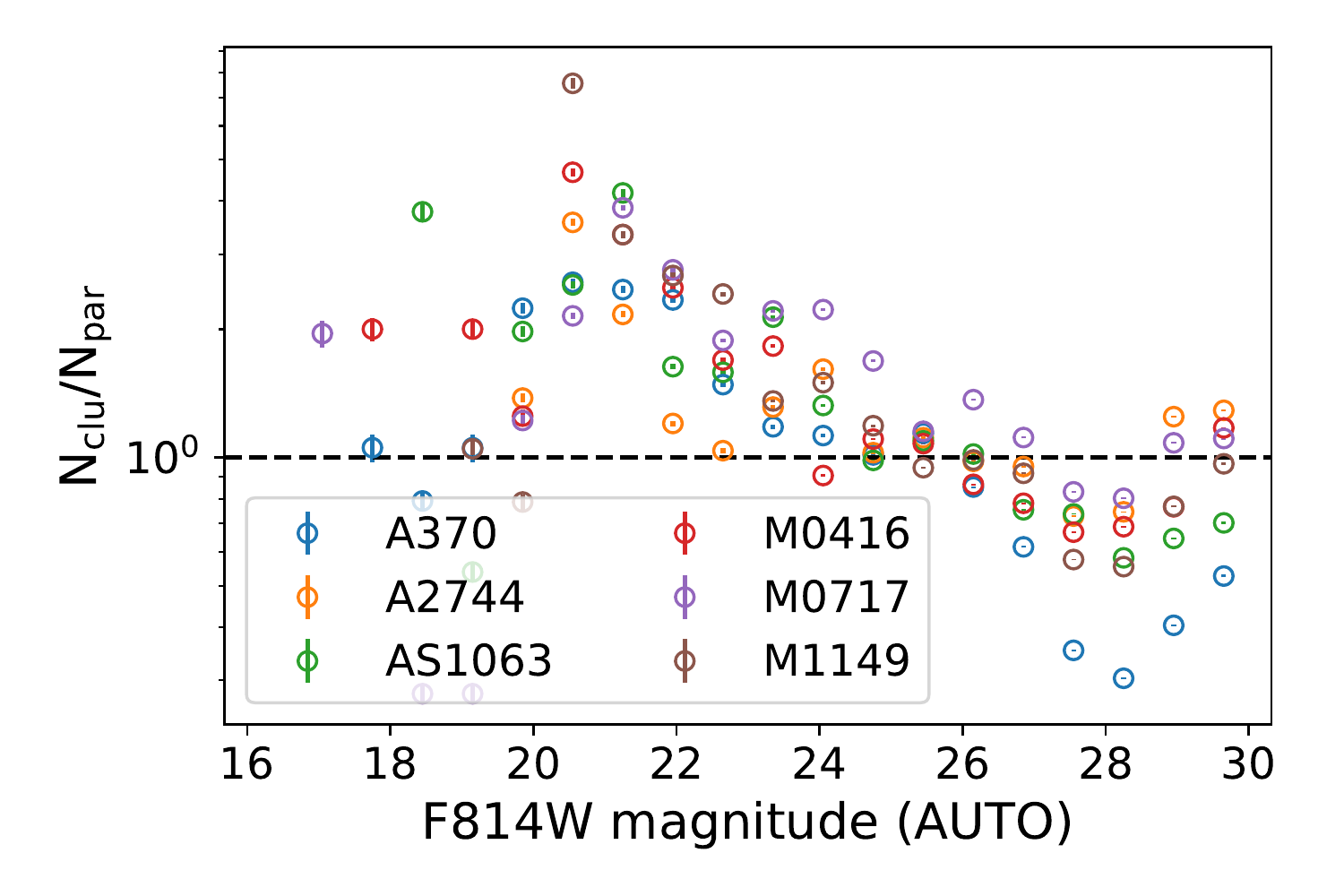}
     \includegraphics[width=0.45\textwidth]{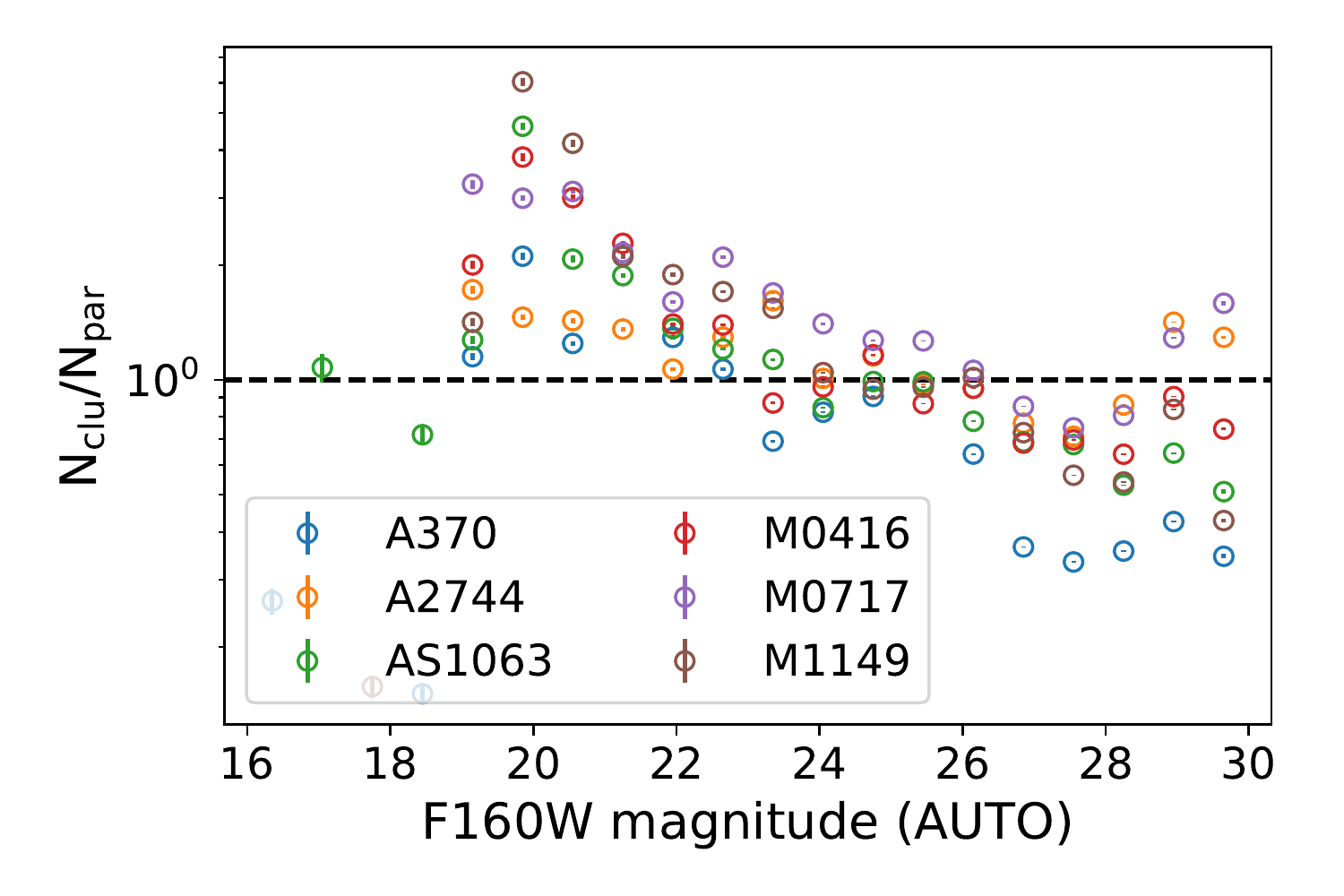}
    \caption{Ratio between magnitude distributions in the parallel and cluster fields for all HFF clusters in the F160W band (top) and F814W band (bottom). The excess of field galaxies over clusters in fainter magnitude bins is likely caused by non-detection of faint sources in clusters due to diffuse light from more luminous objects.}
    \label{fig:ratio_magnitude}
\end{figure}


\section{Source extraction}
\label{sec:source_extraction}

\subsection{HST detections}

After performing the modelling described in the previous section, we homogenize the various HST bands by matching their respective PSFs. 
Images in a panchromatic data set, especially if they sample a wide wavelength range like extra-galactic surveys, are usually affected by different PSF shapes (see Figure \ref{fig:psf_fig}). This means that the fraction of the flux that falls inside an aperture, as well as the resulting noise, varies as a function of bandwidth and pivotal wavelength. 

PSF matching is a convolution procedure to solve these problems by reshaping multi-wavelength images into a common reference system having certain resolution and diffraction properties \citep[i.e., the same ``target'' PSF, see][]{1996AJ....112.2872T,1998ApJ...503..325A}. It has become a standard technique to measure consistent fluxes over a range of wavelengths \citep[e.g.,][]{galametz11} and from different instruments under varied observing conditions \citep[e.g.,][]{laigle16} to effectively characterize the spectral energy distribution (SED) of the observed galaxies.

Specifically, we use the PSF models derived in Section \ref{sec:psf_matching}. 
In our case, the target PSF is the one in the F160W band as it is the reddest HST band. To convert the other HST images we use a convolution kernel obtained by taking (in Fourier space) the ratio between their original and target PSFs.Note that in some cases, the target PSF is marginally smaller than the original. Despite this, we are able to relate target and original PSF since we do not require the convolution kernel to be positive-definite. We show the growth curves for the PSFs extracted from the HST data, normalized by the growth curve of the F160W PSF before and after homogenization, in Figure~\ref{fig:psf_homogenization}. We note that after PSF homogenization, the normalized growth curves are, on average, agree within 2\% for all bands except F275W, which agrees at the 8\% level. PSF matching is not applied to ground-based images nor \textit{Spitzer}: their resolution is appreciably lower than the HST bands and in that case the photometric extraction is performed with a different method (see Section \ref{sec:tphot}).

We then proceed to perform source photometry for each HST image using \texttt{SE}. The strong clustering of the objects in the HFF images pose the additional problem of detection completeness because the objects that are close pairs along the line of sight are not always successfully deblended by the software. To improve the extraction process, we create a co-added IR image from the weighted mean, using their inverse variance as weights, of four WFC3 bands (namely F105W, F125W, F140W, and F160W). Such a stacking enhances the signal-to-noise ratio of faint sources, being effectively deeper than any individual band. We use the stack of infrared images as a detection image, running \texttt{SE} in dual mode.

Moreover, given the variety of scientific goals, it is important to precisely measure both the bright galaxies in the foreground and the fainter galaxies in clusters. However, source extraction with \texttt{SE} is a trade-off between rigorous source deblending of galaxies close to each other on the angular plane, and spurious shredding of structure from a single galaxy. For this reason, previous works \citep[][and references thereon]{caldwell08, gray09, 2013ApJS..206...10G} proposed a ``dual mode", or High Dynamic Range (HDR), approach with a cold mode and a hot mode \texttt{SE} run, where the cold mode detects and performs photometry on the objects contained within visually reliable Kron radii and the hot mode more aggressively deblends and performs photometry on the smallest and faintest objects. Given the aggressive deblending of the hot mode, it can break up larger objects into individual pieces, which we discard when considering the cold catalog. We modify the HDR approach described in~\citet{2013ApJS..206...10G} and discard galaxies from the hot mode that fall within a 0.2 $\times$ Kron radius~\citep{1980ApJS...43..305K} of a cold mode detected object. We show the resulting magnitude distributions of this process in Figure~\ref{fig:mag_hists}.

To alleviate the possibility of fake source detection around the residuals, we utilize the Kron radius of the modeled object and discard objects which fall within it. Furthermore, we remove sources corresponding to diffraction spikes and those susceptible to edge effects on the perimeter of the images. This procedure is performed on both the cluster and parallel fields. 

We present the magnitude histograms in the F814W band for each cluster and their parallel fields (Figure~\ref{fig:allfield_maghists}). It shows the overabundance of bright galaxies in the clusters when compared to the parallel fields. We note that even after ICL+bright galaxy subtraction, the presence of such bright objects diminishes our ability to detect faint objects. We also show the ratio of number counts between the cluster and parallel fields in the F814W and F160W bands (Figure~\ref{fig:ratio_magnitude}). The observed trend here is consistent across all clusters, demonstrating the importance of having parallel fields with comparable depths.


\subsection{Photometry in lower resolution images}
\label{sec:tphot}

In addition to HST, we also rely on imaging observations from Keck, VLT, and \textit{Spitzer}. Those facilities have lower angular resolutions than the HST instruments, increasing the  blending between sources. In order to maximize the information extracted from each image, we use the prior-based code \texttt{T-PHOT}~\citep{2016A&A...595A..97M}. This code uses the high-resolution HST images  and their corresponding \texttt{SE} catalogs as priors to perform dual-mode photometric extraction in the lower-resolution images. This allows measurement of blended sources that are not directly detectable in $K_\mathrm{s}$ or IRAC bands. On the other hand the method misses the so-called ``HST dark'' sources, i.e.\ galaxies so faint in optical and near-IR that are not identified in the prior image, even though they are sufficiently bright to be visible in the low-resolution image. HST dark sources may constitute a hidden population of galaxies providing a non-negligible contribution e.g.\ to the cosmic star formation rate density budget \citep[see][]{wang19-hstdark}. However, the identification of HST dark sources is subsequent to the creation of an HST-based catalog and shall be addressed in future work.

Concerning the \textit{Spitzer} images, before running \texttt{T-PHOT} we perform a series of corrections related to the weight maps (defined as the inverse variance per pixel, i.e.\ $1/\sigma^2$) using the pull diagnostic~\citep{Gross:2018okg}, described as follows. Assuming the distribution is Gaussian, in principle, the ratio between the variable and its standard deviation should produce a Gaussian with a standard deviation ($\sigma$) equal to one. In the limit of large statistics (owing to the Central Limit Theorem), we can approximate the signal on the background pixels to follow a Gaussian distribution. For this purpose, we compute the distribution of the ratio between the background level and its standard deviation, $\sigma_\mathrm{bkg}$. Such a distribution is expected to have a standard deviation $\sigma_\mathrm{pull} \approx 1$ if $\sigma_\mathrm{bkg}$ is properly estimated. A value of $\sigma_\mathrm{pull} < 1$ means that the uncertainties are likely underestimated, and $\sigma_\mathrm{pull} \gg 1$ means that the uncertainties are likely overestimated. In order to have conservative values for the error maps, we multiply these by $\sigma_\mathrm{pull}$ when the latter is greater than one. For Spitzer channels 1, 2, 3, and 4 we only correct the error maps in the largest contiguous region where the cluster and parallel fields are located. We check that this is a good approximation since the resulting corrected background distributions show a single peak. The corrected errors, $\sigma_\mathrm{pull}$, are discussed in Appendix~\ref{app:err_corr}.

As the high resolution prior, we use the science image in the F160W band and the segmentation map created from the weighted IR stack. Previous studies \citep{2007NewA...12..271D,2013ApJS..206...10G} found that \texttt{SE} tends to underestimate the isophotal area of objects in a single-band detection image; such a systematic effect is inversely proportional to the flux. A solution to this issue  \citep{2013ApJS..206...10G,2016A&A...590A..30M} is to dilate the segmentation areas of individual sources. We verify how prominent this effect is by comparing one of the HFF clusters (Abell 370) to the deeper BUFFALO survey, given that the isophotal area increases with increasing signal-to-noise. We choose to compare the HFF IR stack, given its depth, to the BUFFALO F160W band. We find that when using the IR stack the isophotal area of each source is slightly larger, with minimal dependence on the area. The difference is small enough that we found it unnecessary to dilate the segmentation map before feeding it into \texttt{T-PHOT}.

We run \texttt{T-PHOT} for the Ks- and IRAC bands. Because the IRAC PSF varies across the field of view, we take advantage of \texttt{T-PHOT} ``multikernel'' option, which allows for the inclusion of a separate kernel for each object. These kernels are generated from the grid of PRFs produced by \texttt{PRFMap}, as described in Section \ref{sec:psf_matching}.  
We emphasize that the output provided by \texttt{T-PHOT} (namely, the  parameter \texttt{FitQty}) is an estimate of the \textit{total} flux emitted by the given source. 

\begin{figure*}
\includegraphics[width=\textwidth]{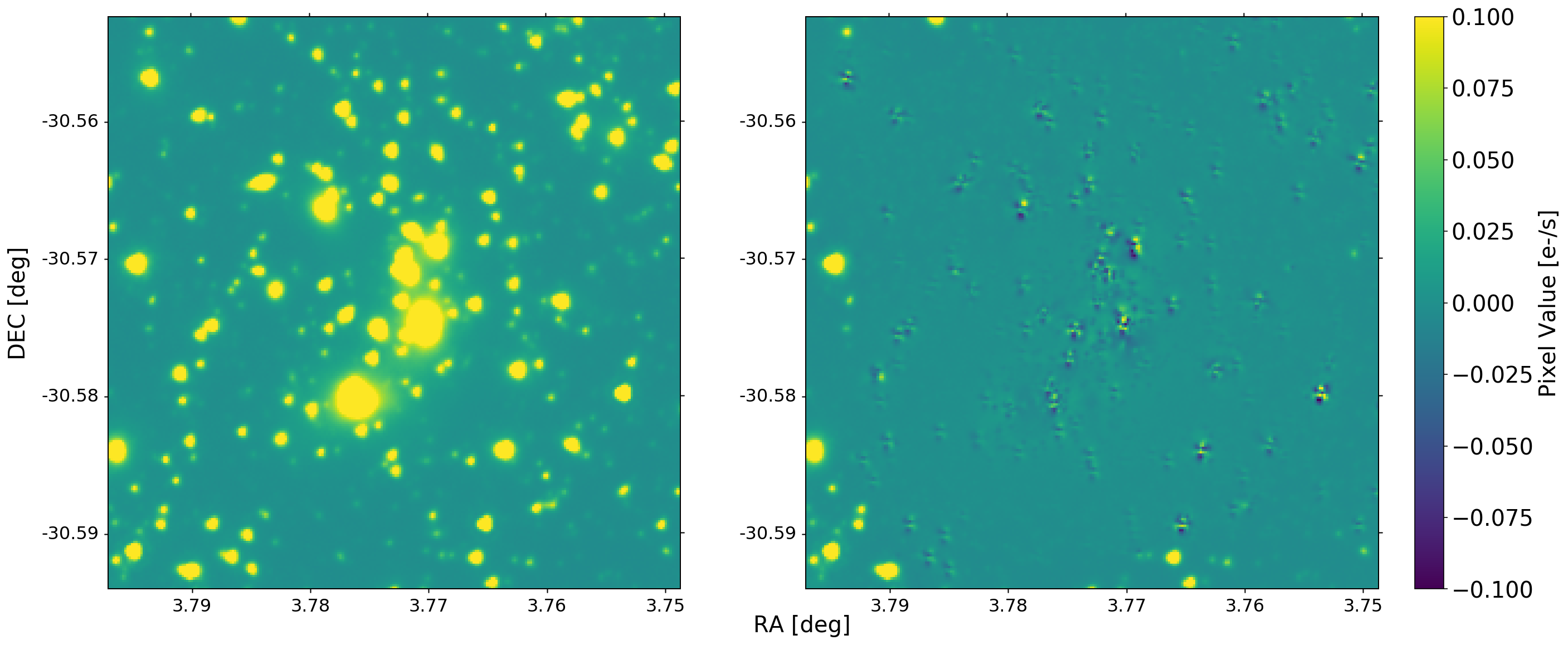}
\caption{(Left to right) Original IRAC Channel 1 image in  Abell 2744 (\textit{left panel}) compared to the residual map (\textit{right panel})  after subtracting \texttt{T-PHOT} galaxy models (see Section \ref{sec:tphot}).}
\label{fig:tphot}
\end{figure*}

\section{Validation of the photometric catalog}



\subsection{Quality and completeness via simulations}
\label{sec:completeness}

 To measure the completeness, we add randomly distributed simulated objects in the observed image and then try to recover them. The number of the simulated objects should be high enough to cover the full field of view and avoid statistical fluctuations, while their surface density should be low enough not to further increase the already large fraction of overlapping sources in the images. This way, we still check the performance of the deblender without introducing an unrealistic number of new blends. 
 
 In order to measure completeness we injected 100 point-like sources in the original images, with magnitudes randomly assigned from a uniform distribution between $23 <mag_{AB}< 31$. The magnitudes of these objects were selected to be the same in all bands, which allows us to compute the IR-weighted magnitude straightforwardly. This was done to simplify interpretation of the results. The processing of these images, was performed using the same steps presented in previous sections. We generated two sets of simulations (hereafter called `flavors'):
\begin{itemize}
    \item one set where we assign the position of the injected objects randomly within the image. We will refer to these simulations as flavor 1;
    \item a set where we assign the centroid position of the injected objects randomly within the empty parts of the image, i.e., where the segmentation map has no detected pixels. We will refer to these simulations as flavor 2.
\end{itemize} 

The reason behind the generation of two sets of simulations is that in the first flavor the fraction of blended objects is overestimated, since we are randomly placing new sources in a field that is already overcrowded. This leads to a pessimistic characterization of our pipeline's performance. However, using the second flavor, we explicitly avoid overlap with other sources. In the latter case the quality assessment will be over-optimistic. 

Ten independent realizations were generated for each of these flavors, 
The injected sources were independently generated for each cluster as 15 arcsec-side noise-free square cutouts. We then formed a collage of the same size of the original image, and added this collage to the original image of each cluster.  Therefore, only the original noise in the HFF mosaics is affecting the present test. This is a good approximation for faint sources, where the background dominates over the shot noise due to the source itself. 


After the injection of simulated sources, these images are analysed by the same software used for the real catalog (see previous Sections). We  compare the output to the list of input (simulated) sources, and their  intrinsic vs.\ recovered properties. We considered a source successfully recovered if there is an entry in the output catalogs whose centroid is within 0.6 arcseconds (10 pixels) from the input location, and the offset between intrinsic and extracted magnitude is $<$0.5\,mag. These thresholds have been chosen to minimize the number of by-chance matches between simulated and real sources, allowing at the same time for some difference between the input and output values. 

\begin{figure*}
    \centering
    \includegraphics[width=0.32\textwidth]{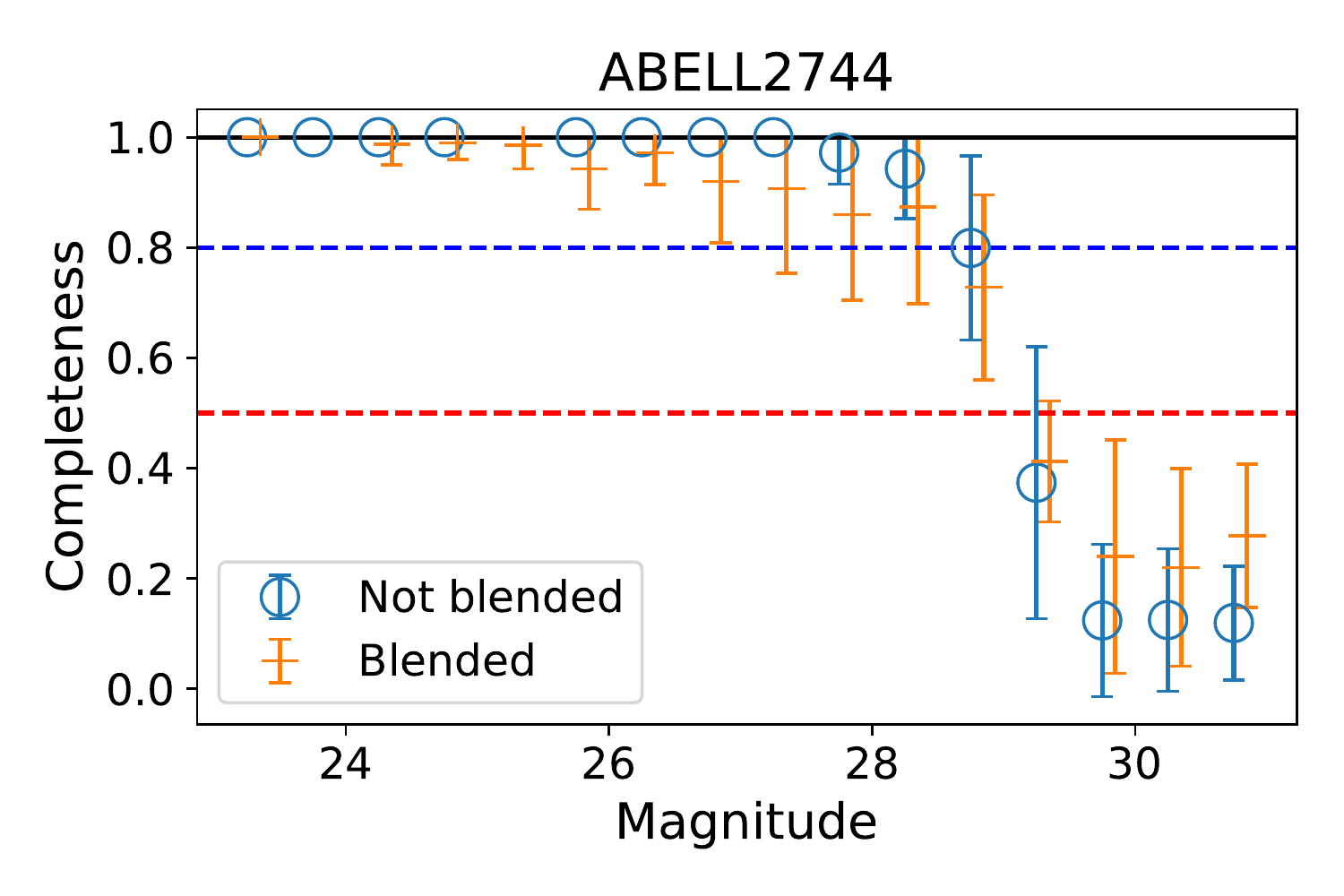}
    \includegraphics[width=0.32\textwidth]{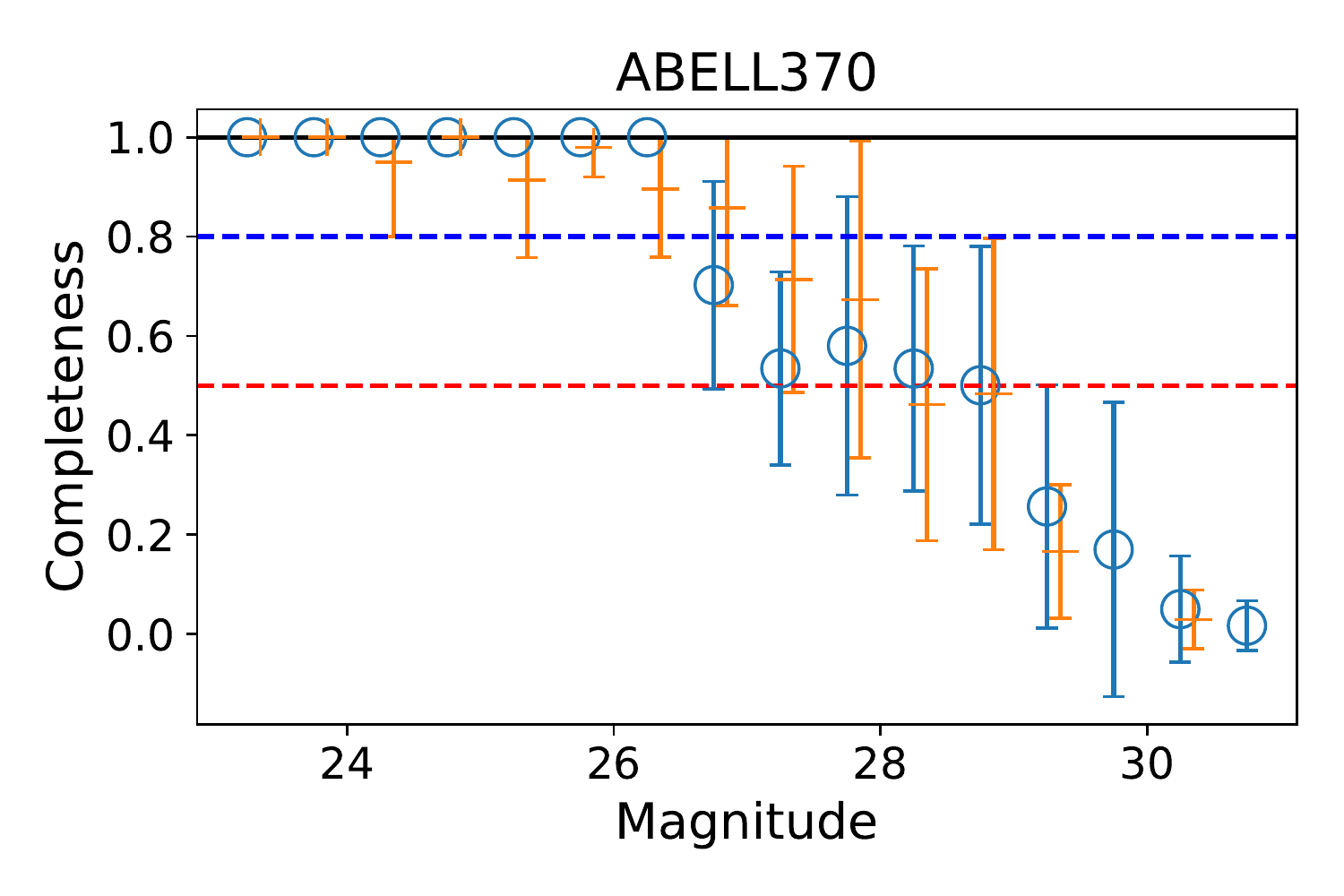}
    \includegraphics[width=0.32\textwidth]{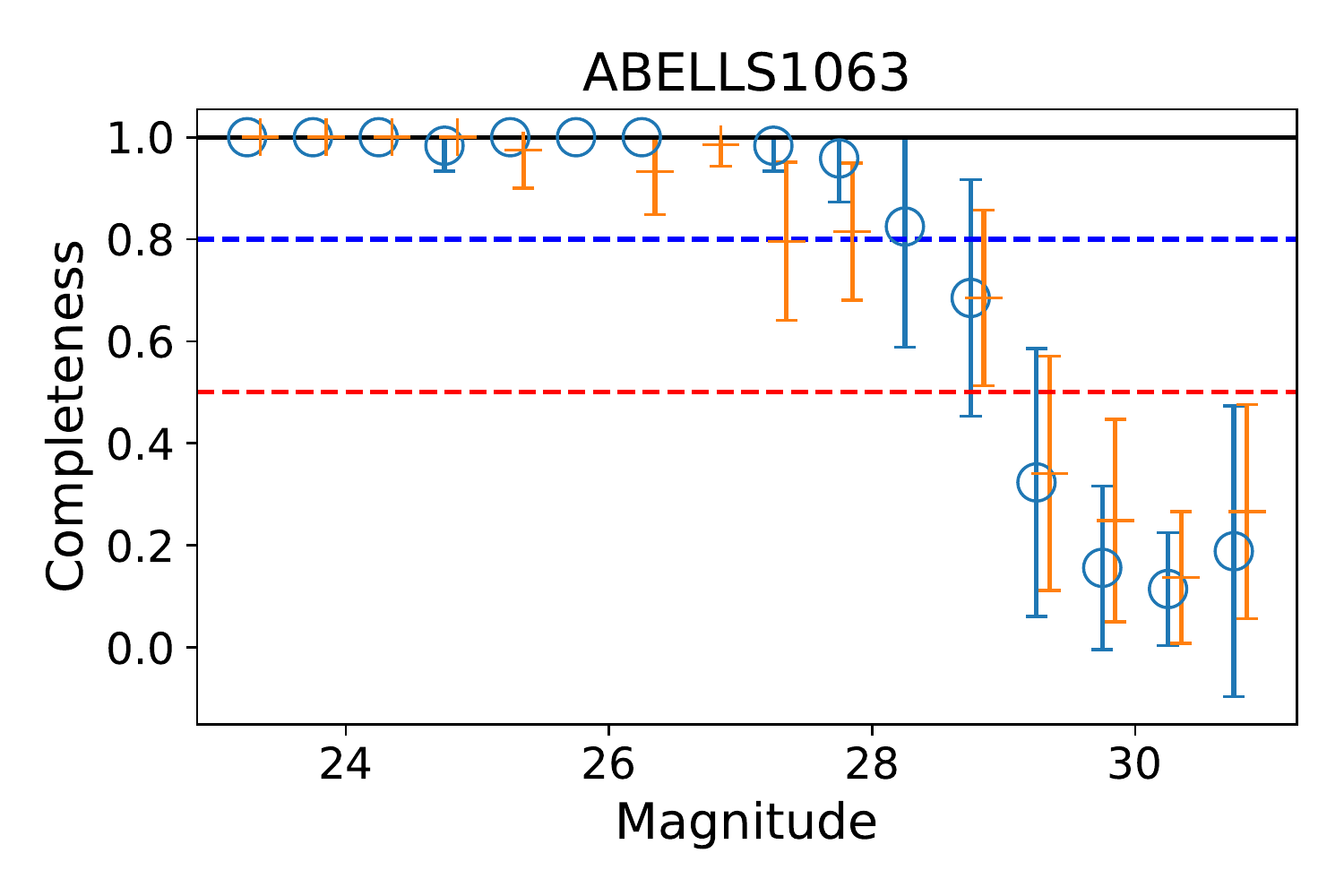}
    \includegraphics[width=0.32\textwidth]{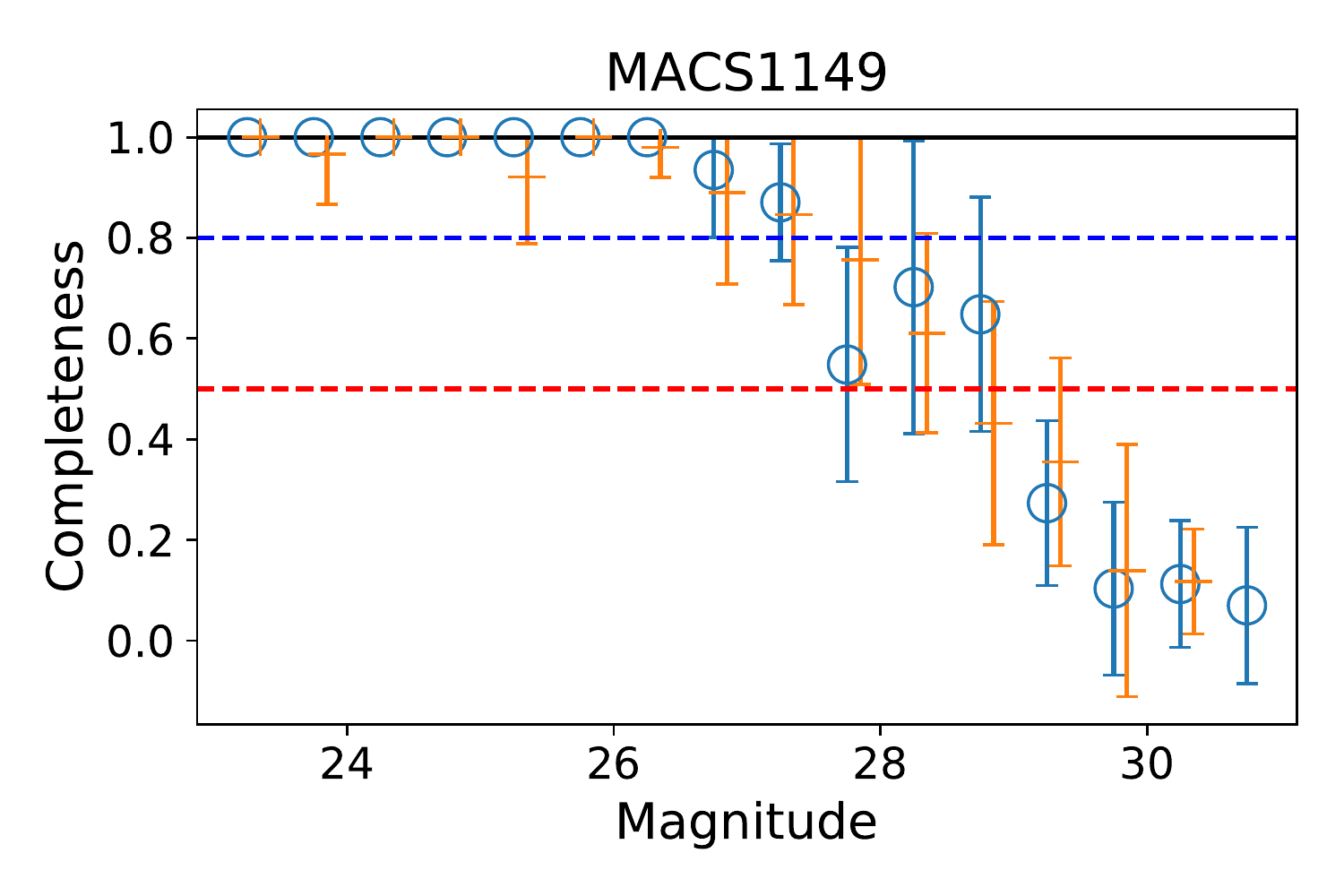}
    \includegraphics[width=0.32\textwidth]{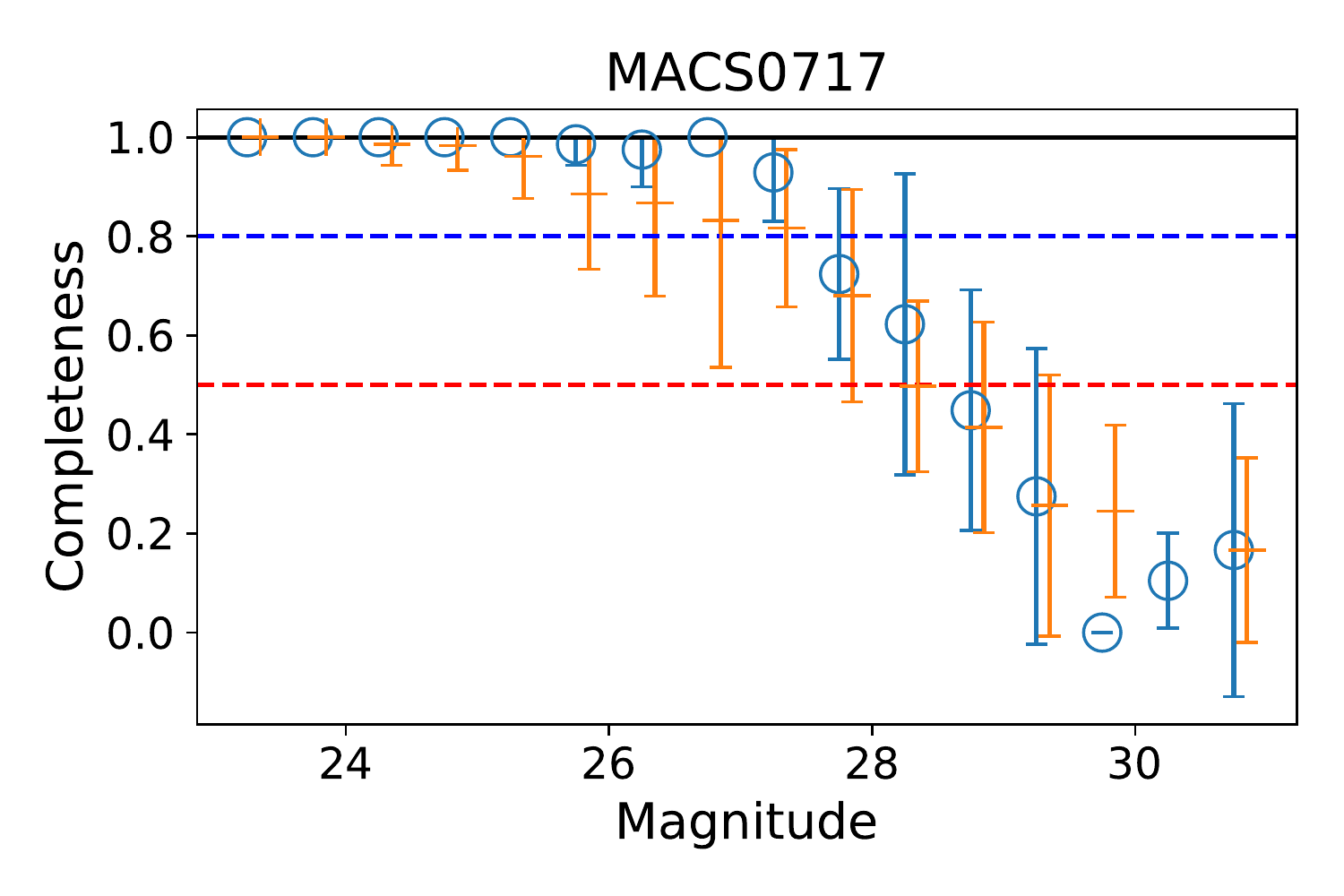}
    \includegraphics[width=0.32\textwidth]{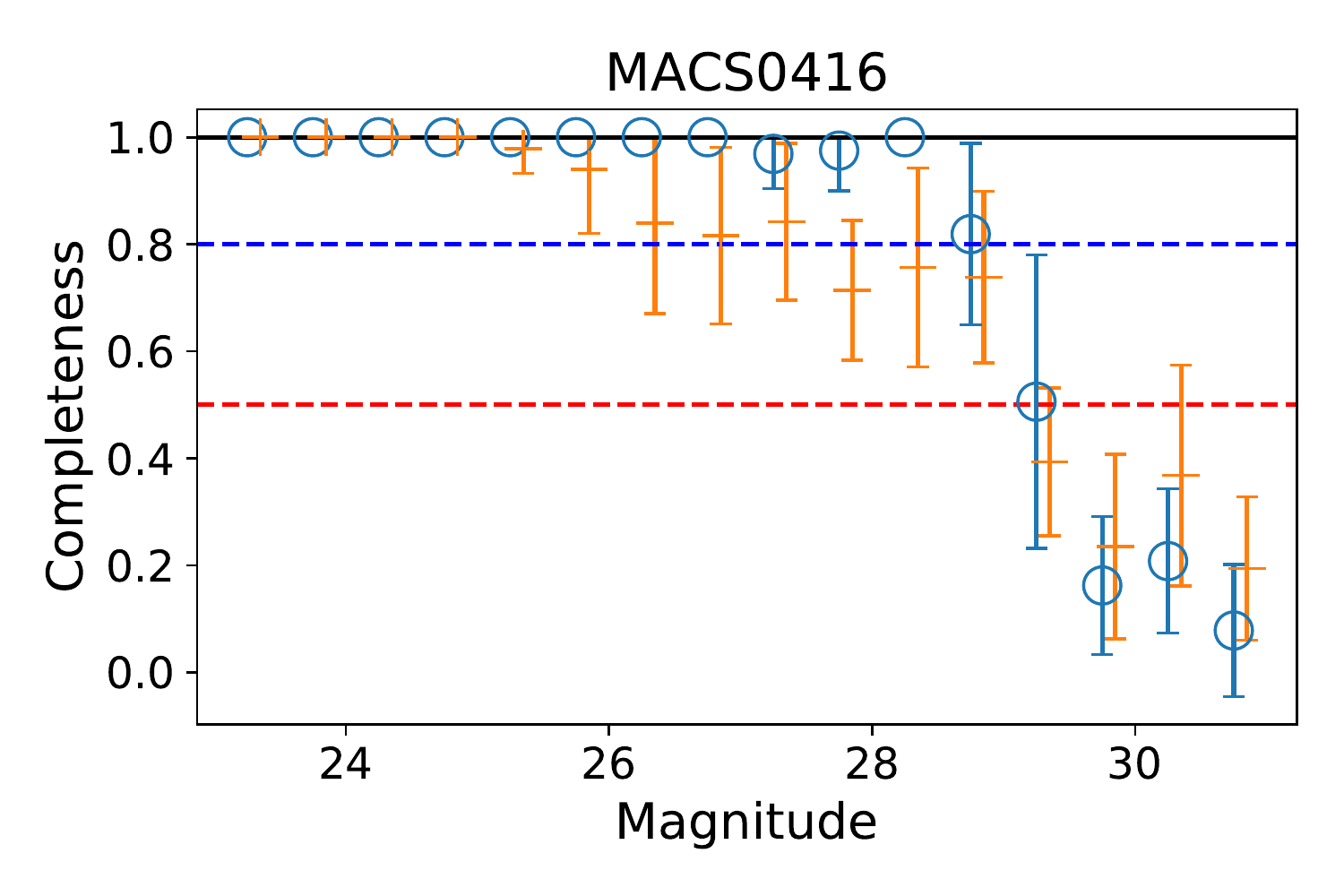}    
    \caption{Completeness as a function of magnitude for injected point-like sources between 24 and 30\,mag in the six analyzed clusters  (\textit{top row}: Abell 2744, Abell 370, Abell S1063; \textit{bottom row}:  MACS1149, MACS0717, MACS0416). The positions of these objects were randomly assigned within the footprint (open circles -- flavor 1, i.e., not blended) or randomly assigned in empty parts of the segmentation map (crosses -- flavor 2). We introduce a small horizontal offset of the different markers and only include the error-bars for the flavor 2 case in order to improve readability. We also add horizontal solid line to mark 100\% completeness, and two dashed lines at 80\% and 60\% completeness as visual help. Detections are made in IR-weighted images.}
    \label{fig:completeness}
\end{figure*}

Results from this test are summarized in Figure~\ref{fig:completeness}. For flavor 2 simulations, we find that completeness is above 60\% for magnitudes brighter than 29, and above 80\% for most of the clusters up to magnitude $\sim 28$.  This confirms the estimated depth of our images and catalogs. When we do not avoid overlap between the objects (i.e., using flavor 1 simulations)  the overall completeness decreases noticeably with respect to the simulation where we avoid blending. Thus, we expect the actual completeness of our catalog to be somewhere in between the pessimistic scenario portrayed by our flavor 1 simulations and the optimistic case of our flavor 2 simulations. 

\subsection{Photometric uncertainties}
\label{ssec:photometric_unc}
Given the complexity of the process to detect objects and measure their flux in different bands, an analytic derivation of their associated uncertainties (in flux, position, and other derived quantities) becomes almost intractable. Forward-modeling will enable us to characterize such uncertainties. To this purpose, we inject another set of simulated objects into the original HFF mosaics. This time the simulated sources are realistic galaxy models from the GREAT3 challenge~\citep{2014JInst...9C4031M}, created by means of  \texttt{GalSim}~\citep{2015A&C....10..121R}. \texttt{GalSim} is a modular, open-source code to render any type of astronomical images. We opted for the GREAT3 library because it includes models with realistic morphology, obtained by fitting deep COSMOS images  with multiple S\'{e}rsic profiles \citep[see][]{2014ApJS..212....5M, 2015MNRAS.450.2963M}. Given that the galaxies from the GREAT3 challenge available in the \texttt{GalSim} catalog only have information for the F140W band, we identify these galaxies in the \citet{2016ApJS..224...24L} catalog using their ID values to obtain the magnitudes in the other bands used in our study. In this way we have realistic colors for our injected galaxies across the entire photometric baseline, with the assumption that their morphology does not vary significantly from band to band. We also ensure that the centroid positions are the same across different bands. We convolve each galaxy profile with the corresponding PSF. The 15 $\times$ 15 square arcsec cutout containing the injected objects (and no additional noise) are then added to the original image. We then repeat the photometric extraction process in a similar fashion to what have been done with our data. In particular, we inject 80 objects randomly placed in each image and generate 12 different realizations (images) of those objects in all the HFF bands.

\begin{figure*}[t]
    \centering
    \includegraphics[width=0.31\textwidth]{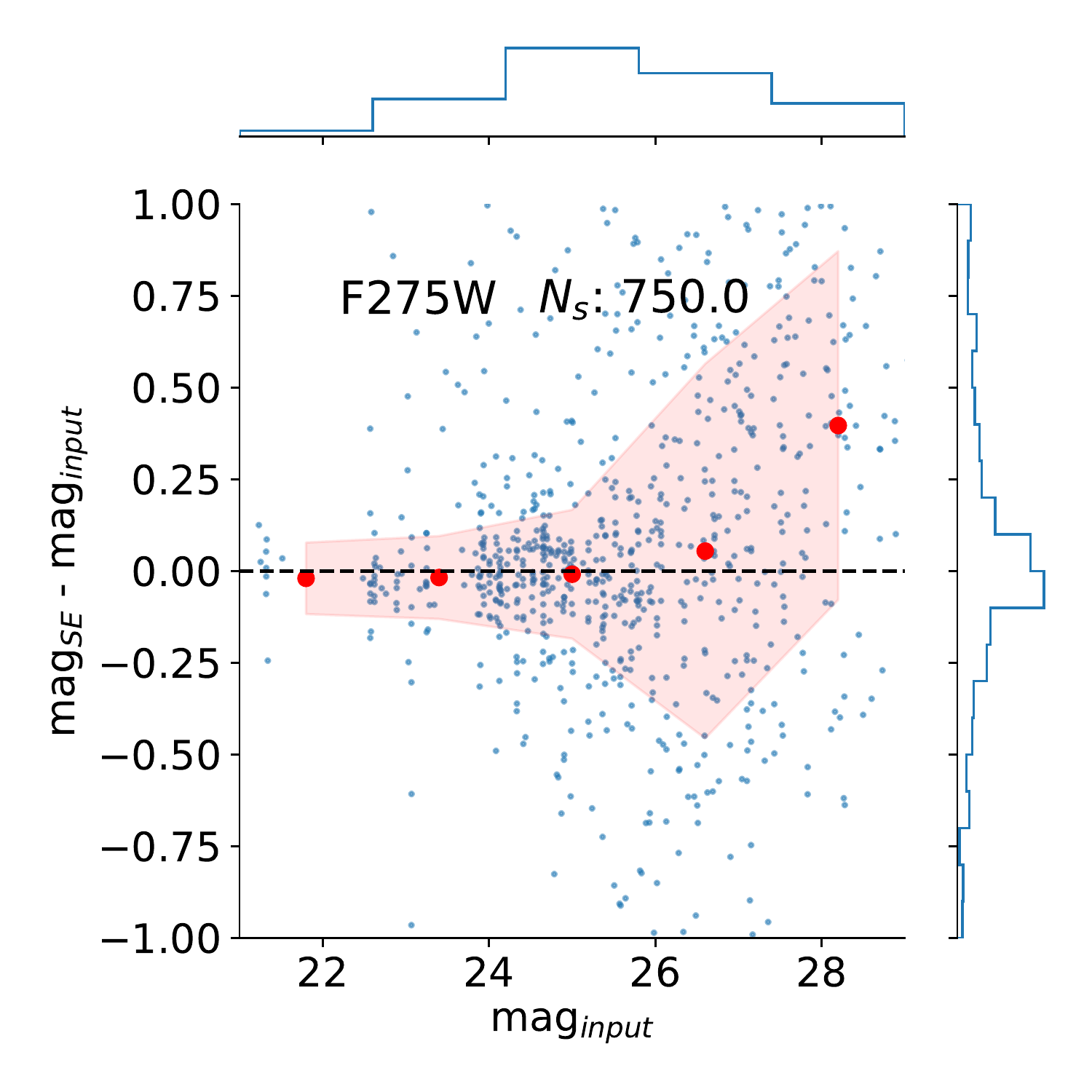}
    \includegraphics[width=0.31\textwidth]{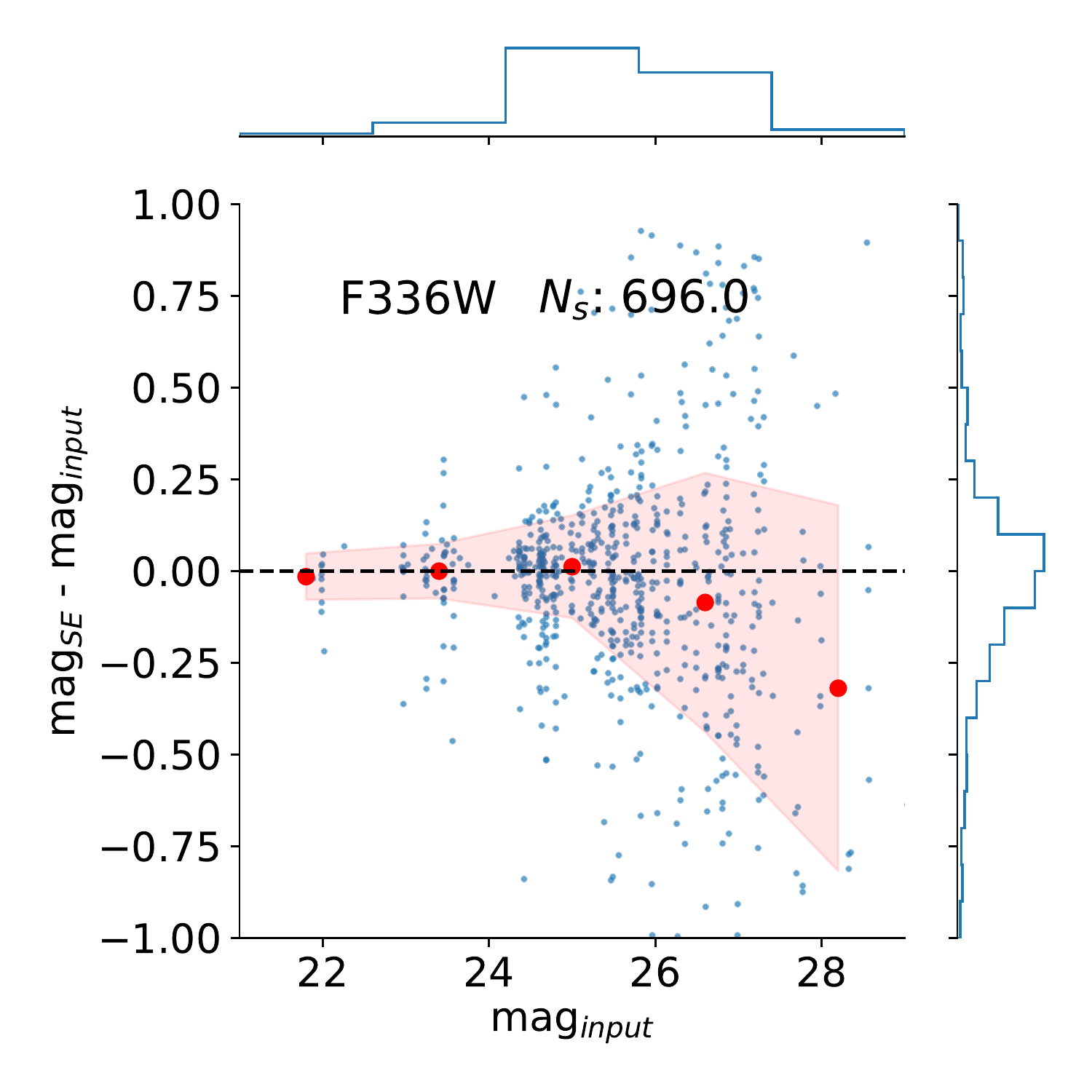}
    \includegraphics[width=0.31\textwidth]{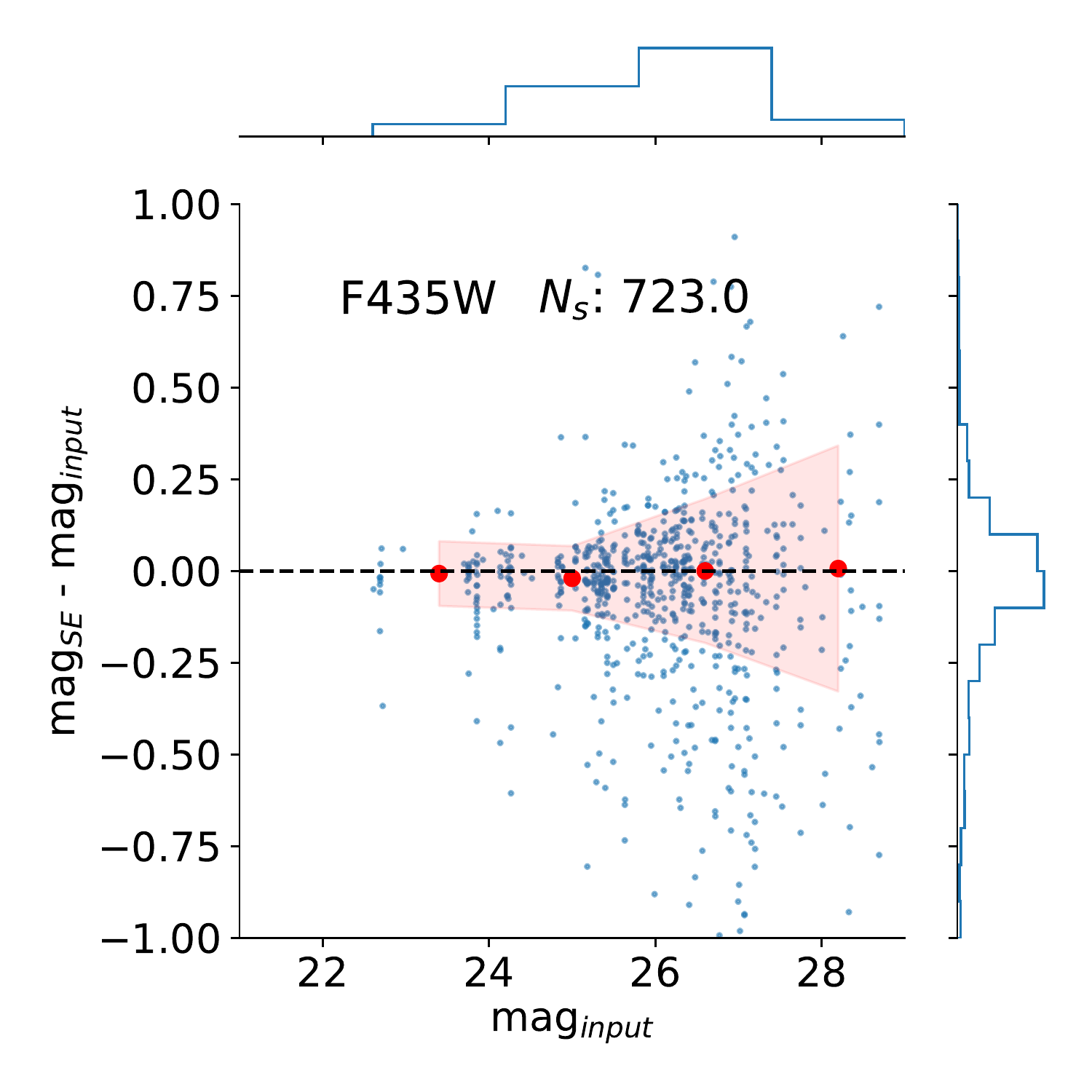}
    \includegraphics[width=0.31\textwidth]{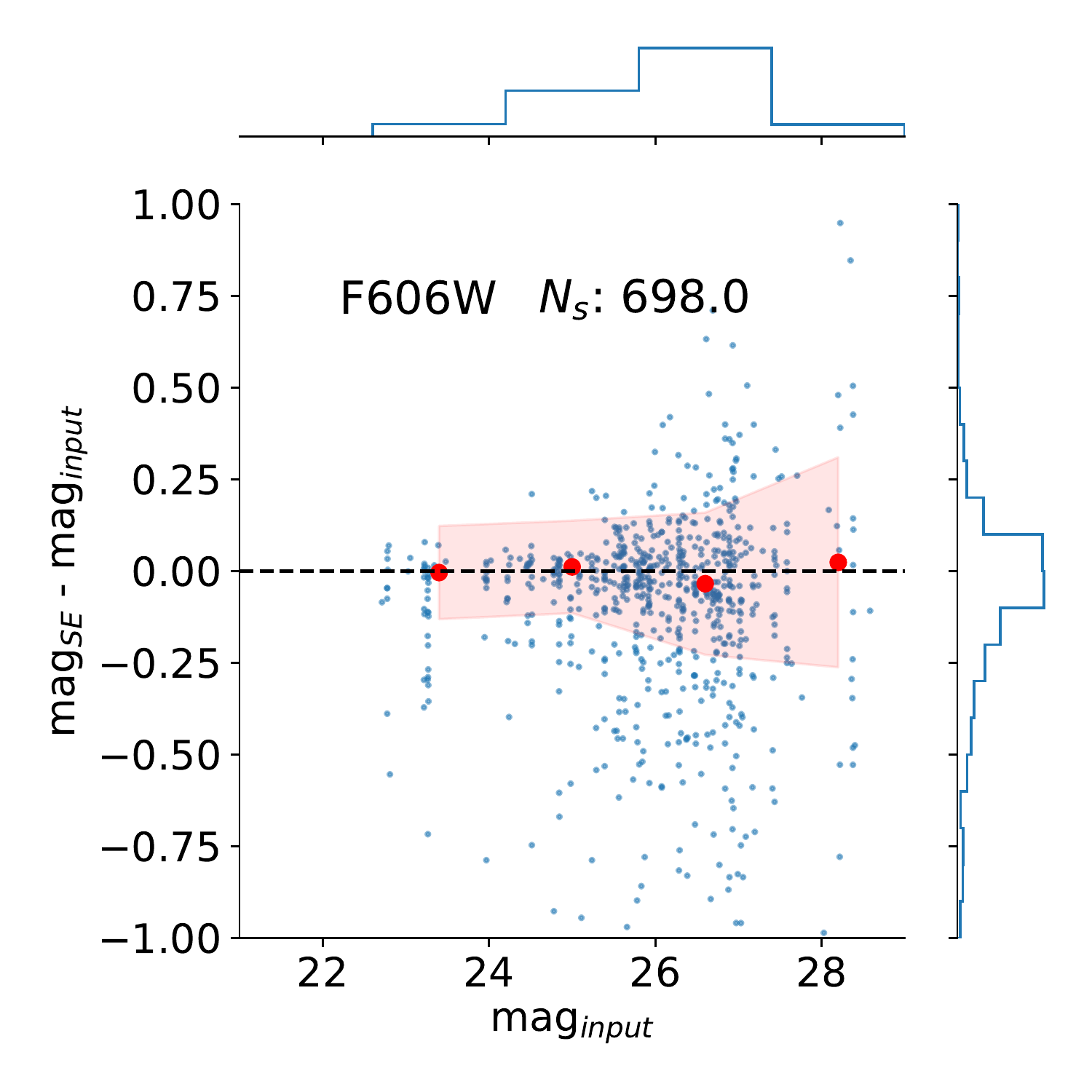}
    \includegraphics[width=0.31\textwidth]{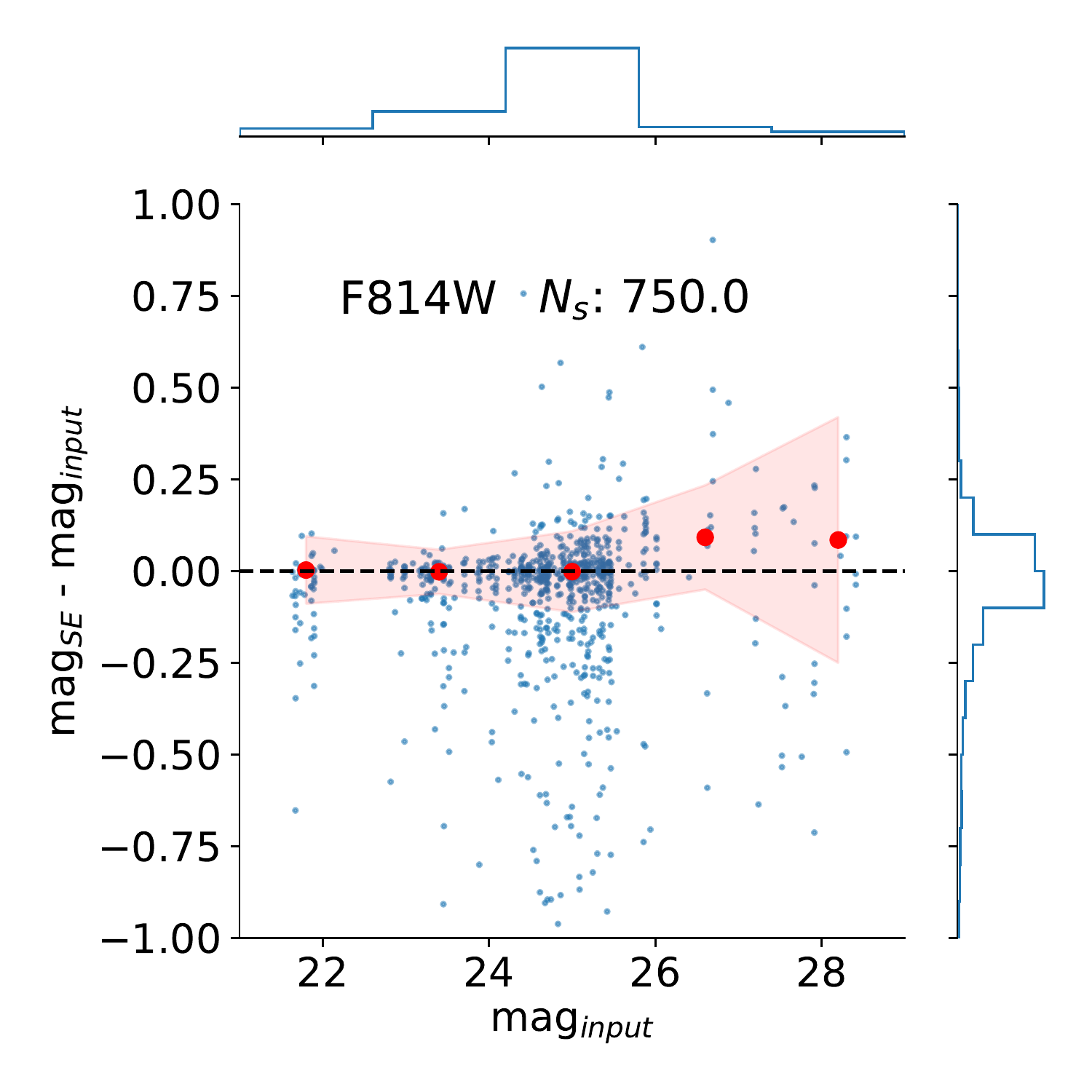}
    \includegraphics[width=0.31\textwidth]{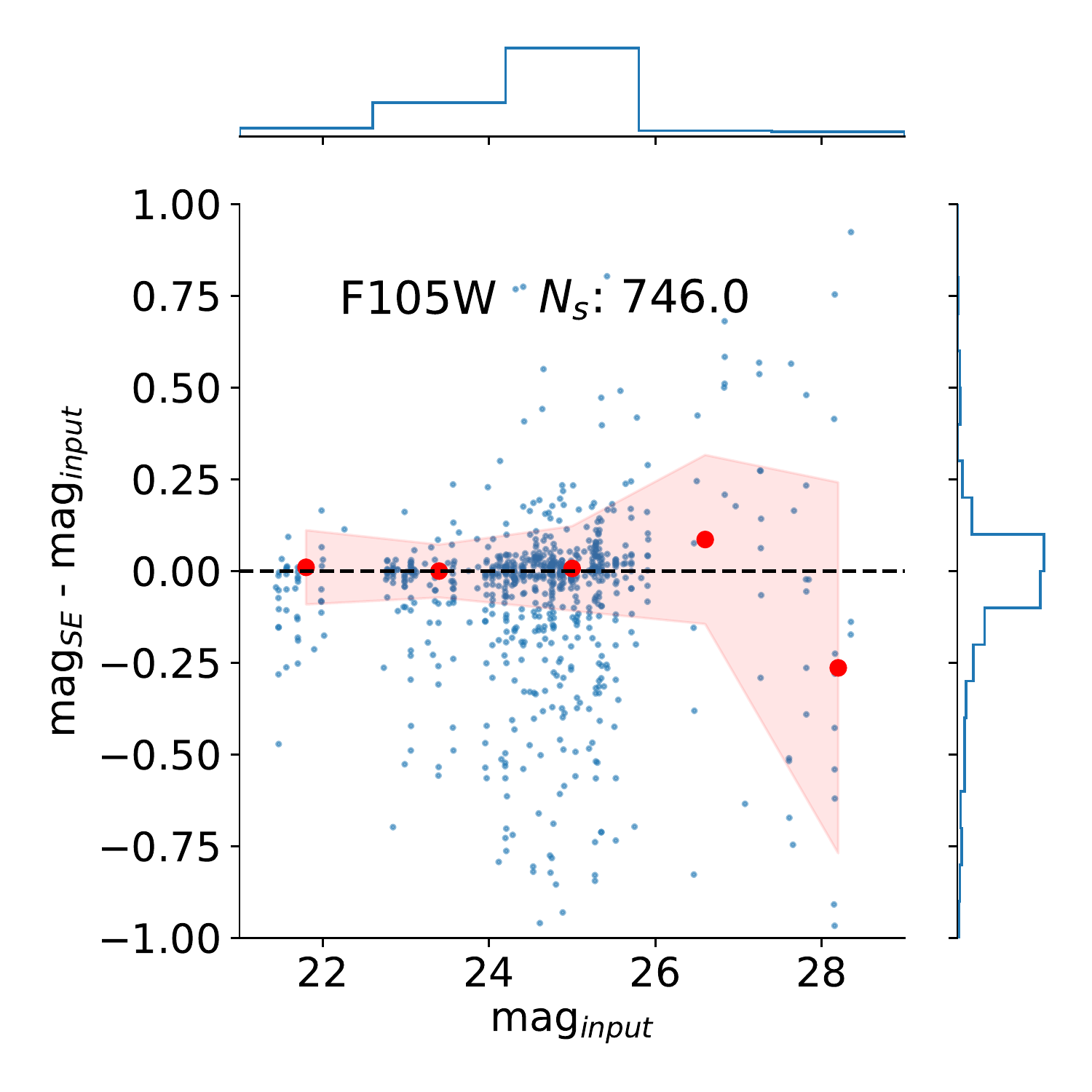}
    \includegraphics[width=0.31\textwidth]{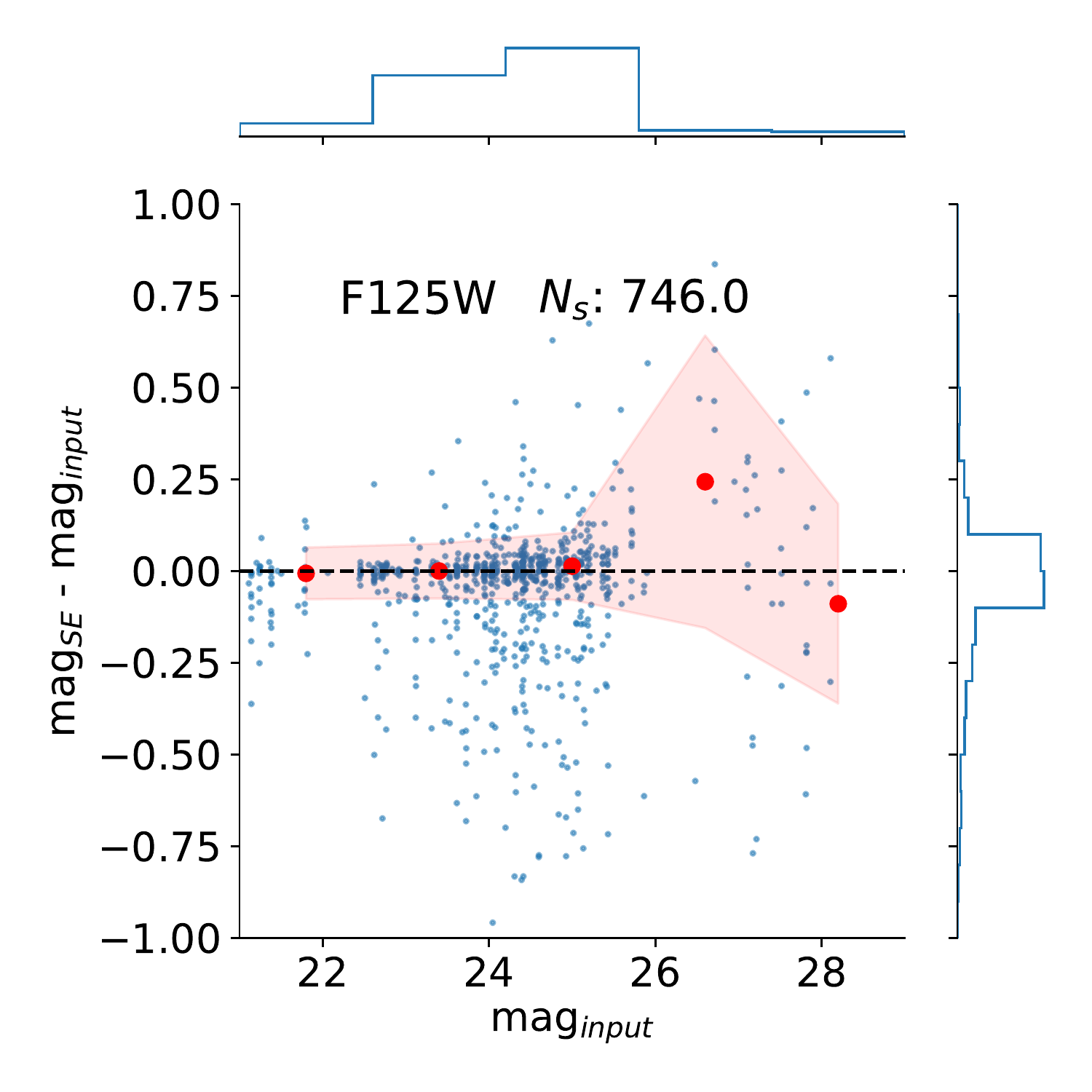}
    \includegraphics[width=0.31\textwidth]{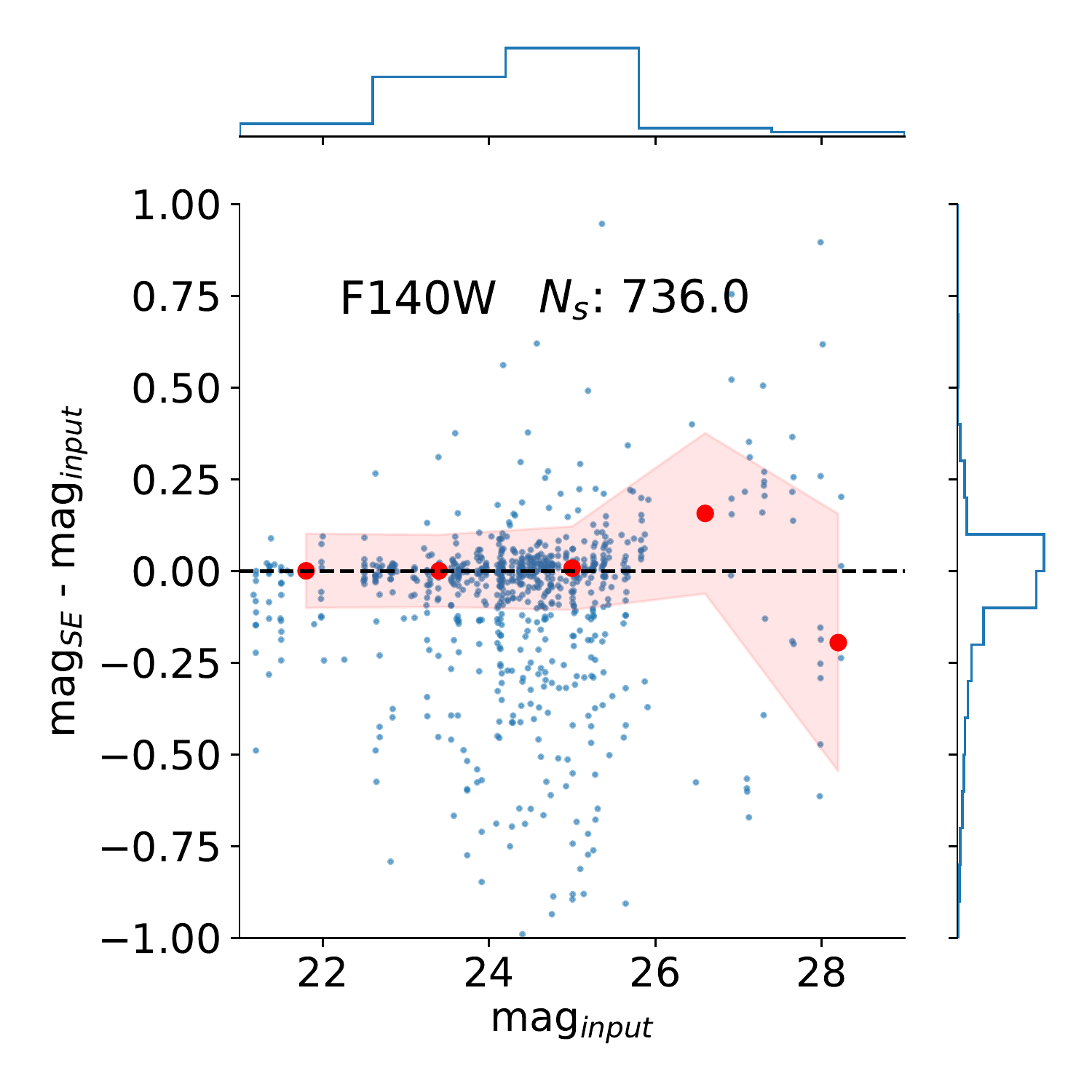}
    \includegraphics[width=0.31\textwidth]{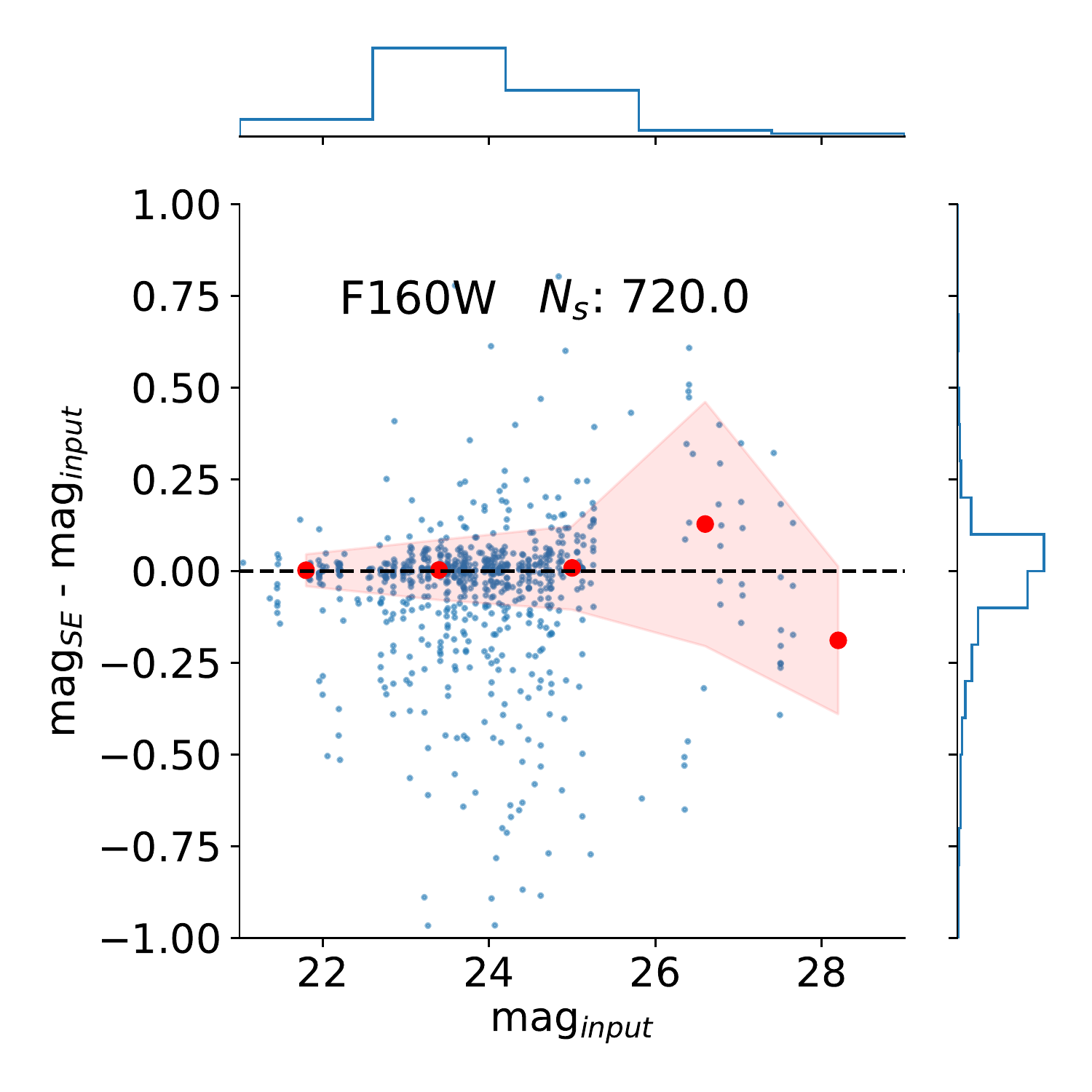}
    \caption{Photometric residuals between the simulated objects injected in the Abell 2744 cluster and the input fluxes, plotted as a function of magnitude for different bands. In each panel, blue small dots are the individual objects residuals, while the red dots represent the median in bins with 1.6\,mag width. The shaded region encompasses the rescaled interquartile range, $0.7413 (Q3 - Q1)$ as a robust estimator of 1$\sigma$ uncertainty~\citep{astroMLText}. The magnitude distribution of each injected sample is projected on the $x$ axis on the top of the corresponding panel, while the photometric residual distribution is shown in the right-hand side of the $y$ axis. }
    \label{fig:error_characterization}
\end{figure*}
\begin{figure*}[t]
    \centering
    \includegraphics[width=0.31\textwidth]{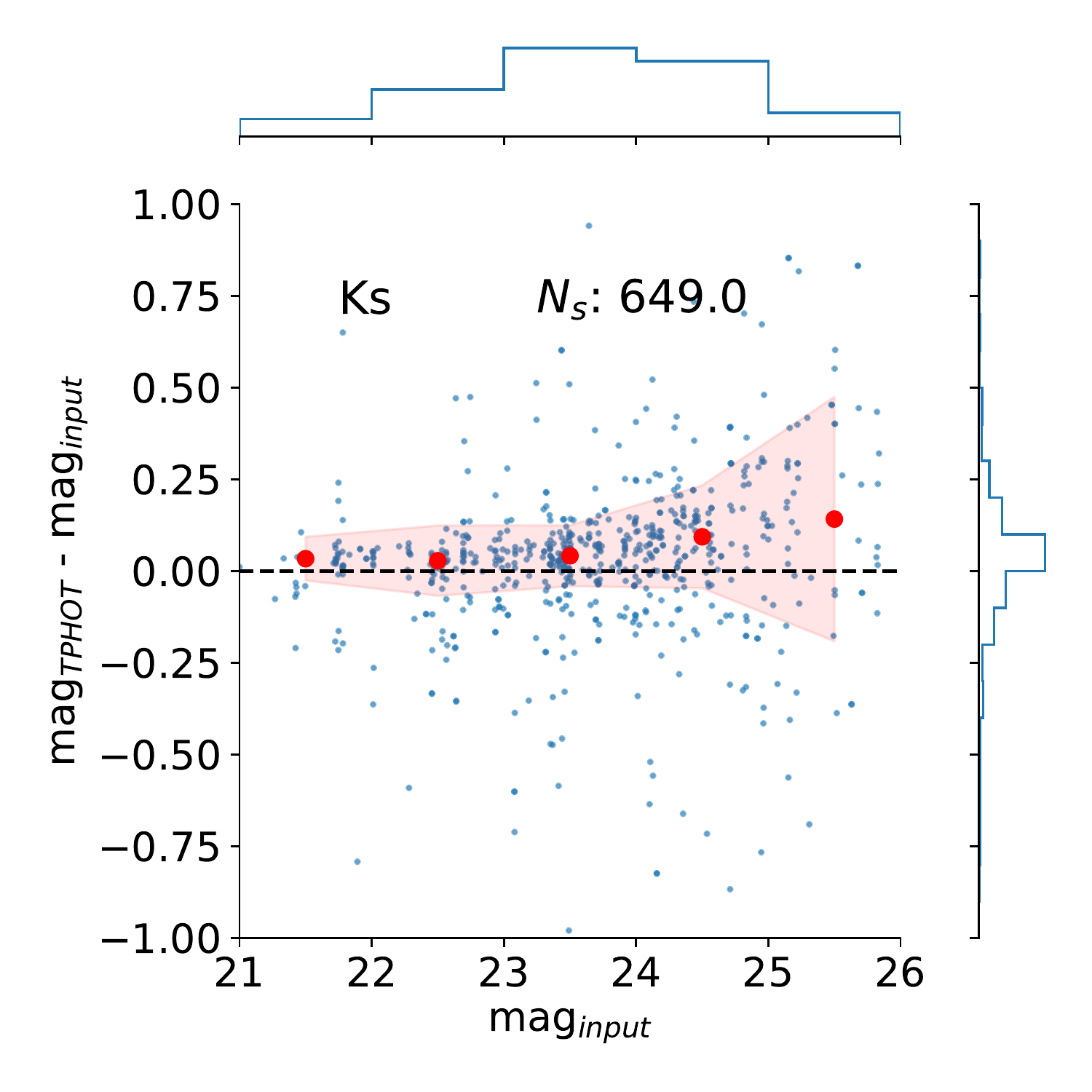}
    \includegraphics[width=0.31\textwidth]{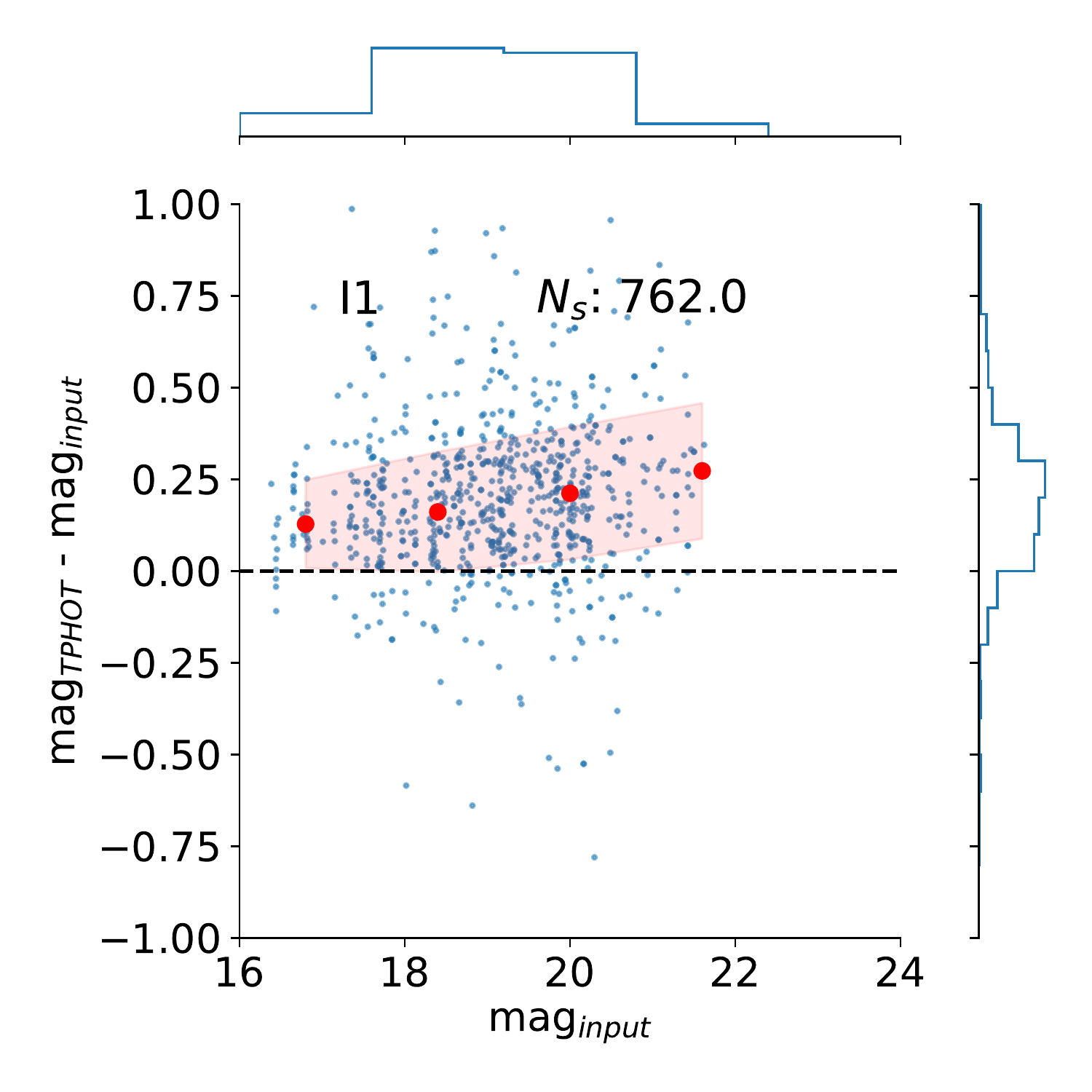}
    \includegraphics[width=0.31\textwidth]{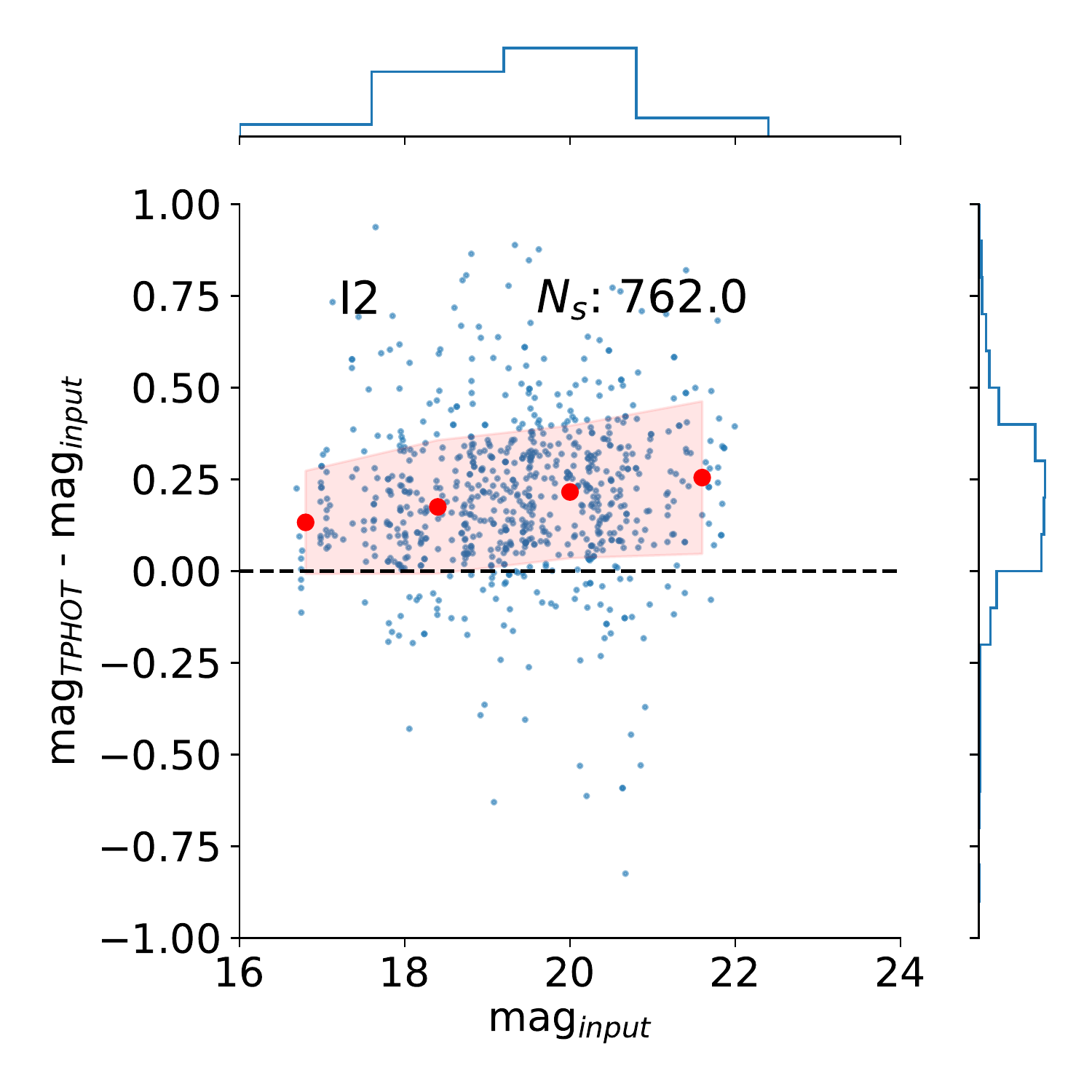}
    \caption{Photometric residuals in ancillary data ($K_\mathrm{s}$ band and \textit{Spitzer}/IRAC channels). See the caption of Figure~\ref{fig:error_characterization} for more details. We find a photometric bias between input and measured magnitude that we correct using the median value of the residual in each band.}
    \label{fig:error_characterization2}
\end{figure*}

Additionally, we also process the cutouts that only contain the injected objects using the same pipeline; we use the outcomes as ``ground truth'' to be compared to the catalog resulting from the injection-recovery procedure.  To match the two samples we use the same strategy as in Section \ref{sec:completeness}. The photometric residuals for each match are then measured. The results are shown in Figures~\ref{fig:error_characterization} and \ref{fig:error_characterization2}, indicating an unbiased estimate of input magnitudes across different HST bands included in this study ($\langle \mathrm{mag}_{meas} - \mathrm{mag}_{input} \rangle < 0.1$ mag), even after ICL subtraction. However, for the IRAC and K bands we find that the recovered objects appear, on average, dimmer than the input. Since the dependence in flux is mild, we decided to include a constant offset to correct for these residuals. In particular, we shift the recovered magnitudes by the median of the residuals found in Figure~\ref{fig:error_characterization2}. These offsets are $\Delta \textrm{mag}= (0.04, 0.189, 0.195)$ for Ks, IRAC channel 1, and 2, respectively, and are subtracted from the measured magnitude. We explored different possibilities for the origin of this correction including: size of segments in the segmentation map, and the smoothing radius of the ICL profile. However, after modifying these we could not find any significant changes in the initial offsets that we measured. Given their mild dependence on the flux, and the fact that the effect seems to be milder in Ks, where the PSF is closer to that in the HST bands, we concluded that the observed offsets may be a consequence of numerical instabilities at the time of using the PSF kernel that relates the low and high-resolution images. We notice that the injected sources cover different magnitude ranges for different bands, due to the fact that we are injecting galaxies with realistic SEDs from COSMOS. Additionally, we see a different behavior when we compare different bands. In particular, the uncertainties for the recovered fluxes in the F160W band seem to be smaller than in most of the other bands. The reason behind these differences is the larger depth of the images in F160W compared to the other bands. It is particularly interesting to notice larger uncertainties derived for F275W compared to the other HST bands. This is a combination of two factors: first, the SNR for F275W is the lowest of all HST bands; second, this is also the bluest band, where we expect the largest morphological differences with respect to our segmentation map.

\begin{figure}
    \centering
    \includegraphics[width=0.48\textwidth]{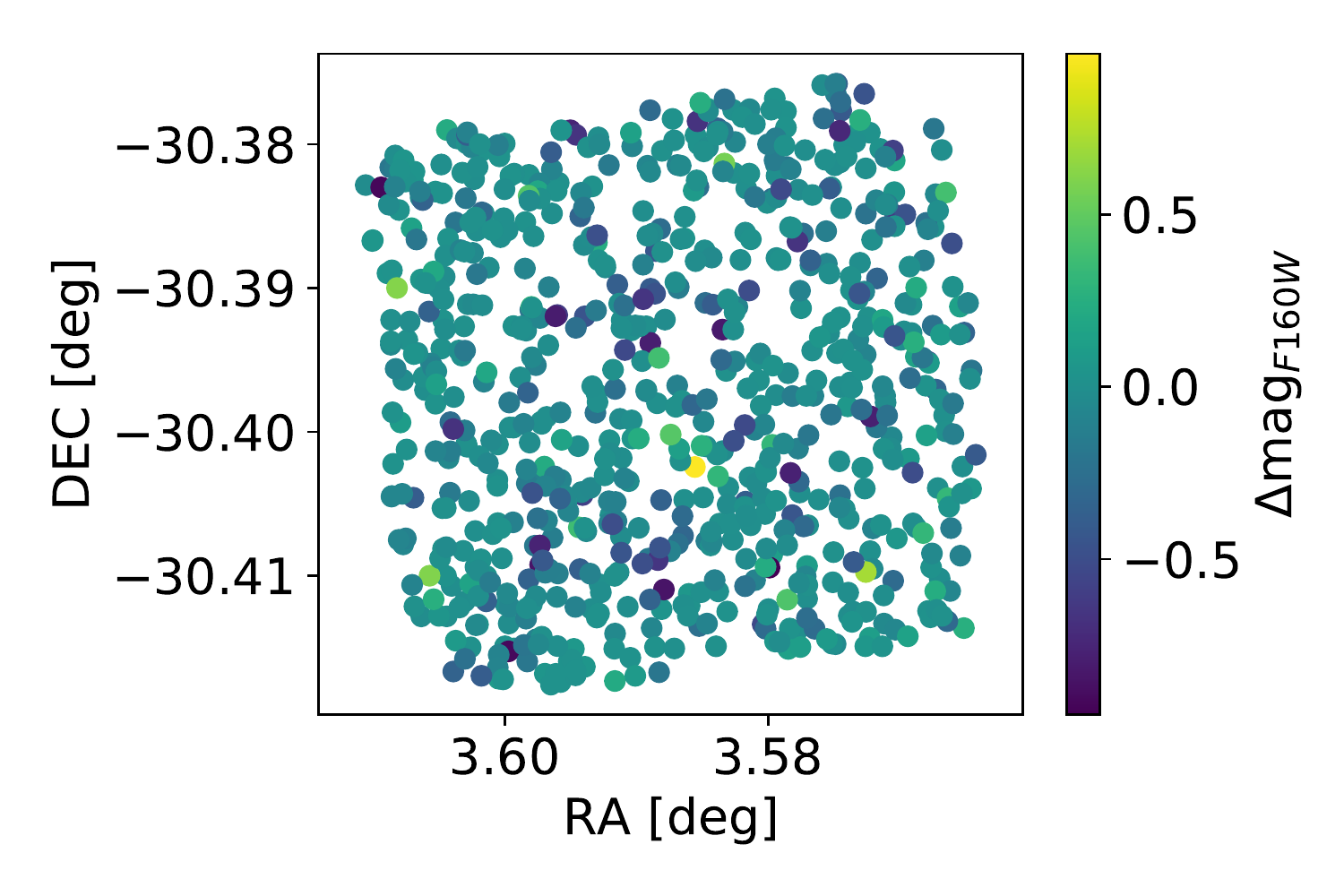}
    \caption{$\Delta \rm{mag} = \rm{mag}_{output} - \rm{mag}_{input}$ for the F160W band as a function of the position of detected injected objects for the Abell 2744 cluster. There is no clear correlation between the residual and the position of the injected object in the field. Similar results are observed for the rest of the bands analyzed.}
    \label{fig:photometry_position}
\end{figure}

We only perform the simulations for the cluster fields. The reason is that we expect these fields to outperform the parallel fields as the later are not affected by the ICL and are less crowded. As a consequence, our measurements for the clusters will set an upper limit for the photometric biases and uncertainties in the parallel fields. As shown in Figure~\ref{fig:photometry_position}, the photometric bias appears to be uncorrelated with position when comparing cluster core offsets with those in the outer regions. Since the source extraction steps in both cluster fields, and parallel fields are the same, the validation tests of the photometric measurements in the cluster fields are also applicable on the parallel fields. 
%

\begin{figure}
    \centering
    \includegraphics[width=0.45\textwidth]{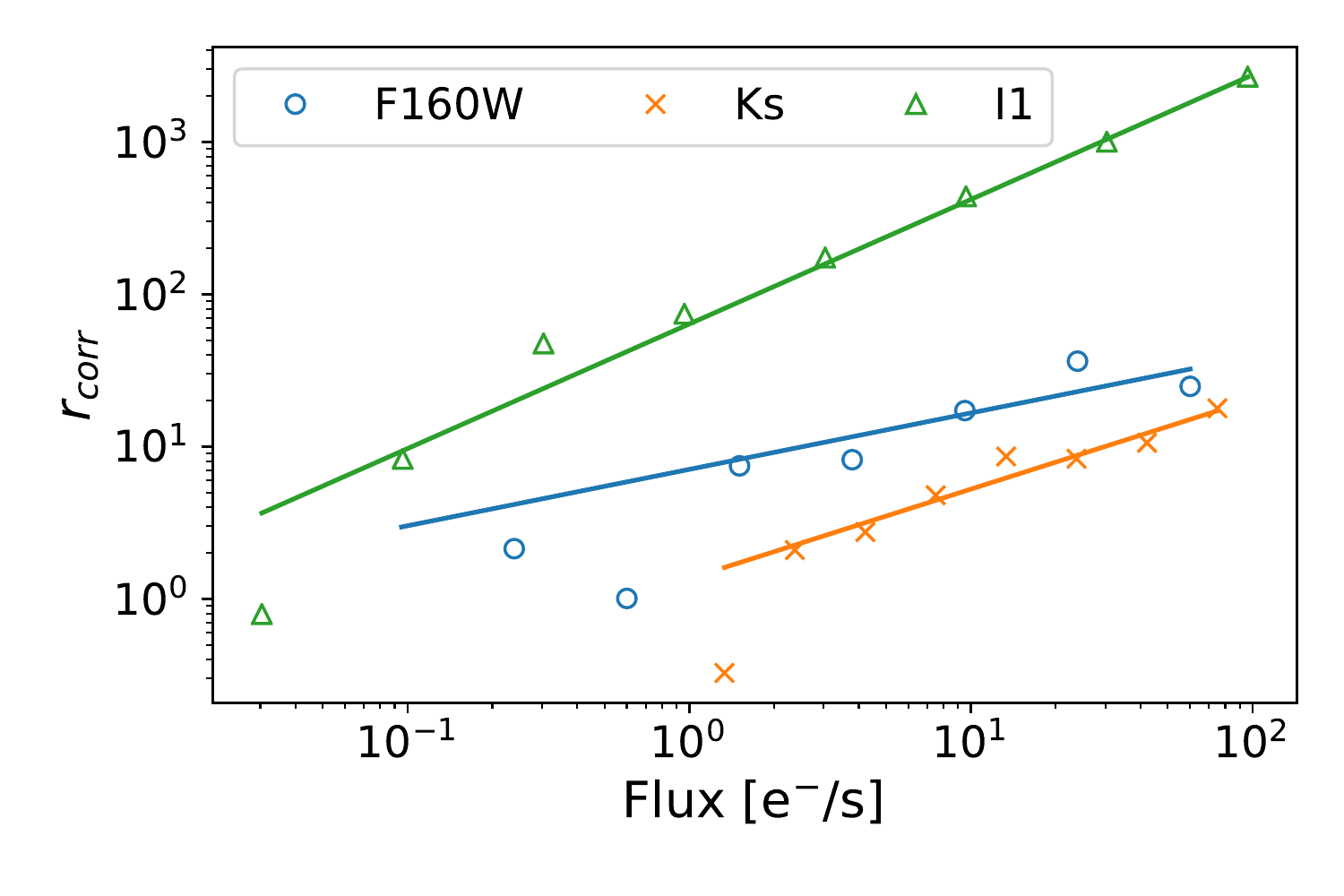}
    \caption{Ratio between the flux uncertainty recovered via forward-modeling and the reported flux uncertainty by Source Extractor in three different bands: F160W (open circles), Ks ($\times$), and IRAC Channel 1 (open triangles).  For each band, a solid line shows the corresponding best-fit model, i.e.\ the power law $r_\mathrm{corr}(F)=a F^{b}$. The best-fit values found for $a$ and $b$ in the different bands can be found in Table~\ref{tab:err-corr}.}
    \label{fig:error_correction}
\end{figure}

Our forward modeling approach allows us to estimate the accuracy of the flux errors as reported by \texttt{SE}. Using the injected galaxies we compute the forward-modeling estimation of the uncertainty in the flux, which we call $\delta F_\mathrm{FM}$, as the RMS of the difference between the measured flux for the injected sources and the input flux, $\delta F_\mathrm{FM} = \sqrt{\langle \left(F_\mathrm{meas} - F_\mathrm{input}\right)^{2} \rangle} $, where $F_\mathrm{meas}$ is the measured flux and $F_\mathrm{input}$ is the injected flux. This estimation is performed in linearly spaced magnitude bins (with a binwidth of 1.6 mag). We then compare $\delta F_\mathrm{FM}$ 
to the mean uncertainties reported by \texttt{SE} in each of the bins, i.e. $\langle \delta F_\mathrm{SE} \rangle$. This is done by  studying the ratio $r_\mathrm{corr} = \frac{\delta F_{FM}}{\delta F_{SE}}$ as a function of flux. We fit the resulting ratio to an exponential model $r_\mathrm{corr} = A F^{b}$ and use this fit to correct the reported uncertainties for each detected object. The ratio and resulting fit are shown in Figure~\ref{fig:error_correction} for F160W, $K_\mathrm{s}$ and IRAC channel 1. The best-fit values for $a$ and $b$ can be found in Table~\ref{tab:err-corr}. By binning the data we lose information about the effects of blending and correlations between flux uncertainties for individual galaxies, but get a more robust estimation of the overall uncertainty of the ensemble.

\begin{deluxetable}{lcc}[h!]
\tablecaption{Flux uncertainty correction factors obtained for the different bands used in this work. }
\label{tab:err-corr}
\tablehead{\colhead{Band} & \colhead{\hspace{.75cm}$a$}\hspace{.5cm} & \colhead{$b$}\hspace{.5cm}}
\startdata
F275W & 11.00 & 0.32 \\
F336W & 3.67 & 0.11 \\
F435W & 3.32 & 0.23 \\
F606W & 3.23 & 0.27 \\
F814W & 7.29 & 0.24 \\
F105W & 10.46 & 0.21 \\
F125W & 7.54 & 0.08 \\
F140W & 8.88 & 0.37 \\
F160W & 5.17 & 0.45 \\
Ks & 1.58 & 0.29 \\
I1 & 42.31 & 0.78 \\
I2 & 46.94 & 0.71 \\
I3 & 3.38 & 0.74 \\
I4 & 1.09 & 1.18 \\
\enddata

\end{deluxetable}

\subsection{Comparison with previous work}

\begin{figure*}
    \centering
    \includegraphics[width=0.32\textwidth]{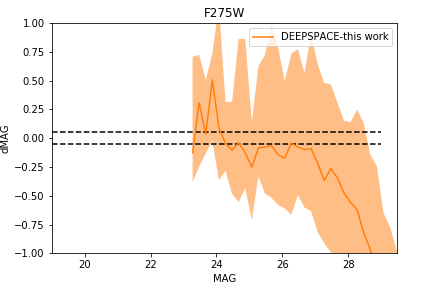}
    \includegraphics[width=0.32\textwidth]{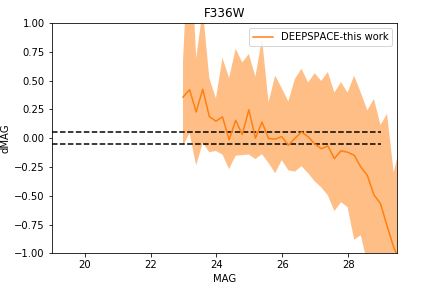}
    \includegraphics[width=0.32\textwidth]{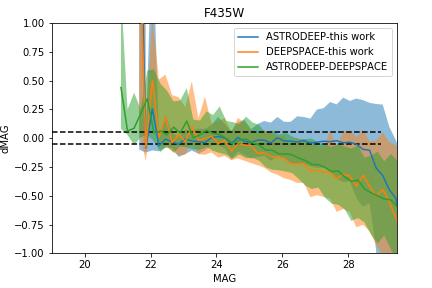}
    \includegraphics[width=0.32\textwidth]{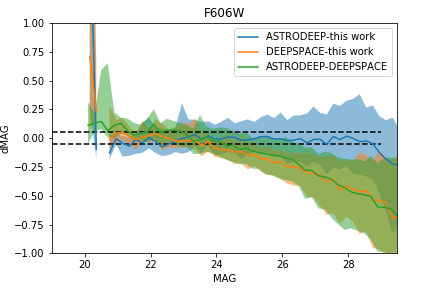}
    \includegraphics[width=0.32\textwidth]{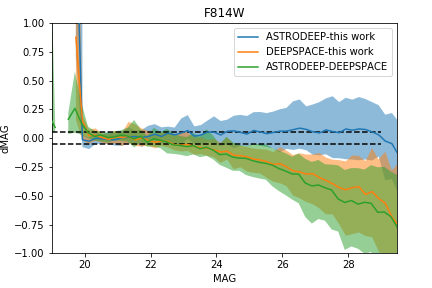}
    \includegraphics[width=0.32\textwidth]{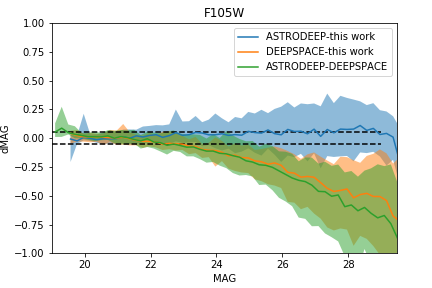}
    \includegraphics[width=0.32\textwidth]{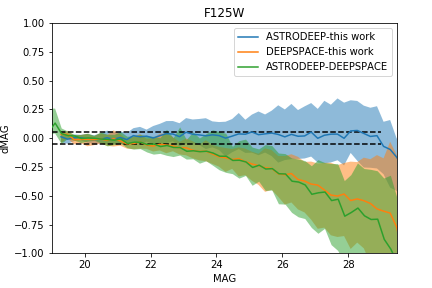}
    \includegraphics[width=0.32\textwidth]{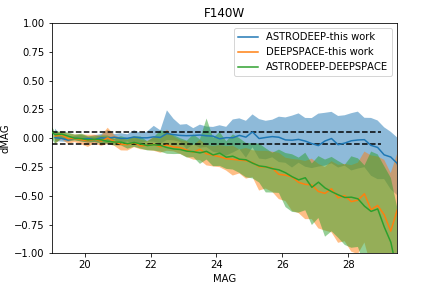}
    \includegraphics[width=0.32\textwidth]{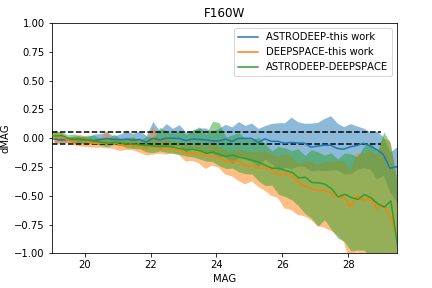}
    \includegraphics[width=0.32\textwidth]{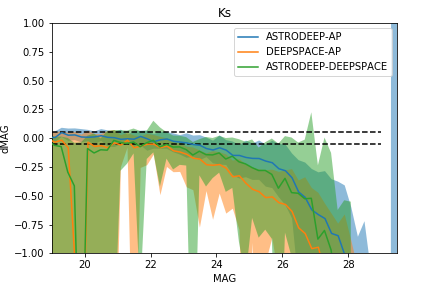}
    \includegraphics[width=0.32\textwidth]{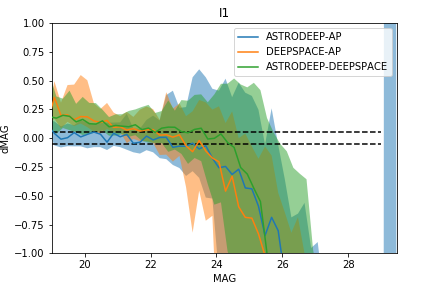}
    \includegraphics[width=0.32\textwidth]{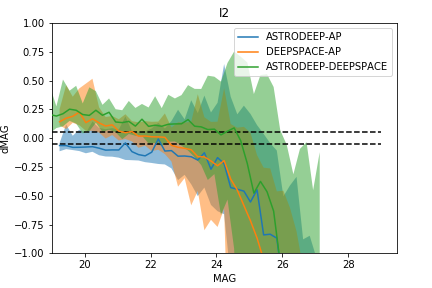}
    \caption{Comparison of our extracted photometry with that of the existing data produced by the ASTRODEEP and DEEPSPACE teams. Shown are the $\Delta$mag where data exists for cluster-only photometry in each band. The black dashed lines represent magnitude offsets of $\pm0.05$. The solid lines represent the running median and the shaded region represents the interquartile range of those measurements. 
    }
    \label{fig:photcomparison}
\end{figure*}

There are other independent studies of the Frontier Fields. Given the complexities of extracting photometry in these deep and crowded fields, it poses a great challenge to determine photometric properties of the objects detected with the desired accuracy. Therefore, we believe there is merit in both exploring the limits of our photometric pipeline and comparing with previous results. Figure \ref{fig:photcomparison} shows how our photometry compares with that of the previous teams, ASTRODEEP (\cite{2016A&A...590A..30M}, \cite{2016A&A...590A..31C}, \cite{2017A&A...607A..30D}, \cite{2019MNRAS.489...99B}) and DeepSpace (\cite{2018ApJS..235...14S}) for intersecting bands. For the HST bands, we note that there is general agreement in the photometry between the three methods within 0.05 magnitudes up to $m_{AB}\sim23$ mag. Beyond this magnitude, our photometry agrees best with the ASTRODEEP team, which could be the result of our reduction procedure being more similar to theirs. Furthermore, we see that the three datasets are statistically compatible up to magnitude $\sim 25.5$ in most bands. However, in the fainter end, there are statistically significant differences between the DeepSpace and ASTRODEEP datasets, and the data presented in this work. These differences are probably due to the difference in modelling, as the measure of fluxes/magnitudes depends on the concrete choices for apertures, etc. However, we check that the colors are consistent among these datasets for the HST and Ks bands, which can be inferred by the consistent behavior found in most bands. Thus, for applications where colors are most important (e.g., photometric redshifts) any of these datasets should give similar results (as evidenced by our photometric redshifts). 



\section{Photometric redshifts}
\label{sec:photz}

Having the self-consistent SEDs for individual galaxies in each of the HFF clusters and their associated parallel fields, we now measure the photometric redshifts using  \texttt{LePhare}  \citep{arnouts99,ilbert06}. The code fits galaxy templates to the observed SEDs to derive a redshift likelihood function ($\mathcal{L}$) for each source. Photometric redshift estimates ($z_\mathrm{phot}$) are then defined as the median of $ \mathcal{L}(z)$. The template library adopted here is the same used in \cite{laigle16} since it was shown that it works efficiently across a wide redshift range (up to $z\sim6$). The library is built by spiral and elliptical galaxy models from \citet{polletta07} along with other ones derived from the stellar population synthesis model of \citet{bruzual&charlot03}. The latter is used to reproduce both young quiescent and starburst galaxies. For each model we produce different templates by changing their dust attenuation between $0<E(B-V)\leq0.5$ and assuming either \citet{prevot84} or \citet{calzetti2000} extinction laws depending on the galaxy type (see \citealp{laigle16} for more details). When \citet{calzetti2000} is chosen, we produce three alternate versions by adding different parametrization of the 2,175\,\AA  bump \citep[see][]{ilbert09}. Absorption by the intervening inter-galactic medium is also implemented as prescribed in ~\citet{madau95}. We also add the main nebular emission lines in rest-frame optical to the templates corresponding to the star-forming SEDs, modelling line fluxes and their ratios as in ~\citet{shun20}. 

Before running \texttt{LePhare}, a modification is required  to the HST bands: the fraction of the flux lost by \texttt{SE} isophotal measurements (FLUX\_ISO) must be taken into account. FLUX\_ISO provide more accurate colors than FLUX\_AUTO, which is another \texttt{SE} photometric measurement specifically designed to recover total flux by means of an adaptive aperture \citep[cf.][]{kron80}; therefore the former are prefreed for $z_\mathrm{phot}$ computation.  However, our ancillary photometry is extracted with \texttt{T-PHOT}, which does not provide an equivalent to FLUX\_ISO. Therefore, we include in our baseline $K_\mathrm{s}$ and IRAC total fluxes  while HST bands fluxes are rescaled by a factor 
\begin{equation}
    f_\mathrm{tot} =  \frac{\sum_i w_i (\mathrm{FLUX\_AUTO}/\mathrm{FLUX\_ISO})_i}{\sum_i w_i},
\end{equation}
i.e.\ the weighted mean of the AUTO-to-ISO flux ratio summed over the observed HST bands. Weights are defined as $w= \sqrt{\sigma_\mathrm{AUTO}^2 +\sigma_\mathrm{ISO}^2}$, i.e.\ square root of the sum in quadrature of the corresponding \texttt{SE} errors. 

In order to empirically correct for systematic effects we perform a calibration that relies on the spectroscopic redshifts ($z_\mathrm{spec}$) available in each field. The sample includes spectroscopic data from several programmes retrieved from the NASA/IPAC Extragalactic Database\footnote{The NASA/IPAC Extragalactic Database (NED) is funded by the National Aeronautics and Space Administration and operated by the California Institute of Technology (\url{http://ned.ipac.caltech.edu}).} 
The spectroscopic campaigns contributing to the majority of the data set are described in \citet[][]{owers11,ebeling14,richard14,balestra16,treu15,schmidt14,lagattuta17,mahler18}. We also include more recent observations carried out by \citet{lagattuta19} in Abell 370, which were not included in previous photometric redshift analysis in the literature. 
The calibration procedure is as follows. First, we run \texttt{LePhare} on the HFF galaxies with known spectroscopic redshifts after fixing their redshifts to their spectroscopic $z_\mathrm{spec}$ value. We only consider galaxies with a reliable spectroscopic redshift, i.e.\ with a ``quality flag'' $>$3.\footnote{It is common practice that the persons responsible for the $z_\mathrm{spec}$ measurements assign a quality flag, ranging from 1 to 4, following the scheme proposed for the first time in \citet{lefevre04}, where a flag equal to 4 corresponds to the highest confidence level} 
We do not apply any magnitude cut. Once obtained the best-fit solution from \texttt{LePhare} (i.e., the fitting model resulting in the smallest $\chi^2$) we measure in each photometric band the difference between the observed flux and the prediction by the best-fit template. To find the systematic offset (in a given band) we compute the average difference in the spectroscopic sample.  Since results are similar in the six fields, we eventually consider only the offsets measured for the Abell 370 cluster because it has the largest number of spectroscopic sources (320 of them reliable spectroscopic redshifts). These  offsets, when applied to the photometric baseline, will compensate for a possible bias in the template library and/or for calibration issues in data reduction. In the HST bands the corrections are between 2 and 5\%, except for the F425W filter which is 8\%; a similar value is found for $K_\mathrm{s}$ while in the IRAC channels 1 and 2 the correction is a factor 1.11. We note that similar offsets (namely, 0.1\,mag in both IRAC channels 1 and 2) have been found by \texttt{LePhare} also in another extragalactic survey \citep{mehta18}.  All offsets are quoted in Table~\ref{tab:photo-calib}, with the exception of IRAC channels 3 and 4 since the systematics here would be mainly driven by the absence of dust re-emission in the templates \citep[see][]{mehta18}. To take such a limitation into account we increased the error bars in these two channels by adding 0.5\,mag in quadrature, under-weighing in practice their contribution to the fit.

\begin{deluxetable}{lc}[h!]

\tablecaption{Multiplicative factors applied to each band in the photo-z calibration step. The offsets found in the HST bands are consistent with the median magnitude residual values found in Figure~\ref{fig:error_characterization}, showing the robustness of our forward-modelling procedure.}
\label{tab:photo-calib}
\tablehead{\colhead{Band} & \colhead{Multiplicative Factor}\hspace{.5cm}}

\startdata
F275W & 1.04 \\
F336W & 1.02 \\
F435W & 1.07 \\
F606W & 0.99 \\
F814W & 0.96 \\
F105W & 0.98 \\
F125W & 1.01 \\
F140W & 1.02 \\
F160W & 1.01 \\
Ks & 1.00 \\
I1 & 1.11 \\
I2 & 1.11 \\
\enddata

\end{deluxetable}



\begin{figure}
\includegraphics[width=0.49\textwidth]{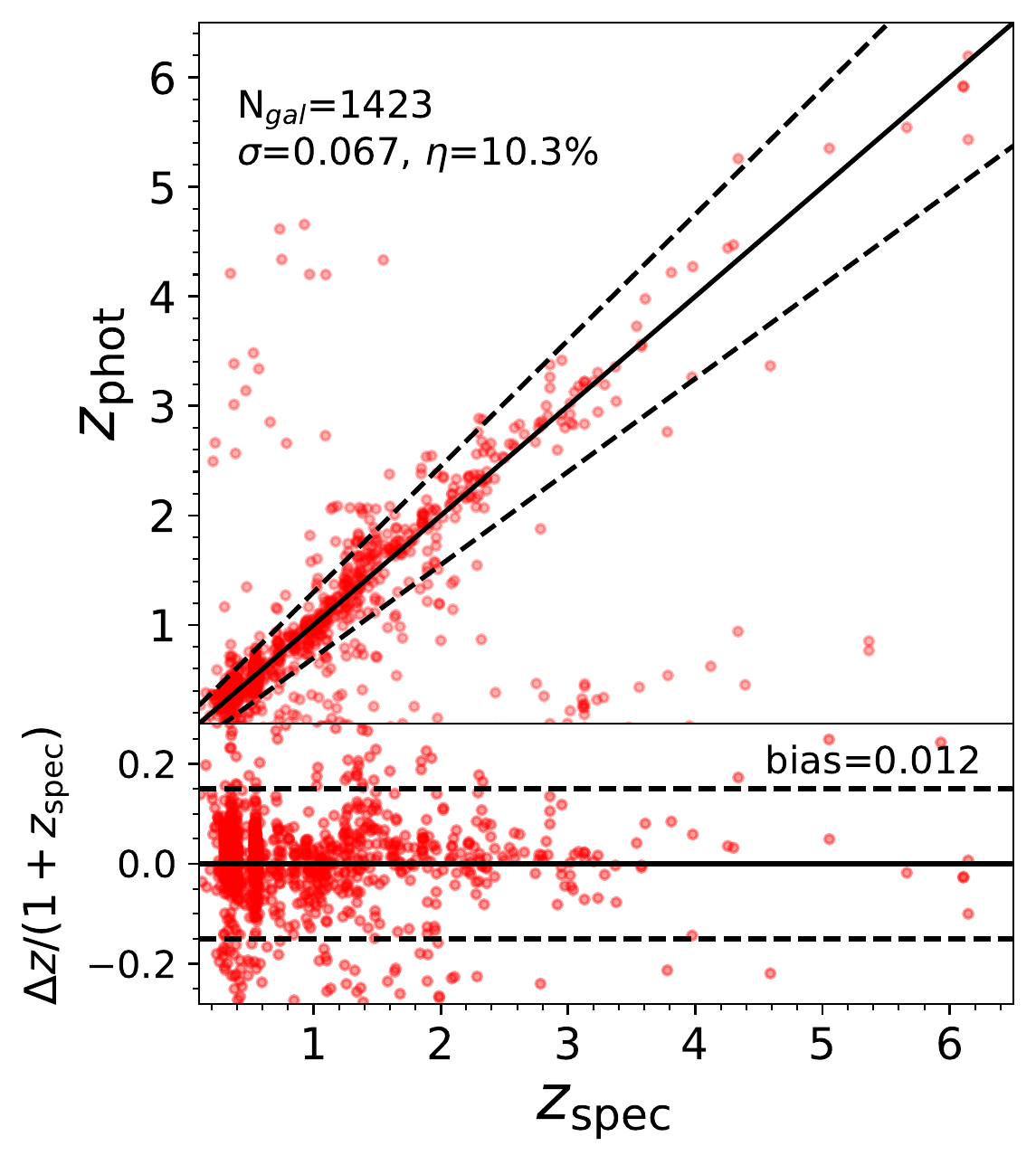}
\caption{Assessing the quality of photometric redshifts estimated through  SED fitting. \textit{Upper panel:}  $z_\mathrm{phot}$ vs.\ $z_\mathrm{spec}$ comparison. Red dots are 1,423 spectroscopic redshifts with $16<F160W<26$,   the solid line shows the 1:1 relationship, and the dashed lines encloses the $z_\mathrm{phot} = z_\mathrm{spec} \pm0.15(1+z_\mathrm{spec})$ threshold used to identify outliers (i.e., catastrophic errors). NMAD scatter ($\sigma$) and outlier fraction ($\eta$) are reported on the top-left corner.  \textit{Lower panel:}  $\Delta z \equiv z_\mathrm{phot}-z_\mathrm{spec}$ scatter (red dots are spectrosopic objects) with the median bias indicated by a solid line; dashed lines represent the threshold for catastrophic errors as in the upper panel.  See Figure \ref{fig:zspec-zphot2} for detailed performance in each individual cluster. }
\label{fig:zspec-zphot1}
\end{figure}

\begin{figure*}
 \centering
\includegraphics[width=0.85\textwidth]{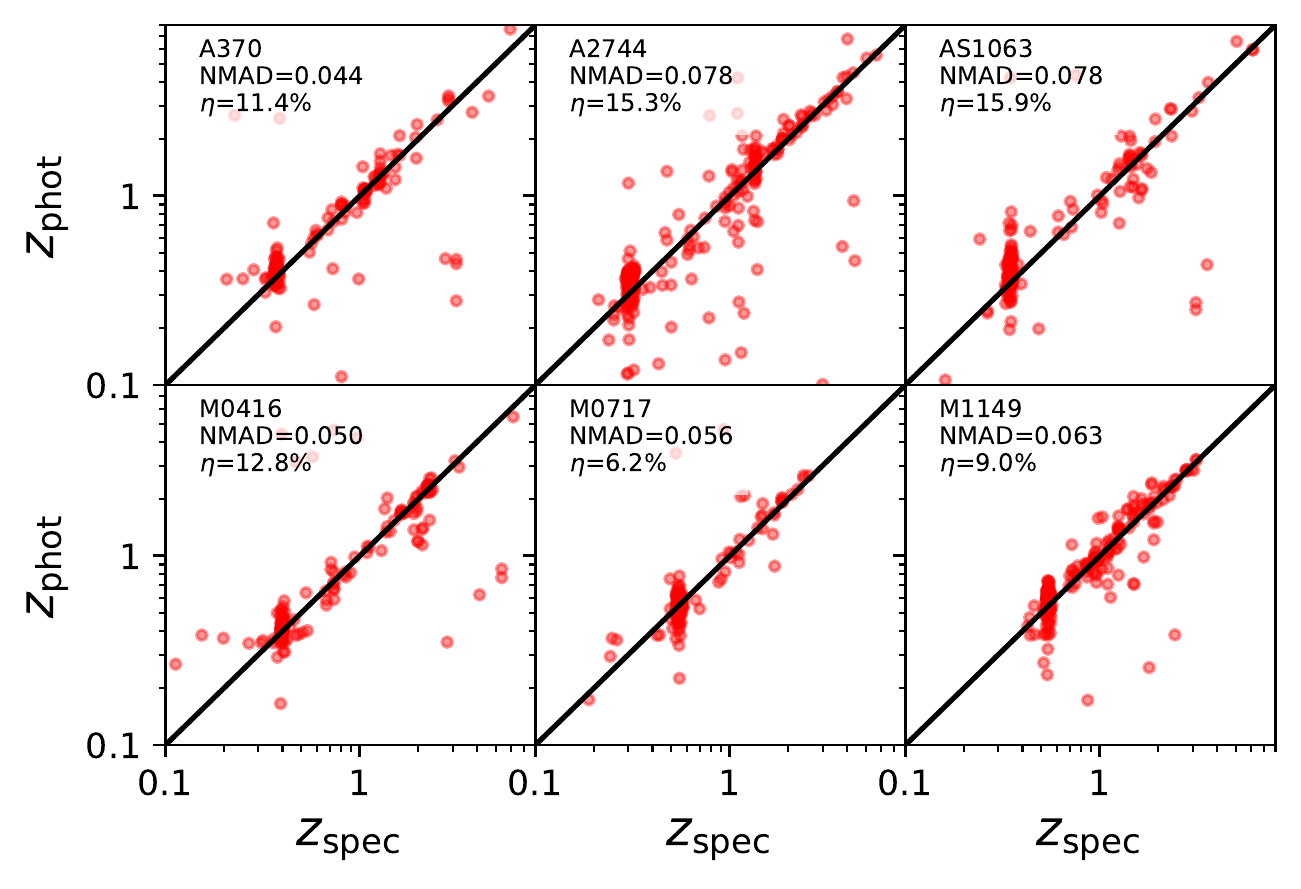}
\caption{Comparison between $z_\mathrm{phot}$ and $z_\mathrm{spec}$ using the most reliable spectroscopic redshift (quality flag equal to 4).}
\label{fig:zspec-zphot2}
\end{figure*}

\begin{figure*}
 \centering
\includegraphics[width=0.85\textwidth]{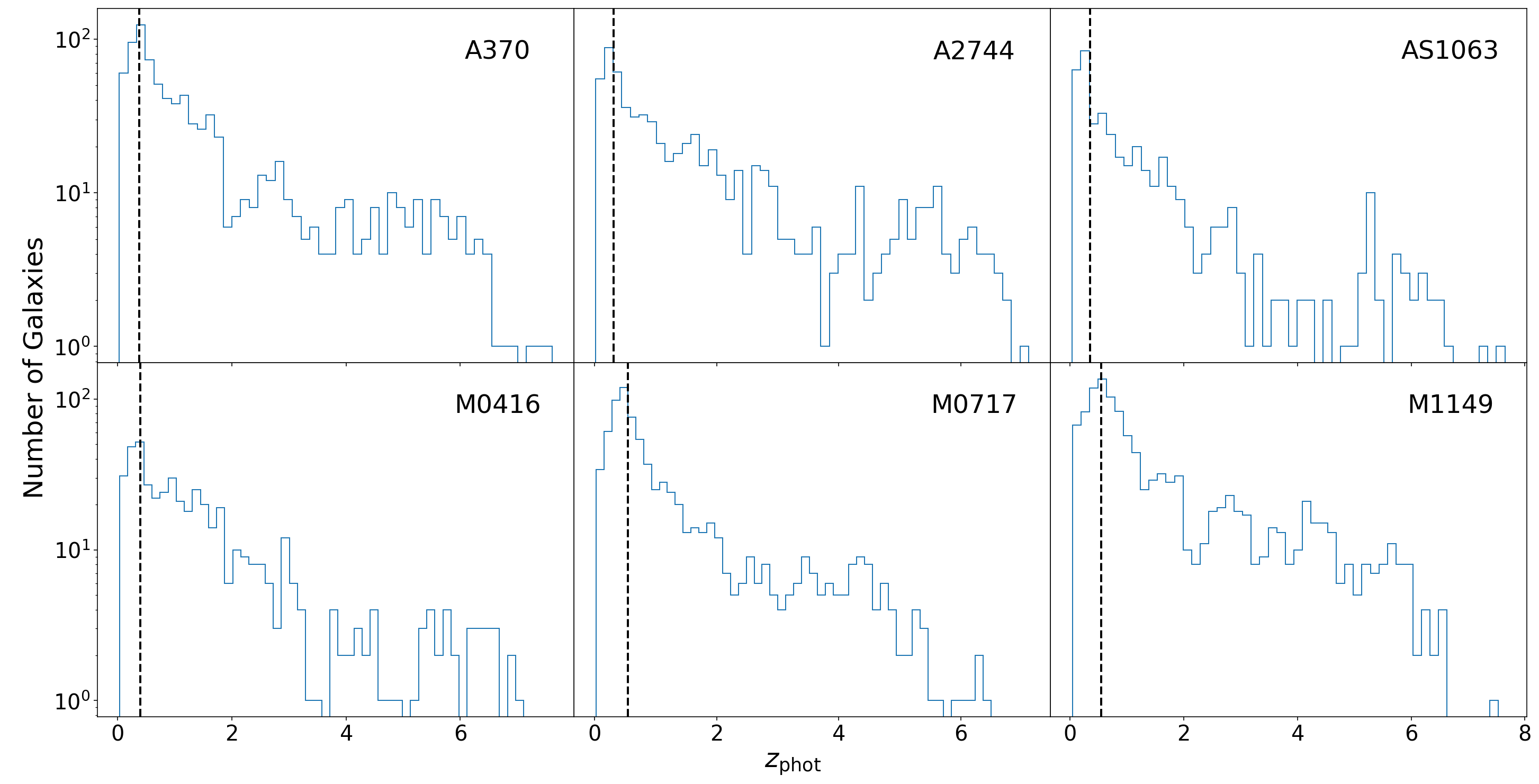}
\caption{Histogram of the photometric redshift distributions for individual clusters. The black dashed line corresponds to the redshift of the cluster. Objects selected have SED fits with a reduced ${\chi}^{2} < 10$}
\label{fig:zhists}
\end{figure*}

We fit the SED of 1,423 spectroscopic galaxies with $16<F160W<26$ across the six clusters to assess $z_\mathrm{phot}$ quality. The calibration offsets (Table \ref{tab:photo-calib}) are taken into account. Similar to the previous work \citep[e.g.,][]{brammer08}, the  scatter is defined as the normalized median absolute deviation \citep[NMAD][]{hoaglin83}, i.e.
\begin{equation}
\sigma_{\rm NMAD}=1.48\times\mathrm{median}\left(\frac{\vert \Delta z-\mathrm{median}(\Delta z)\vert}{1+z_\mathrm{spec}}\right),
\label{eq:MAD}
\end{equation}
with $\Delta z \equiv z_\mathrm{phot}-z_\mathrm{spec}$. The incidence of catastrophic errors $\eta$ is defined as the fraction of redshift outliers having $\vert \Delta z\vert>0.15(1+z_\mathrm{spec})$. For pathological PDF(z) where the median is not a good approximation of the main peak of the PDF, we used the redshift corresponding to the minimum $\chi^2$ solution (as in \citep{davidzon17}).
For the entire sample $\sigma=0.067$ and $\eta=10.3$\% (Figure \ref{fig:zspec-zphot1}, upper panel). After excluding the outliers we also recompute the mean of $\Delta z$ to asses the so-called redshift bias \citep[e.g.,][]{masters17}, which is 0.012 in our case. We repeat the same procedure for the parallel fields, in which a total of only 62 spectroscopic redshifts are available. In that case we estimate $\sigma=0.044$ with no outliers.  Individual cluster field redshift comparisons are shown in Figure \ref{fig:zspec-zphot2}. In this figure, the comparison is made using the highest quality spectroscopic redshifts. Figure \ref{fig:zhists} also shows the photometric redshift distribution for objects in each cluster SED fits with a reduced ${\chi}^{2} < 10$. 

\section{De-magnification}

Leveraging the fact that these HFF clusters exhibit some of the strongest lensing currently observed, we calculate and include lens model magnification factors for all relevant objects in our final catalog (i.e. for those objects with redshifts larger than the clusters' mean redshifts). We use the lens models provided by Brada\v{c}~\citep[][and references therein]{2016ApJ...831..182H}, Caminha~\citep{2017A&A...600A..90C}, CATS~\citep[][and references therein]{2014MNRAS.443.1549J}, DIEGO~\citep[][and references therein]{2015MNRAS.446..683D}, GLAFIC~\citep[][and references therein]{2018ApJ...855....4K}, Keeton~\citep[][and references therein]{2014MNRAS.443.3631M}, Merten \& Zitrin~\citep[][and references therein]{2011MNRAS.417..333M}, Sharon~\citep{2007NJPh....9..447J, 2014ApJ...797...48J}, and Williams~\citep[][and references therein]{2014MNRAS.443.1549J} lens modeling teams. 
We use a script (D. Coe, private comm.) that functions in a similar way to the Interactive Model Magnification Web Tool\footnote{\url{https://archive.stsci.edu/prepds/frontier/lensmodels/\#magcalc}} by utilizing the available mass (kappa, $\kappa$) and shear (gamma, $\gamma$) maps to directly calculate the magnification factors, and provide the value of the magnification factor, $\mu$, for each lens model in separate columns included in the photometry catalog (see Appendix~\ref{app:catalog} for details). Concretely, we compute $\mu$ as follows:
\begin{equation}
    \mu = \frac{1}{(1-\kappa)^{2}-\gamma^{2}}.
\end{equation}
For more information about how these lens models are constructed, we refer the reader to the HFF Lens Modeling webpage\footnote{\url{https://archive.stsci.edu/prepds/frontier/lensmodels/}}.

\section{Data Products}
\label{sec:datarelease}

The final data catalog resulting from this work is accompanied by: (i) the PSF models; (ii) the ICL maps and the bright galaxy models; (iii) median filtered science images with the ICL and galaxy models subtracted; and (iv) photometry, redshift, and magnification catalogs. A brief description of the produces is given below:
\renewcommand{\theenumi}{\roman{enumi}}%
\begin{enumerate}
  \item The PSF models for the HST and Ks bands are released as 2.34x2.34 square arcsecond (39x39 square pixel) fits file arrays. The total flux density of the arrays representing the PSF model is set to unity. IRAC PRFs used in this work are available upon request since they vary across the focal plane.
  \item The ICL maps and the bright galaxy models are fits arrays of the same resolution and dimensions as the science images. They also use the same units as the science images, that is, ADU/s. 
  \item The ICL+bright galaxy subtracted median filtered images are also fits arrays of the same resolution and dimensions as the science images, in ADU/s units. These are the files on which detection was applied.
  \item The photometric and redshift catalogs are ascii files for each cluster. For a list of column names, we refer the reader to Appendix~\ref{app:catalog}. We also include the photometric offsets used in the photometric redshift calculation in a README file in case the user would like to apply these offsets to the given photometries. Galactic extinction has been applied when calculating photometric redshifts using the $E(B-V)$ values derived from the dust maps presented at ~\citet{2011ApJ...737..103S}. However, the photometric catalogs do not have galactic extinction applied.
\end{enumerate}



\section{Conclusions}
\label{sec:conclusions}
In this paper we present and publicly release catalogs of the six Frontier Field clusters and their parallel fields. This includes a total of $>$ 33,500 objects across a wide wavelength range, 0.275 - 8 $\mu m$. We release maps of the intra-cluster light and bright galaxy models. We have outlined the complexity involved in performing this analysis in a crowded field with bright cluster members and intra-cluster light that could bias the measurements. We have successfully removed the contamination from these sources to further reduce the image into a ``blank field" on which we can detect the faintest objects in the field without biasing the flux measurements of objects on the perimeter of the field of view. We perform an error analysis using \texttt{GALSIM} by injecting COSMOS-like galaxies to both estimate the errors and to analyse the validity of our pipeline. We note there is no evident photometric bias close to the cluster core as compared to the outskirts.

We estimate the photometric redshifts for objects in all six clusters and their parallel fields as measured by \texttt{LePhare}. When compared to the available spectroscopic redshifts for the HFF clusters, this gives a combined outlier fraction of 10.3\% and a redshift bias of 0.012 after excluding the outliers. 

We find some differences in our photometry measurements with those from the literature and note the importance of cross-checks between datasets in such crowded and complex environments. We demonstrate the utility of performing source injection, where knowledge of ground truth is accessible to calibrate measurements and characterize biases and uncertainties.

We have scripted a pipeline to analyze each of the Frontier Fields in an efficient, streamlined, and reproducible manner. We plan to apply a similar version of the pipeline to the next generation survey of the Frontier Fields, namely the BUFFALO survey \citep{steinhardt20} which expands these same fields in area by a factor of 3-4 and pushes the 5-$\sigma$ depth $\sim$ 1 magnitude fainter.




\acknowledgments

The authors would like to thank the anonymous referee for their useful comments and suggestions that improved the overall quality of this manuscript. AP would like to thank A. Alavi, B. H\"{a}u\ss ler, E. Merlin, and M. Nowinski for useful discussion and comments; AP would also like to thank A. Faisst for providing the code to generate \textit{Spitzer} PRFs that are used in this work. AP acknowledges support from the NASA MUREP Institutional Opportunity (MIRO) through the grant NNX15AP99A.
FJS acknowledges that this document was prepared using the resources of the Fermi National Accelerator Laboratory (Fermilab), a U.S. Department of Energy, Office of Science, HEP User Facility. Fermilab is managed by Fermi Research Alliance, LLC (FRA), acting under Contract No. DE-AC02-07CH11359. 
ID acknowledges the support received from the European Union's Horizon 2020 research and innovation programme under the Marie Sk\l{}odowska-Curie grant agreement No. 896225. 
This work utilizes gravitational lensing models produced by PIs Brada\v{c}, Natarajan \& Kneib (CATS), Merten \& Zitrin, Sharon, Williams, Keeton, Bernstein and Diego, and the GLAFIC group. This lens modeling was partially funded by the HST Frontier Fields program conducted by STScI. STScI is operated by the Association of Universities for Research in Astronomy, Inc. under NASA contract NAS 5-26555. The lens models were obtained from the Mikulski Archive for Space Telescopes (MAST). This work has made use of the CANDIDE Cluster at the Institut d'Astrophysique de Paris and made possible by grants from the PNCG and the DIM-ACAV. The Cosmic Dawn Center is funded by the Danish National Research Foundation under grant No. 140.


\appendix

\section{Error correction}
\label{app:err_corr}

In this section we show correction factors applied at various stages in the pipeline. The values for the error map correction from pull-plots calculated in Section~\ref{sec:tphot} are shown in Table~\ref{tab:pull-plot}. 

\begin{deluxetable}{lcccc}[h!]
\tablecaption{This table contains the correction factors that we apply to Spitzer RMS images for each cluster. The procedure to obtain these correction factors is described in Section~\ref{sec:tphot}. The different columns refer to different IRAC Channels.}
\label{tab:pull-plot}
\tablehead{\colhead{Cluster} & \colhead{\hspace{.75cm}$I1$}\hspace{.5cm} & \colhead{\hspace{.75cm}$I2$}\hspace{.5cm} &
\colhead{\hspace{.75cm}$I3$}\hspace{.5cm} &
\colhead{$I4$}\hspace{.5cm}}
\startdata
Abell 370 & 3.12 & 2.68 & 1.03 & 1.02 \\
MACS J0717.5+3745 & 2.98 & 2.74 & 0.90 & 0.83 \\
MACS J0416.1-2403 & 3.04 & 2.88 & 0.91 & 0.86 \\
Abell S1063 & 2.96 & 2.73 & - & - \\
Abell 2744 & 5.83 & 4.44 & 0.88 & 0.85 \\
MACS J1149.5+2223 & 2.82 & 2.56 & - & - \\
\enddata
\end{deluxetable}

\bigbreak

%





\section{Catalog Columns}
\label{app:catalog}


Each cluster's photometric catalog has the following columns, with the descriptions taken from each software's respective documentation \citep{bertin96,arnouts99,2016A&A...595A..97M}.:
\begin{itemize}
\item \texttt{IDENT}: unique galaxy identifier
\item \texttt{ALPHA\_J2000}: J2000 right ascension of the isophotal image centroid
\item \texttt{DELTA\_J2000}: J2000 declination of the isophotal image centroid
\item \texttt{FLUX\_TOT\_BAND}: Weighted mean of the AUTO-to-ISO flux ratio
\item \texttt{FLUXERR\_TOT\_BAND}: RMS error estimate for the weighted mean of the AUTO-to-ISO flux ratio
\item \texttt{FLUX\_AUTO\_BAND}: Kron-like automated aperture flux
\item \texttt{FLUXERR\_AUTO\_BAND}: RMS error estimate for Kron-like automated aperture flux
\item \texttt{FLUX\_ISO\_BAND}: Isophotal flux
\item \texttt{FLUXERR\_ISO\_BAND}: RMS error estimate for isophotal flux
\item \texttt{CLASS\_STAR}: star/galaxy classifier
\item \texttt{FLAGS}: source extraction flags
\item \texttt{MAGNIFICATION\_\{MODEL\}}: Magnification factors, $\mu$, for each lens map/model. Demagnified fluxes for a given model, \texttt{MODEL}, can be obtained by dividing the preferred flux estimate (e.g., \texttt{ISO, AUTO}) by the value given by \texttt{MAGNIFICATION\_\{MODEL\}}.
\end{itemize}

There are two photometric redshift catalogs; one where the redshifts were calculated with all bands presented in this work, and one where only HST bands are used. The column order is as follows:
\begin{itemize}
\item \texttt{IDENT}: unique galaxy identifier
\item \texttt{Z\_BEST}: Zphot Best
\item \texttt{Z\_BEST68\_LOW}: Zphot min from $\Delta\chi^2 = 1.0(68\%)$
\item \texttt{Z\_BEST68\_HIGH}: Zphot max from $\Delta\chi^2 = 1.0(68\%)$
\item \texttt{Z\_ML}: Zphot from Median of ML distribution
\item \texttt{Z\_ML68\_LOW}: Zphot min at $68\%$ of ML distribution
\item \texttt{Z\_ML68\_HIGH}: Zphot max at $68\%$ of ML distribution
\item \texttt{CHI\_BEST}: lowest galaxy $\chi^2$ for galaxy
\item \texttt{MOD\_BEST}: galaxy model for best $\chi^2$ 
\item \texttt{EXTLAW\_BEST}: Extinction law  
\item \texttt{EBV\_BEST}: E(B-V)
\item \texttt{AGE\_BEST}: Age (yr)
\item \texttt{PDZ\_BEST}: Probability per Z bins
\item \texttt{Z\_SEC}: Galaxy secondary Zphot solution from $F(z)$ function  
\item \texttt{CHI\_SEC}: $\chi^2$ evaluated at secondary Zphot from $F(z)$ function    
\item \texttt{MOD\_SEC}: Galaxy model for \texttt{CHI\_SEC}
\item \texttt{EBV\_SEC}: E(B-V)
\item \texttt{PDZ\_SEC}: Probability per Z bins for the secondary model
\item \texttt{MOD\_STAR}: Star solution  
\item \texttt{CHI\_STAR}: $\chi^2$ for star solution
\item \texttt{SCALE\_BEST}: Scaling factor  
\item \texttt{NBAND\_USED}: Number of bands used for each object
\item \texttt{CONTEXT}: Bands to be considered in fit
\item \texttt{ZSPEC}: Spectroscopic redshift if available  
\item \texttt{MAG\_OBS\_N}: Observed magnitude of band N
\item \texttt{MAG\_MOD\_N}: Predicted model magnitude
\item \texttt{AGE\_MED}: Age (yr)
\item \texttt{MASS\_MED}: Rescale Mass ($M_{\odot}$)
\item \texttt{SFR\_MED}: Rescaled SFR ($M_{\odot}/$yr)  
\item \texttt{ZSPEC\_FLAG}: Quality flag for Zspec
\item \texttt{ZSPEC\_REF}: Reference for Zspec  
\item \texttt{FIELD}: cluster name  
\item \texttt{RA}: J2000 right ascension of the isophotal image centroid
\item \texttt{DEC}: J2000 declination of the isophotal image centroid  
\end{itemize}











\bibliography{bib}{}

\newcommand{\noop}[1]{}
\begin{thebibliography}{}
\expandafter\ifx\csname natexlab\endcsname\relax\def\natexlab#1{#1}\fi
\providecommand{\url}[1]{\href{#1}{#1}}
\providecommand{\dodoi}[1]{doi:~\href{http://doi.org/#1}{\nolinkurl{#1}}}
\providecommand{\doeprint}[1]{\href{http://ascl.net/#1}{\nolinkurl{http://ascl.net/#1}}}
\providecommand{\doarXiv}[1]{\href{https://arxiv.org/abs/#1}{\nolinkurl{https://arxiv.org/abs/#1}}}

\bibitem[{{Alard} \& {Lupton}(1998)}]{1998ApJ...503..325A}
{Alard}, C., \& {Lupton}, R.~H. 1998, \apj, 503, 325, \dodoi{10.1086/305984}

\bibitem[{{Alavi} {et~al.}(2016){Alavi}, {Siana}, {Richard}, {Rafelski},
  {Jauzac}, {Limousin}, {Freeman}, {Scarlata}, {Robertson}, {Stark}, {Teplitz},
  \& {Desai}}]{2016ApJ...832...56A}
{Alavi}, A., {Siana}, B., {Richard}, J., {et~al.} 2016, \apj, 832, 56,
  \dodoi{10.3847/0004-637X/832/1/56}

\bibitem[{{Arnouts} {et~al.}(1999){Arnouts}, {Cristiani}, {Moscardini},
  {Matarrese}, {Lucchin}, {Fontana}, \& {Giallongo}}]{arnouts99}
{Arnouts}, S., {Cristiani}, S., {Moscardini}, L., {et~al.} 1999, \mnras, 310,
  540, \dodoi{10.1046/j.1365-8711.1999.02978.x}

\bibitem[{{Atek} {et~al.}(2018){Atek}, {Richard}, {Kneib}, \&
  {Schaerer}}]{atek18}
{Atek}, H., {Richard}, J., {Kneib}, J.-P., \& {Schaerer}, D. 2018, \mnras, 479,
  5184, \dodoi{10.1093/mnras/sty1820}

\bibitem[{{Balestra} {et~al.}(2016){Balestra}, {Mercurio}, {Sartoris},
  {Girardi}, {Grillo}, {Nonino}, {Rosati}, {Biviano}, {Ettori}, {Forman},
  {Jones}, {Koekemoer}, {Medezinski}, {Merten}, {Ogrean}, {Tozzi}, {Umetsu},
  {Vanzella}, {van Weeren}, {Zitrin}, {Annunziatella}, {Caminha}, {Broadhurst},
  {Coe}, {Donahue}, {Fritz}, {Frye}, {Kelson}, {Lombardi}, {Maier},
  {Meneghetti}, {Monna}, {Postman}, {Scodeggio}, {Seitz}, \&
  {Ziegler}}]{balestra16}
{Balestra}, I., {Mercurio}, A., {Sartoris}, B., {et~al.} 2016, \apjs, 224, 33,
  \dodoi{10.3847/0067-0049/224/2/33}

\bibitem[{{Barden} {et~al.}(2012){Barden}, {H{\"a}u{\ss}ler}, {Peng},
  {McIntosh}, \& {Guo}}]{2012MNRAS.422..449B}
{Barden}, M., {H{\"a}u{\ss}ler}, B., {Peng}, C.~Y., {McIntosh}, D.~H., \&
  {Guo}, Y. 2012, \mnras, 422, 449, \dodoi{10.1111/j.1365-2966.2012.20619.x}

\bibitem[{{Barkana} \& {Loeb}(2001)}]{barkana01}
{Barkana}, R., \& {Loeb}, A. 2001, \physrep, 349, 125,
  \dodoi{10.1016/S0370-1573(01)00019-9}

\bibitem[{{Beckwith} {et~al.}(2006){Beckwith}, {Stiavelli}, {Koekemoer},
  {Caldwell}, {Ferguson}, {Hook}, {Lucas}, {Bergeron}, {Corbin}, {Jogee},
  {Panagia}, {Robberto}, {Royle}, {Somerville}, \&
  {Sosey}}]{2006AJ....132.1729B}
{Beckwith}, S.~V.~W., {Stiavelli}, M., {Koekemoer}, A.~M., {et~al.} 2006, \aj,
  132, 1729, \dodoi{10.1086/507302}

\bibitem[{{Bertin} \& {Arnouts}(1996)}]{1996A&AS..117..393B}
{Bertin}, E., \& {Arnouts}, S. 1996, \aaps, 117, 393,
  \dodoi{10.1051/aas:1996164}

\bibitem[{Bertin \& Arnouts(1996)}]{bertin96}
Bertin, E., \& Arnouts, S. 1996, \apjs, 117, 393

\bibitem[{{Blandford} \& {Narayan}(1986)}]{blandford_narayan86}
{Blandford}, R., \& {Narayan}, R. 1986, \apj, 310, 568, \dodoi{10.1086/164709}

\bibitem[{{Brada{\v{c}}} {et~al.}(2019){Brada{\v{c}}}, {Huang}, {Fontana},
  {Castellano}, {Merlin}, {Amor{\'\i}n}, {Hoag}, {Strait}, {Santini}, {Ryan},
  {Casertano}, {Lemaux}, {Lubin}, {Schmidt}, {Schrabback}, {Treu}, {von der
  Linden}, {Mason}, \& {Wang}}]{2019MNRAS.489...99B}
{Brada{\v{c}}}, M., {Huang}, K.-H., {Fontana}, A., {et~al.} 2019, \mnras, 489,
  99, \dodoi{10.1093/mnras/stz2119}

\bibitem[{{Brammer} {et~al.}(2008){Brammer}, {van Dokkum}, \&
  {Coppi}}]{brammer08}
{Brammer}, G.~B., {van Dokkum}, P.~G., \& {Coppi}, P. 2008, \apj, 686, 1503,
  \dodoi{10.1086/591786}

\bibitem[{{Brammer} {et~al.}(2016){Brammer}, {Marchesini}, {Labb{\'e}},
  {Spitler}, {Lange-Vagle}, {Barker}, {Tanaka}, {Fontana}, {Galametz},
  {Ferr{\'e}-Mateu}, {Kodama}, {Lundgren}, {Martis}, {Muzzin}, {Stefanon},
  {Toft}, {van der Wel}, {Vulcani}, \& {Whitaker}}]{KIFF}
{Brammer}, G.~B., {Marchesini}, D., {Labb{\'e}}, I., {et~al.} 2016, The
  Astrophysical Journal Supplement Series, 226, 6,
  \dodoi{10.3847/0067-0049/226/1/6}

\bibitem[{{Bruzual} \& {Charlot}(2003)}]{bruzual&charlot03}
{Bruzual}, G., \& {Charlot}, S. 2003, \mnras, 344, 1000,
  \dodoi{10.1046/j.1365-8711.2003.06897.x}

\bibitem[{{Caldwell} {et~al.}(2008){Caldwell}, {McIntosh}, {Rix}, {Barden},
  {Beckwith}, {Bell}, {Borch}, {Heymans}, {H{\"a}u{\ss}ler}, {Jahnke}, {Jogee},
  {Meisenheimer}, {Peng}, {S{\'a}nchez}, {Somerville}, {Wisotzki}, \&
  {Wolf}}]{caldwell08}
{Caldwell}, J. A.~R., {McIntosh}, D.~H., {Rix}, H.-W., {et~al.} 2008, \apjs,
  174, 136, \dodoi{10.1086/521080}

\bibitem[{Calzetti {et~al.}(2000)Calzetti, Armus, Bohlin, Kinney, Koornneef, \&
  Storchi-Bergmann}]{calzetti2000}
Calzetti, D., Armus, L., Bohlin, R.~C., {et~al.} 2000, \apj, 533, 682

\bibitem[{{Caminha} {et~al.}(2017){Caminha}, {Grillo}, {Rosati}, {Balestra},
  {Mercurio}, {Vanzella}, {Biviano}, {Caputi}, {Delgado-Correal}, {Karman},
  {Lombardi}, {Meneghetti}, {Sartoris}, \& {Tozzi}}]{2017A&A...600A..90C}
{Caminha}, G.~B., {Grillo}, C., {Rosati}, P., {et~al.} 2017, \aap, 600, A90,
  \dodoi{10.1051/0004-6361/201629297}

\bibitem[{{Castellano} {et~al.}(2016){Castellano}, {Amor{\'\i}n}, {Merlin},
  {Fontana}, {McLure}, {M{\'a}rmol-Queralt{\'o}}, {Mortlock}, {Parsa},
  {Dunlop}, {Elbaz}, {Balestra}, {Boucaud}, {Bourne}, {Boutsia}, {Brammer},
  {Bruce}, {Buitrago}, {Capak}, {Cappelluti}, {Ciesla}, {Comastri}, {Cullen},
  {Derriere}, {Faber}, {Giallongo}, {Grazian}, {Grillo}, {Mercurio},
  {Micha{\l}owski}, {Nonino}, {Paris}, {Pentericci}, {Pilo}, {Rosati},
  {Santini}, {Schreiber}, {Shu}, \& {Wang}}]{2016A&A...590A..31C}
{Castellano}, M., {Amor{\'\i}n}, R., {Merlin}, E., {et~al.} 2016, \aap, 590,
  A31, \dodoi{10.1051/0004-6361/201527514}

\bibitem[{{Davidzon} {et~al.}(2017){Davidzon}, {Ilbert}, {Laigle}, {Coupon},
  {McCracken}, {Delvecchio}, {Masters}, {Capak}, {Hsieh}, {Le F{\`e}vre},
  {Tresse}, {Bethermin}, {Chang}, {Faisst}, {Le Floc'h}, {Steinhardt}, {Toft},
  {Aussel}, {Dubois}, {Hasinger}, {Salvato}, {Sanders}, {Scoville}, \&
  {Silverman}}]{davidzon17}
{Davidzon}, I., {Ilbert}, O., {Laigle}, C., {et~al.} 2017, \aap, 605, A70,
  \dodoi{10.1051/0004-6361/201730419}

\bibitem[{{De Santis} {et~al.}(2007){De Santis}, {Grazian}, {Fontana}, \&
  {Santini}}]{2007NewA...12..271D}
{De Santis}, C., {Grazian}, A., {Fontana}, A., \& {Santini}, P. 2007, \na, 12,
  271, \dodoi{10.1016/j.newast.2006.10.004}

\bibitem[{{Di Criscienzo} {et~al.}(2017){Di Criscienzo}, {Merlin},
  {Castellano}, {Santini}, {Fontana}, {Amorin}, {Boutsia}, {Derriere},
  {Dunlop}, {Elbaz}, {Grazian}, {McLure}, {M{\'a}rmol-Queralt{\'o}},
  {Michalowski}, {Mortlock}, {Parsa}, \& {Pentericci}}]{2017A&A...607A..30D}
{Di Criscienzo}, M., {Merlin}, E., {Castellano}, M., {et~al.} 2017, \aap, 607,
  A30, \dodoi{10.1051/0004-6361/201731172}

\bibitem[{{Diego} {et~al.}(2015){Diego}, {Broadhurst}, {Benitez}, {Umetsu},
  {Coe}, {Sendra}, {Sereno}, {Izzo}, \& {Covone}}]{2015MNRAS.446..683D}
{Diego}, J.~M., {Broadhurst}, T., {Benitez}, N., {et~al.} 2015, \mnras, 446,
  683, \dodoi{10.1093/mnras/stu2064}

\bibitem[{{Diolaiti} {et~al.}(2000){Diolaiti}, {Bendinelli}, {Bonaccini},
  {Close}, {Currie}, \& {Parmeggiani}}]{2000SPIE.4007..879D}
{Diolaiti}, E., {Bendinelli}, O., {Bonaccini}, D., {et~al.} 2000, in \procspie,
  Vol. 4007, Adaptive Optical Systems Technology, ed. P.~L. {Wizinowich},
  879--888

\bibitem[{{Ebeling} {et~al.}(2014){Ebeling}, {Ma}, \& {Barrett}}]{ebeling14}
{Ebeling}, H., {Ma}, C.-J., \& {Barrett}, E. 2014, \apjs, 211, 21,
  \dodoi{10.1088/0067-0049/211/2/21}

\bibitem[{{Galametz} {et~al.}(2013){Galametz}, {Grazian}, {Fontana},
  {Ferguson}, {Ashby}, {Barro}, {Castellano}, {Dahlen}, {Donley}, {Faber},
  {Grogin}, {Guo}, {Huang}, {Kocevski}, {Koekemoer}, {Lee}, {McGrath}, {Peth},
  {Willner}, {Almaini}, {Cooper}, {Cooray}, {Conselice}, {Dickinson}, {Dunlop},
  {Fazio}, {Foucaud}, {Gardner}, {Giavalisco}, {Hathi}, {Hartley}, {Koo},
  {Lai}, {de Mello}, {McLure}, {Lucas}, {Paris}, {Pentericci}, {Santini},
  {Simpson}, {Sommariva}, {Targett}, {Weiner}, {Wuyts}, \& {the CANDELS
  Team}}]{2013ApJS..206...10G}
{Galametz}, A., {Grazian}, A., {Fontana}, A., {et~al.} 2013, \apjs, 206, 10,
  \dodoi{10.1088/0067-0049/206/2/10}

\bibitem[{{Galametz} {et~al.}(2011){Galametz}, {Madden}, {Galliano}, {Hony},
  {Bendo}, \& {Sauvage}}]{galametz11}
{Galametz}, M., {Madden}, S.~C., {Galliano}, F., {et~al.} 2011, \aap, 532, A56,
  \dodoi{10.1051/0004-6361/201014904}

\bibitem[{{Gao} {et~al.}(2005){Gao}, {Springel}, \&
  {White}}]{2005MNRAS.363L..66G}
{Gao}, L., {Springel}, V., \& {White}, S. D.~M. 2005, \mnras, 363, L66,
  \dodoi{10.1111/j.1745-3933.2005.00084.x}

\bibitem[{{Gonzaga} \& {et al.}(2012)}]{drizzlepac}
{Gonzaga}, S., \& {et al.} 2012, {The DrizzlePac Handbook}

\bibitem[{Gray {et~al.}(2009)Gray, Wolf, Barden, Peng, Häußler, Bell,
  McIntosh, Guo, Caldwell, Bacon, Balogh, Barazza, Böhm, Heymans, Jahnke,
  Jogee, Van~Kampen, Lane, Meisenheimer, Sánchez, Taylor, Wisotzki, Zheng,
  Green, Beswick, Saikia, Gilmour, Johnson, \& Papovich}]{gray09}
Gray, M.~E., Wolf, C., Barden, M., {et~al.} 2009, Monthly Notices of the Royal
  Astronomical Society, 393, 1275, \dodoi{10.1111/j.1365-2966.2008.14259.x}

\bibitem[{Gross(2018)}]{Gross:2018okg}
Gross, E. 2018, CERN Yellow Rep. School Proc., 3, 199,
  \dodoi{10.23730/CYRSP-2018-003.199}

\bibitem[{{H{\"a}u{\ss}ler} {et~al.}(2013){H{\"a}u{\ss}ler}, {Bamford}, {Vika},
  {Rojas}, {Barden}, {Kelvin}, {Alpaslan}, {Robotham}, {Driver}, {Baldry},
  {Brough}, {Hopkins}, {Liske}, {Nichol}, {Popescu}, \&
  {Tuffs}}]{2013MNRAS.430..330H}
{H{\"a}u{\ss}ler}, B., {Bamford}, S.~P., {Vika}, M., {et~al.} 2013, \mnras,
  430, 330, \dodoi{10.1093/mnras/sts633}

\bibitem[{{Hoag} {et~al.}(2016){Hoag}, {Huang}, {Treu}, {Brada{\v{c}}},
  {Schmidt}, {Wang}, {Brammer}, {Broussard}, {Amorin}, {Castellano}, {Fontana},
  {Merlin}, {Schrabback}, {Trenti}, \& {Vulcani}}]{2016ApJ...831..182H}
{Hoag}, A., {Huang}, K.~H., {Treu}, T., {et~al.} 2016, \apj, 831, 182,
  \dodoi{10.3847/0004-637X/831/2/182}

\bibitem[{{Hoaglin} {et~al.}(1983){Hoaglin}, {Mosteller}, \&
  {Tukey}}]{hoaglin83}
{Hoaglin}, D.~C., {Mosteller}, F., \& {Tukey}, J.~W. 1983, {Understanding
  robust and exploratory data anlysis}

\bibitem[{{Hoyos} {et~al.}(2012){Hoyos}, {Arag{\'o}n-Salamanca}, {Gray},
  {Maltby}, {Bell}, {Barazza}, {B{\"o}hm}, {H{\"a}u{\ss}ler}, {Jahnke},
  {Jogee}, {Lane}, {McIntosh}, \& {Wolf}}]{2012MNRAS.419.2703H}
{Hoyos}, C., {Arag{\'o}n-Salamanca}, A., {Gray}, M.~E., {et~al.} 2012, \mnras,
  419, 2703, \dodoi{10.1111/j.1365-2966.2011.19918.x}

\bibitem[{Ilbert {et~al.}(2006)Ilbert, Arnouts, McCracken, Bolzonella, Bertin,
  Le~F{\`e}vre, Mellier, Zamorani, Pell{\`o}, Iovino, Tresse, Le~Brun, Bottini,
  Garilli, Maccagni, Picat, Scaramella, Scodeggio, Vettolani, Zanichelli,
  Adami, Bardelli, Cappi, Charlot, Ciliegi, Contini, Cucciati, Foucaud,
  Franzetti, Gavignaud, Guzzo, Marano, Marinoni, Mazure, Meneux, Merighi,
  Paltani, Pollo, Pozzetti, Radovich, Zucca, Bondi, Bongiorno, Busarello, de~la
  Torre, Gregorini, Lamareille, Mathez, Merluzzi, Ripepi, Rizzo, \&
  Vergani}]{ilbert06}
Ilbert, O., Arnouts, S., McCracken, H.~J., {et~al.} 2006, \aap, 457, 841

\bibitem[{{Ilbert} {et~al.}(2009){Ilbert}, {Capak}, {Salvato}, {Aussel},
  {McCracken}, {Sanders}, {Scoville}, {Kartaltepe}, {Arnouts}, {Le Floc'h},
  {Mobasher}, {Taniguchi}, {Lamareille}, {Leauthaud}, {Sasaki}, {Thompson},
  {Zamojski}, {Zamorani}, {Bardelli}, {Bolzonella}, {Bongiorno}, {Brusa},
  {Caputi}, {Carollo}, {Contini}, {Cook}, {Coppa}, {Cucciati}, {de la Torre},
  {de Ravel}, {Franzetti}, {Garilli}, {Hasinger}, {Iovino}, {Kampczyk},
  {Kneib}, {Knobel}, {Kovac}, {Le Borgne}, {Le Brun}, {F{\`e}vre}, {Lilly},
  {Looper}, {Maier}, {Mainieri}, {Mellier}, {Mignoli}, {Murayama}, {Pell{\`o}},
  {Peng}, {P{\'e}rez-Montero}, {Renzini}, {Ricciardelli}, {Schiminovich},
  {Scodeggio}, {Shioya}, {Silverman}, {Surace}, {Tanaka}, {Tasca}, {Tresse},
  {Vergani}, \& {Zucca}}]{ilbert09}
{Ilbert}, O., {Capak}, P., {Salvato}, M., {et~al.} 2009, \apj, 690, 1236,
  \dodoi{10.1088/0004-637X/690/2/1236}

\bibitem[{{Ivezi{\'c}} {et~al.}(2014){Ivezi{\'c}}, {Connolly}, {Vanderplas}, \&
  {Gray}}]{astroMLText}
{Ivezi{\'c}}, {\v Z}., {Connolly}, A., {Vanderplas}, J., \& {Gray}, A. 2014,
  Statistics, Data Mining and Machine Learning in Astronomy (Princeton
  University Press)

\bibitem[{{Jauzac} {et~al.}(2014){Jauzac}, {Cl{\'e}ment}, {Limousin},
  {Richard}, {Jullo}, {Ebeling}, {Atek}, {Kneib}, {Knowles}, {Natarajan},
  {Eckert}, {Egami}, {Massey}, \& {Rexroth}}]{2014MNRAS.443.1549J}
{Jauzac}, M., {Cl{\'e}ment}, B., {Limousin}, M., {et~al.} 2014, \mnras, 443,
  1549, \dodoi{10.1093/mnras/stu1355}

\bibitem[{{Jauzac} {et~al.}(2015){Jauzac}, {Richard}, {Jullo}, {Cl{\'e}ment},
  {Limousin}, {Kneib}, {Ebeling}, {Natarajan}, {Rodney}, {Atek}, {Massey},
  {Eckert}, {Egami}, \& {Rexroth}}]{2015MNRAS.452.1437J}
{Jauzac}, M., {Richard}, J., {Jullo}, E., {et~al.} 2015, \mnras, 452, 1437,
  \dodoi{10.1093/mnras/stv1402}

\bibitem[{{Johnson} {et~al.}(2014){Johnson}, {Sharon}, {Bayliss}, {Gladders},
  {Coe}, \& {Ebeling}}]{2014ApJ...797...48J}
{Johnson}, T.~L., {Sharon}, K., {Bayliss}, M.~B., {et~al.} 2014, \apj, 797, 48,
  \dodoi{10.1088/0004-637X/797/1/48}

\bibitem[{{Jullo} {et~al.}(2007){Jullo}, {Kneib}, {Limousin},
  {El{\'\i}asd{\'o}ttir}, {Marshall}, \& {Verdugo}}]{2007NJPh....9..447J}
{Jullo}, E., {Kneib}, J.~P., {Limousin}, M., {et~al.} 2007, New Journal of
  Physics, 9, 447, \dodoi{10.1088/1367-2630/9/12/447}

\bibitem[{{Kawamata} {et~al.}(2018){Kawamata}, {Ishigaki}, {Shimasaku},
  {Oguri}, {Ouchi}, \& {Tanigawa}}]{2018ApJ...855....4K}
{Kawamata}, R., {Ishigaki}, M., {Shimasaku}, K., {et~al.} 2018, \apj, 855, 4,
  \dodoi{10.3847/1538-4357/aaa6cf}

\bibitem[{{Kneib} \& {Natarajan}(2011)}]{kneib_natarajan11}
{Kneib}, J.-P., \& {Natarajan}, P. 2011, \aapr, 19, 47,
  \dodoi{10.1007/s00159-011-0047-3}

\bibitem[{Koekemoer {et~al.}(2011)Koekemoer, Faber, Ferguson, Grogin, Kocevski,
  Koo, Lai, Lotz, Lucas, McGrath, Ogaz, Rajan, Riess, Rodney, Strolger,
  Casertano, Castellano, Dahlen, Dickinson, Dolch, Fontana, Giavalisco,
  Grazian, Guo, Hathi, Huang, van~der Wel, Yan, Acquaviva, Alexander, Almaini,
  Ashby, Barden, Bell, Bournaud, Brown, Caputi, Cassata, Challis, Chary,
  Cheung, Cirasuolo, Conselice, Cooray, Croton, Daddi, Dav{\'e}, De~Mello,
  de~Ravel, Dekel, Donley, Dunlop, Dutton, Elbaz, Fazio, Filippenko,
  Finkelstein, Frazer, Gardner, Garnavich, Gawiser, Gruetzbauch, Hartley,
  H{\"a}ussler, Herrington, Hopkins, Huang, Jha, Johnson, Kartaltepe,
  Khostovan, Kirshner, Lani, Lee, Li, Madau, McCarthy, McIntosh, Mclure,
  Mcpartland, Mobasher, Moreira, Mortlock, Moustakas, Mozena, Nandra, Newman,
  Nielsen, Niemi, Noeske, Papovich, Pentericci, Pope, Primack, Ravindranath,
  Reddy, Renzini, Rix, Robaina, Rosario, Rosati, Salimbeni, Scarlata, Siana,
  Simard, Smidt, Snyder, Somerville, Spinrad, Straughn, Telford, Teplitz,
  Trump, Vargas, Villforth, Wagner, Wandro, Wechsler, Weiner, Wiklind, Wild,
  Wilson, Wuyts, \& Yun}]{Koekemoer:2011p12718}
Koekemoer, A.~M., Faber, S.~M., Ferguson, H.~C., {et~al.} 2011, \apjs, 197, 36

\bibitem[{{Kron}(1980)}]{1980ApJS...43..305K}
{Kron}, R.~G. 1980, \apjs, 43, 305, \dodoi{10.1086/190669}

\bibitem[{Kron(1980)}]{kron80}
Kron, R.~G. 1980, \apjs, 43, 305

\bibitem[{{Lagattuta} {et~al.}(2017){Lagattuta}, {Richard}, {Cl{\'e}ment},
  {Mahler}, {Patr{\'\i}cio}, {Pell{\'o}}, {Soucail}, {Schmidt}, {Wisotzki},
  {Martinez}, \& {Bina}}]{lagattuta17}
{Lagattuta}, D.~J., {Richard}, J., {Cl{\'e}ment}, B., {et~al.} 2017, \mnras,
  469, 3946, \dodoi{10.1093/mnras/stx1079}

\bibitem[{{Lagattuta} {et~al.}(2019){Lagattuta}, {Richard}, {Bauer},
  {Cl{\'e}ment}, {Mahler}, {Soucail}, {Carton}, {Kneib}, {Laporte}, {Martinez},
  {Patr{\'\i}cio}, {Payne}, {Pell{\'o}}, {Schmidt}, \& {de la
  Vieuville}}]{lagattuta19}
{Lagattuta}, D.~J., {Richard}, J., {Bauer}, F.~E., {et~al.} 2019, \mnras, 485,
  3738, \dodoi{10.1093/mnras/stz620}

\bibitem[{{Laigle} {et~al.}(2016{\natexlab{a}}){Laigle}, {McCracken}, {Ilbert},
  {Hsieh}, {Davidzon}, {Capak}, {Hasinger}, {Silverman}, {Pichon}, {Coupon},
  {Aussel}, {Le Borgne}, {Caputi}, {Cassata}, {Chang}, {Civano}, {Dunlop},
  {Fynbo}, {Kartaltepe}, {Koekemoer}, {Le F{\`e}vre}, {Le Floc'h}, {Leauthaud},
  {Lilly}, {Lin}, {Marchesi}, {Milvang-Jensen}, {Salvato}, {Sanders},
  {Scoville}, {Smolcic}, {Stockmann}, {Taniguchi}, {Tasca}, {Toft}, {Vaccari},
  \& {Zabl}}]{laigle16}
{Laigle}, C., {McCracken}, H.~J., {Ilbert}, O., {et~al.} 2016{\natexlab{a}},
  \apjs, 224, 24, \dodoi{10.3847/0067-0049/224/2/24}

\bibitem[{{Laigle} {et~al.}(2016{\natexlab{b}}){Laigle}, {McCracken}, {Ilbert},
  {Hsieh}, {Davidzon}, {Capak}, {Hasinger}, {Silverman}, {Pichon}, {Coupon},
  {Aussel}, {Le Borgne}, {Caputi}, {Cassata}, {Chang}, {Civano}, {Dunlop},
  {Fynbo}, {Kartaltepe}, {Koekemoer}, {Le F{\`e}vre}, {Le Floc'h}, {Leauthaud},
  {Lilly}, {Lin}, {Marchesi}, {Milvang-Jensen}, {Salvato}, {Sanders},
  {Scoville}, {Smolcic}, {Stockmann}, {Taniguchi}, {Tasca}, {Toft}, {Vaccari},
  \& {Zabl}}]{2016ApJS..224...24L}
---. 2016{\natexlab{b}}, \apjs, 224, 24, \dodoi{10.3847/0067-0049/224/2/24}

\bibitem[{{Laureijs} {et~al.}(2011){Laureijs}, {Amiaux}, {Arduini},
  {Augu{\`e}res}, {Brinchmann}, {Cole}, {Cropper}, {Dabin}, {Duvet}, {Ealet},
  {Garilli}, {Gondoin}, {Guzzo}, {Hoar}, {Hoekstra}, {Holmes}, {Kitching},
  {Maciaszek}, {Mellier}, {Pasian}, {Percival}, {Rhodes}, {Saavedra Criado},
  {Sauvage}, {Scaramella}, {Valenziano}, {Warren}, {Bender}, {Castander},
  {Cimatti}, {Le F{\`e}vre}, {Kurki-Suonio}, {Levi}, {Lilje}, {Meylan},
  {Nichol}, {Pedersen}, {Popa}, {Rebolo Lopez}, {Rix}, {Rottgering},
  {Zeilinger}, {Grupp}, {Hudelot}, {Massey}, {Meneghetti}, {Miller}, {Paltani},
  {Paulin-Henriksson}, {Pires}, {Saxton}, {Schrabback}, {Seidel}, {Walsh},
  {Aghanim}, {Amendola}, {Bartlett}, {Baccigalupi}, {Beaulieu}, {Benabed},
  {Cuby}, {Elbaz}, {Fosalba}, {Gavazzi}, {Helmi}, {Hook}, {Irwin}, {Kneib},
  {Kunz}, {Mannucci}, {Moscardini}, {Tao}, {Teyssier}, {Weller}, {Zamorani},
  {Zapatero Osorio}, {Boulade}, {Foumond}, {Di Giorgio}, {Guttridge}, {James},
  {Kemp}, {Martignac}, {Spencer}, {Walton}, {Bl{\"u}mchen}, {Bonoli},
  {Bortoletto}, {Cerna}, {Corcione}, {Fabron}, {Jahnke}, {Ligori}, {Madrid},
  {Martin}, {Morgante}, {Pamplona}, {Prieto}, {Riva}, {Toledo}, {Trifoglio},
  {Zerbi}, {Abdalla}, {Douspis}, {Grenet}, {Borgani}, {Bouwens}, {Courbin},
  {Delouis}, {Dubath}, {Fontana}, {Frailis}, {Grazian}, {Koppenh{\"o}fer},
  {Mansutti}, {Melchior}, {Mignoli}, {Mohr}, {Neissner}, {Noddle}, {Poncet},
  {Scodeggio}, {Serrano}, {Shane}, {Starck}, {Surace}, {Taylor},
  {Verdoes-Kleijn}, {Vuerli}, {Williams}, {Zacchei}, {Altieri}, {Escudero
  Sanz}, {Kohley}, {Oosterbroek}, {Astier}, {Bacon}, {Bardelli}, {Baugh},
  {Bellagamba}, {Benoist}, {Bianchi}, {Biviano}, {Branchini}, {Carbone},
  {Cardone}, {Clements}, {Colombi}, {Conselice}, {Cresci}, {Deacon}, {Dunlop},
  {Fedeli}, {Fontanot}, {Franzetti}, {Giocoli}, {Garcia-Bellido}, {Gow},
  {Heavens}, {Hewett}, {Heymans}, {Holland}, {Huang}, {Ilbert}, {Joachimi},
  {Jennins}, {Kerins}, {Kiessling}, {Kirk}, {Kotak}, {Krause}, {Lahav}, {van
  Leeuwen}, {Lesgourgues}, {Lombardi}, {Magliocchetti}, {Maguire}, {Majerotto},
  {Maoli}, {Marulli}, {Maurogordato}, {McCracken}, {McLure}, {Melchiorri},
  {Merson}, {Moresco}, {Nonino}, {Norberg}, {Peacock}, {Pello}, {Penny},
  {Pettorino}, {Di Porto}, {Pozzetti}, {Quercellini}, {Radovich}, {Rassat},
  {Roche}, {Ronayette}, {Rossetti}, {Sartoris}, {Schneider}, {Semboloni},
  {Serjeant}, {Simpson}, {Skordis}, {Smadja}, {Smartt}, {Spano}, {Spiro},
  {Sullivan}, {Tilquin}, {Trotta}, {Verde}, {Wang}, {Williger}, {Zhao},
  {Zoubian}, \& {Zucca}}]{laureijs11}
{Laureijs}, R., {Amiaux}, J., {Arduini}, S., {et~al.} 2011, arXiv e-prints,
  arXiv:1110.3193.
\newblock \doarXiv{1110.3193}

\bibitem[{{Le F{\`e}vre} {et~al.}(2004){Le F{\`e}vre}, {Vettolani}, {Paltani},
  {Tresse}, {Zamorani}, {Le Brun}, {Moreau}, {Bottini}, {Maccagni}, {Picat},
  {Scaramella}, {Scodeggio}, {Zanichelli}, {Adami}, {Arnouts}, {Bardelli},
  {Bolzonella}, {Cappi}, {Charlot}, {Contini}, {Foucaud}, {Franzetti},
  {Garilli}, {Gavignaud}, {Guzzo}, {Ilbert}, {Iovino}, {McCracken}, {Mancini},
  {Marano}, {Marinoni}, {Mathez}, {Mazure}, {Meneux}, {Merighi}, {Pell{\`o}},
  {Pollo}, {Pozzetti}, {Radovich}, {Zucca}, {Arnaboldi}, {Bondi}, {Bongiorno},
  {Busarello}, {Ciliegi}, {Gregorini}, {Mellier}, {Merluzzi}, {Ripepi}, \&
  {Rizzo}}]{lefevre04}
{Le F{\`e}vre}, O., {Vettolani}, G., {Paltani}, S., {et~al.} 2004, \aap, 428,
  1043, \dodoi{10.1051/0004-6361:20048072}

\bibitem[{{Lotz} {et~al.}(2017){Lotz}, {Koekemoer}, {Coe}, {Grogin}, {Capak},
  {Mack}, {Anderson}, {Avila}, {Barker}, {Borncamp}, {Brammer}, {Durbin},
  {Gunning}, {Hilbert}, {Jenkner}, {Khandrika}, {Levay}, {Lucas}, {MacKenty},
  {Ogaz}, {Porterfield}, {Reid}, {Robberto}, {Royle}, {Smith},
  {Storrie-Lombardi}, {Sunnquist}, {Surace}, {Taylor}, {Williams}, {Bullock},
  {Dickinson}, {Finkelstein}, {Natarajan}, {Richard}, {Robertson}, {Tumlinson},
  {Zitrin}, {Flanagan}, {Sembach}, {Soifer}, \& {Mountain}}]{lotz17}
{Lotz}, J.~M., {Koekemoer}, A., {Coe}, D., {et~al.} 2017, ApJ, 837, 97,
  \dodoi{10.3847/1538-4357/837/1/97}

\bibitem[{{Madau}(1995)}]{madau95}
{Madau}, P. 1995, \apj, 441, 18, \dodoi{10.1086/175332}

\bibitem[{{Mahler} {et~al.}(2018){Mahler}, {Richard}, {Cl{\'e}ment},
  {Lagattuta}, {Schmidt}, {Patr{\'\i}cio}, {Soucail}, {Bacon}, {Pello},
  {Bouwens}, {Maseda}, {Martinez}, {Carollo}, {Inami}, {Leclercq}, \&
  {Wisotzki}}]{mahler18}
{Mahler}, G., {Richard}, J., {Cl{\'e}ment}, B., {et~al.} 2018, \mnras, 473,
  663, \dodoi{10.1093/mnras/stx1971}

\bibitem[{{Mandelbaum} {et~al.}(2014){Mandelbaum}, {Rowe}, {Bosch}, {Chang},
  {Courbin}, {Gill}, {Jarvis}, {Kannawadi}, {Kacprzak}, {Lackner}, {Leauthaud},
  {Miyatake}, {Nakajima}, {Rhodes}, {Simet}, {Zuntz}, {Armstrong}, {Bridle},
  {Coupon}, {Dietrich}, {Gentile}, {Heymans}, {Jurling}, {Kent}, {Kirkby},
  {Margala}, {Massey}, {Melchior}, {Peterson}, {Roodman}, \&
  {Schrabback}}]{2014ApJS..212....5M}
{Mandelbaum}, R., {Rowe}, B., {Bosch}, J., {et~al.} 2014, \apjs, 212, 5,
  \dodoi{10.1088/0067-0049/212/1/5}

\bibitem[{{Mandelbaum} {et~al.}(2015){Mandelbaum}, {Rowe}, {Armstrong}, {Bard},
  {Bertin}, {Bosch}, {Boutigny}, {Courbin}, {Dawson}, {Donnarumma}, {Fenech
  Conti}, {Gavazzi}, {Gentile}, {Gill}, {Hogg}, {Huff}, {Jee}, {Kacprzak},
  {Kilbinger}, {Kuntzer}, {Lang}, {Luo}, {March}, {Marshall}, {Meyers},
  {Miller}, {Miyatake}, {Nakajima}, {Ngol{\'e} Mboula}, {Nurbaeva}, {Okura},
  {Paulin-Henriksson}, {Rhodes}, {Schneider}, {Shan}, {Sheldon}, {Simet},
  {Starck}, {Sureau}, {Tewes}, {Zarb Adami}, {Zhang}, \&
  {Zuntz}}]{2015MNRAS.450.2963M}
{Mandelbaum}, R., {Rowe}, B., {Armstrong}, R., {et~al.} 2015, \mnras, 450,
  2963, \dodoi{10.1093/mnras/stv781}

\bibitem[{{Masters} {et~al.}(2017){Masters}, {Stern}, {Cohen}, {Capak},
  {Rhodes}, {Castander}, \& {Paltani}}]{masters17}
{Masters}, D.~C., {Stern}, D.~K., {Cohen}, J.~G., {et~al.} 2017, \apj, 841,
  111, \dodoi{10.3847/1538-4357/aa6f08}

\bibitem[{{McCully} {et~al.}(2014){McCully}, {Keeton}, {Wong}, \&
  {Zabludoff}}]{2014MNRAS.443.3631M}
{McCully}, C., {Keeton}, C.~R., {Wong}, K.~C., \& {Zabludoff}, A.~I. 2014,
  \mnras, 443, 3631, \dodoi{10.1093/mnras/stu1316}

\bibitem[{{McLeod} {et~al.}(2016){McLeod}, {McLure}, \&
  {Dunlop}}]{2016MNRAS.459.3812M}
{McLeod}, D.~J., {McLure}, R.~J., \& {Dunlop}, J.~S. 2016, \mnras, 459, 3812,
  \dodoi{10.1093/mnras/stw904}

\bibitem[{{Mehta} {et~al.}(2018){Mehta}, {Scarlata}, {Capak}, {Davidzon},
  {Faisst}, {Hsieh}, {Ilbert}, {Jarvis}, {Laigle}, {Phillips}, {Silverman},
  {Strauss}, {Tanaka}, {Bowler}, {Coupon}, {Foucaud}, {Hemmati}, {Masters},
  {McCracken}, {Mobasher}, {Ouchi}, {Shibuya}, \& {Wang}}]{mehta18}
{Mehta}, V., {Scarlata}, C., {Capak}, P., {et~al.} 2018, \apjs, 235, 36,
  \dodoi{10.3847/1538-4365/aab60c}

\bibitem[{{Merlin} {et~al.}(2016{\natexlab{a}}){Merlin}, {Bourne},
  {Castellano}, {Ferguson}, {Wang}, {Derriere}, {Dunlop}, {Elbaz}, \&
  {Fontana}}]{2016A&A...595A..97M}
{Merlin}, E., {Bourne}, N., {Castellano}, M., {et~al.} 2016{\natexlab{a}},
  \aap, 595, A97, \dodoi{10.1051/0004-6361/201628751}

\bibitem[{{Merlin} {et~al.}(2016{\natexlab{b}}){Merlin}, {Amor{\'\i}n},
  {Castellano}, {Fontana}, {Buitrago}, {Dunlop}, {Elbaz}, {Boucaud}, {Bourne},
  {Boutsia}, {Brammer}, {Bruce}, {Capak}, {Cappelluti}, {Ciesla}, {Comastri},
  {Cullen}, {Derriere}, {Faber}, {Ferguson}, {Giallongo}, {Grazian}, {Lotz},
  {Micha{\l}owski}, {Paris}, {Pentericci}, {Pilo}, {Santini}, {Schreiber},
  {Shu}, \& {Wang}}]{2016A&A...590A..30M}
{Merlin}, E., {Amor{\'\i}n}, R., {Castellano}, M., {et~al.} 2016{\natexlab{b}},
  \aap, 590, A30, \dodoi{10.1051/0004-6361/201527513}

\bibitem[{{Merten} {et~al.}(2011){Merten}, {Coe}, {Dupke}, {Massey}, {Zitrin},
  {Cypriano}, {Okabe}, {Frye}, {Braglia}, {Jim{\'e}nez-Teja}, {Ben{\'\i}tez},
  {Broadhurst}, {Rhodes}, {Meneghetti}, {Moustakas}, {Sodr{\'e}}, {Krick}, \&
  {Bregman}}]{2011MNRAS.417..333M}
{Merten}, J., {Coe}, D., {Dupke}, R., {et~al.} 2011, \mnras, 417, 333,
  \dodoi{10.1111/j.1365-2966.2011.19266.x}

\bibitem[{{Miralda-Escud{\'e}}(2003)}]{miralda-escude03}
{Miralda-Escud{\'e}}, J. 2003, Science, 300, 1904,
  \dodoi{10.1126/science.1085325}

\bibitem[{{Miyatake} {et~al.}(2014){Miyatake}, {Mandelbaum}, \&
  {Rowe}}]{2014JInst...9C4031M}
{Miyatake}, H., {Mandelbaum}, R., \& {Rowe}, B. 2014, Journal of
  Instrumentation, 9, C04031, \dodoi{10.1088/1748-0221/9/04/C04031}

\bibitem[{{Montes} \& {Trujillo}(2019)}]{montes19}
{Montes}, M., \& {Trujillo}, I. 2019, \mnras, 482, 2838,
  \dodoi{10.1093/mnras/sty2858}

\bibitem[{{Morishita} {et~al.}(2017){Morishita}, {Abramson}, {Treu}, {Schmidt},
  {Vulcani}, \& {Wang}}]{2017ApJ...846..139M}
{Morishita}, T., {Abramson}, L.~E., {Treu}, T., {et~al.} 2017, \apj, 846, 139,
  \dodoi{10.3847/1538-4357/aa8403}

\bibitem[{{Owers} {et~al.}(2011){Owers}, {Randall}, {Nulsen}, {Couch}, {David},
  \& {Kempner}}]{owers11}
{Owers}, M.~S., {Randall}, S.~W., {Nulsen}, P. E.~J., {et~al.} 2011, \apj, 728,
  27, \dodoi{10.1088/0004-637X/728/1/27}

\bibitem[{{Paranjape} {et~al.}(2018){Paranjape}, {Hahn}, \&
  {Sheth}}]{2018MNRAS.476.3631P}
{Paranjape}, A., {Hahn}, O., \& {Sheth}, R.~K. 2018, \mnras, 476, 3631,
  \dodoi{10.1093/mnras/sty496}

\bibitem[{{Peng} {et~al.}(2010){Peng}, {Ho}, {Impey}, \&
  {Rix}}]{2010AJ....139.2097P}
{Peng}, C.~Y., {Ho}, L.~C., {Impey}, C.~D., \& {Rix}, H.-W. 2010, \aj, 139,
  2097, \dodoi{10.1088/0004-6256/139/6/2097}

\bibitem[{Polletta {et~al.}(2007)Polletta, Tajer, Maraschi, Trinchieri,
  Lonsdale, Chiappetti, Andreon, Pierre, Le~F{\`e}vre, Zamorani, Maccagni,
  Garcet, Surdej, Franceschini, Alloin, Shupe, Surace, Fang, Rowan-Robinson,
  Smith, \& Tresse}]{polletta07}
Polletta, M., Tajer, M., Maraschi, L., {et~al.} 2007, \apj, 663, 81

\bibitem[{{Postman} {et~al.}(2012){Postman}, {Coe}, {Ben{\'\i}tez}, {Bradley},
  {Broadhurst}, {Donahue}, {Ford}, {Graur}, {Graves}, {Jouvel}, {Koekemoer},
  {Lemze}, {Medezinski}, {Molino}, {Moustakas}, {Ogaz}, {Riess}, {Rodney},
  {Rosati}, {Umetsu}, {Zheng}, {Zitrin}, {Bartelmann}, {Bouwens}, {Czakon},
  {Golwala}, {Host}, {Infante}, {Jha}, {Jimenez-Teja}, {Kelson}, {Lahav},
  {Lazkoz}, {Maoz}, {McCully}, {Melchior}, {Meneghetti}, {Merten}, {Moustakas},
  {Nonino}, {Patel}, {Reg{\"o}s}, {Sayers}, {Seitz}, \& {Van der
  Wel}}]{postman12}
{Postman}, M., {Coe}, D., {Ben{\'\i}tez}, N., {et~al.} 2012, \apjs, 199, 25,
  \dodoi{10.1088/0067-0049/199/2/25}

\bibitem[{Prevot {et~al.}(1984)Prevot, Lequeux, Prevot, Maurice, \&
  Rocca-Volmerange}]{prevot84}
Prevot, M.~L., Lequeux, J., Prevot, L., Maurice, E., \& Rocca-Volmerange, B.
  1984, Astronomy and Astrophysics (ISSN 0004-6361), 132, 389

\bibitem[{{Richard} {et~al.}(2014){Richard}, {Jauzac}, {Limousin}, {Jullo},
  {Cl{\'e}ment}, {Ebeling}, {Kneib}, {Atek}, {Natarajan}, {Egami}, {Livermore},
  \& {Bower}}]{richard14}
{Richard}, J., {Jauzac}, M., {Limousin}, M., {et~al.} 2014, \mnras, 444, 268,
  \dodoi{10.1093/mnras/stu1395}

\bibitem[{{Rix} {et~al.}(2004){Rix}, {Barden}, {Beckwith}, {Bell}, {Borch},
  {Caldwell}, {H{\"a}ussler}, {Jahnke}, {Jogee}, {McIntosh}, {Meisenheimer},
  {Peng}, {Sanchez}, {Somerville}, {Wisotzki}, \& {Wolf}}]{hanswalter04}
{Rix}, H.-W., {Barden}, M., {Beckwith}, S. V.~W., {et~al.} 2004, \apjs, 152,
  163, \dodoi{10.1086/420885}

\bibitem[{{Rowe} {et~al.}(2015){Rowe}, {Jarvis}, {Mandelbaum}, {Bernstein},
  {Bosch}, {Simet}, {Meyers}, {Kacprzak}, {Nakajima}, {Zuntz}, {Miyatake},
  {Dietrich}, {Armstrong}, {Melchior}, \& {Gill}}]{2015A&C....10..121R}
{Rowe}, B.~T.~P., {Jarvis}, M., {Mandelbaum}, R., {et~al.} 2015, Astronomy and
  Computing, 10, 121, \dodoi{10.1016/j.ascom.2015.02.002}

\bibitem[{{Saito} {et~al.}(2020){Saito}, {de la Torre}, {Ilbert}, {Dubois},
  {Yabe}, \& {Coupon}}]{shun20}
{Saito}, S., {de la Torre}, S., {Ilbert}, O., {et~al.} 2020, \mnras, 494, 199,
  \dodoi{10.1093/mnras/staa727}

\bibitem[{{Sampaio-Santos} {et~al.}(2020){Sampaio-Santos}, {Zhang}, {Ogando},
  {Shin}, {Golden-Marx}, {Yanny}, {Herner}, {Hilton}, {Choi}, {Gatti}, {Gruen},
  {Hoyle}, {Rau}, {De Vicente}, {Zuntz}, {Abbott}, {Aguena}, {Allam}, {Annis},
  {Avila}, {Bertin}, {Brooks}, {Burke}, {Carrasco Kind}, {Carretero}, {Chang},
  {Costanzi}, {da Costa}, {Diehl}, {Doel}, {Everett}, {Evrard}, {Flaugher},
  {Fosalba}, {Frieman}, {Garcia-Bellido}, {Gaztanaga}, {Gerdes}, {Gruendl},
  {Gschwend}, {Gutierrez}, {Hinton}, {Hollowood}, {Honscheid}, {James},
  {Jarvis}, {Jeltema}, {Kuehn}, {Kuropatkin}, {Lahav}, {Maia}, {March},
  {Marshall}, {Miquel}, {Palmese}, {Paz-Chinchon}, {Plazas}, {Sanchez},
  {Santiago}, {Scarpine}, {Schubnell}, {Smith}, {Suchyta}, {Tarle}, {Tucker},
  {Varga}, \& {Wechsler}}]{sampaio-santos20}
{Sampaio-Santos}, H., {Zhang}, Y., {Ogando}, R.~L.~C., {et~al.} 2020, arXiv
  e-prints, arXiv:2005.12275.
\newblock \doarXiv{2005.12275}

\bibitem[{{Schlafly} \& {Finkbeiner}(2011)}]{2011ApJ...737..103S}
{Schlafly}, E.~F., \& {Finkbeiner}, D.~P. 2011, \apj, 737, 103,
  \dodoi{10.1088/0004-637X/737/2/103}

\bibitem[{{Schmidt} {et~al.}(2014){Schmidt}, {Treu}, {Brammer}, {Brada{\v{c}}},
  {Wang}, {Dijkstra}, {Dressler}, {Fontana}, {Gavazzi}, {Henry}, {Hoag},
  {Jones}, {Kelly}, {Malkan}, {Mason}, {Pentericci}, {Poggianti}, {Stiavelli},
  {Trenti}, {von der Linden}, \& {Vulcani}}]{schmidt14}
{Schmidt}, K.~B., {Treu}, T., {Brammer}, G.~B., {et~al.} 2014, \apjl, 782, L36,
  \dodoi{10.1088/2041-8205/782/2/L36}

\bibitem[{{Schneider}(1984)}]{schneider84}
{Schneider}, P. 1984, \aap, 140, 119

\bibitem[{Scoville {et~al.}(2007)Scoville, Aussel, Brusa, Capak, Carollo,
  Elvis, Giavalisco, Guzzo, Hasinger, Impey, Kneib, Lefevre, Lilly, Mobasher,
  Renzini, Rich, Sanders, Schinnerer, Schminovich, Shopbell, Taniguchi, \&
  Tyson}]{Scoville:2007p12720}
Scoville, N., Aussel, H., Brusa, M., {et~al.} 2007, \apjs, 172, 1

\bibitem[{{Shipley} {et~al.}(2018){Shipley}, {Lange-Vagle}, {Marchesini},
  {Brammer}, {Ferrarese}, {Stefanon}, {Kado-Fong}, {Whitaker}, {Oesch},
  {Feinstein}, {Labb{\'e}}, {Lundgren}, {Martis}, {Muzzin}, {Nedkova},
  {Skelton}, \& {van der Wel}}]{2018ApJS..235...14S}
{Shipley}, H.~V., {Lange-Vagle}, D., {Marchesini}, D., {et~al.} 2018, \apjs,
  235, 14, \dodoi{10.3847/1538-4365/aaacce}

\bibitem[{{Spergel} {et~al.}(2015){Spergel}, {Gehrels}, {Baltay}, {Bennett},
  {Breckinridge}, {Donahue}, {Dressler}, {Gaudi}, {Greene}, {Guyon}, {Hirata},
  {Kalirai}, {Kasdin}, {Macintosh}, {Moos}, {Perlmutter}, {Postman},
  {Rauscher}, {Rhodes}, {Wang}, {Weinberg}, {Benford}, {Hudson}, {Jeong},
  {Mellier}, {Traub}, {Yamada}, {Capak}, {Colbert}, {Masters}, {Penny},
  {Savransky}, {Stern}, {Zimmerman}, {Barry}, {Bartusek}, {Carpenter}, {Cheng},
  {Content}, {Dekens}, {Demers}, {Grady}, {Jackson}, {Kuan}, {Kruk}, {Melton},
  {Nemati}, {Parvin}, {Poberezhskiy}, {Peddie}, {Ruffa}, {Wallace}, {Whipple},
  {Wollack}, \& {Zhao}}]{spergel15}
{Spergel}, D., {Gehrels}, N., {Baltay}, C., {et~al.} 2015, arXiv e-prints,
  arXiv:1503.03757.
\newblock \doarXiv{1503.03757}

\bibitem[{{Steinhardt} {et~al.}(2020){Steinhardt}, {Jauzac}, {Acebron}, {Atek},
  {Capak}, {Davidzon}, {Eckert}, {Harvey}, {Koekemoer}, {Lagos}, {Mahler},
  {Montes}, {Niemiec}, {Nonino}, {Oesch}, {Richard}, {Rodney}, {Schaller},
  {Sharon}, {Strolger}, {Allingham}, {Amara}, {Bah{\'e}}, {B{\oe}hm}, {Bose},
  {Bouwens}, {Bradley}, {Brammer}, {Broadhurst}, {Ca{\~n}as}, {Cen},
  {Cl{\'e}ment}, {Clowe}, {Coe}, {Connor}, {Darvish}, {Diego}, {Ebeling},
  {Edge}, {Egami}, {Ettori}, {Faisst}, {Frye}, {Furtak}, {G{\'o}mez-Guijarro},
  {Remolina Gonz{\'a}lez}, {Gonzalez}, {Graur}, {Gruen}, {Harvey}, {Hensley},
  {Hovis-Afflerbach}, {Jablonka}, {Jha}, {Jullo}, {Kneib}, {Kokorev},
  {Lagattuta}, {Limousin}, {von der Linden}, {Linzer}, {Lopez}, {Magdis},
  {Massey}, {Masters}, {Maturi}, {McCully}, {McGee}, {Meneghetti}, {Mobasher},
  {Moustakas}, {Murphy}, {Natarajan}, {Neyrinck}, {O'Connor}, {Oguri}, {Pagul},
  {Rhodes}, {Rich}, {Robertson}, {Sereno}, {Shan}, {Smith}, {Sneppen},
  {Squires}, {Tam}, {Tchernin}, {Toft}, {Umetsu}, {Weaver}, {van Weeren},
  {Williams}, {Wilson}, {Yan}, \& {Zitrin}}]{steinhardt20}
{Steinhardt}, C.~L., {Jauzac}, M., {Acebron}, A., {et~al.} 2020, \apjs, 247,
  64, \dodoi{10.3847/1538-4365/ab75ed}

\bibitem[{{Tomaney} \& {Crotts}(1996)}]{1996AJ....112.2872T}
{Tomaney}, A.~B., \& {Crotts}, A. P.~S. 1996, \aj, 112, 2872,
  \dodoi{10.1086/118228}

\bibitem[{{Treu} {et~al.}(2015){Treu}, {Schmidt}, {Brammer}, {Vulcani}, {Wang},
  {Brada{\v{c}}}, {Dijkstra}, {Dressler}, {Fontana}, {Gavazzi}, {Henry},
  {Hoag}, {Huang}, {Jones}, {Kelly}, {Malkan}, {Mason}, {Pentericci},
  {Poggianti}, {Stiavelli}, {Trenti}, \& {von der Linden}}]{treu15}
{Treu}, T., {Schmidt}, K.~B., {Brammer}, G.~B., {et~al.} 2015, \apj, 812, 114,
  \dodoi{10.1088/0004-637X/812/2/114}

\bibitem[{{Wang} {et~al.}(2019){Wang}, {Schreiber}, {Elbaz}, {Yoshimura},
  {Kohno}, {Shu}, {Yamaguchi}, {Pannella}, {Franco}, {Huang}, {Lim}, \&
  {Wang}}]{wang19-hstdark}
{Wang}, T., {Schreiber}, C., {Elbaz}, D., {et~al.} 2019, \nat, 572, 211,
  \dodoi{10.1038/s41586-019-1452-4}

\bibitem[{{Williams}(1995)}]{williams95}
{Williams}, R. 1995, {The Hubble Deep Field}, HST Proposal

\end{thebibliography}
\bibliographystyle{aasjournal}
\end{document}